\documentclass[12pt]{article}
\pdfoutput=1

\usepackage[a4paper,text={16.8cm,22.4cm}]{geometry}
\usepackage{amsmath,amsfonts,slashed,amssymb,tikz,bm,psfrag,graphicx,color,dsfont}
\usepackage{multicol}
\usepackage{float}

\RequirePackage[sort&compress,square,comma,numbers]{natbib}
\allowdisplaybreaks
\begin{document}

\begin{titlepage}

\begin{flushright}
\normalsize
 UWTHPH 2016-29 \\
May 11, 2017
\end{flushright}

\vspace{0.1cm}
\begin{center}
\Large\bf
Perturbative corrections to $B \to D$ form factors in QCD
\end{center}

\vspace{0.5cm}
\begin{center}
{\bf Yu-Ming Wang$^{a,b}$, Yan-Bing Wei$^{c,d}$, Yue-Long Shen$^{e}$, Cai-Dian L\"{u}$^{c,d}$} \\
\vspace{0.7cm}
{\sl   ${}^a$ \, School of Physics, Nankai University, 300071 Tianjin, China  \\
 ${}^b$ \,  Fakult\"{a}t f\"{u}r Physik, Universit\"{a}t Wien, Boltzmanngasse 5, 1090 Vienna, Austria \\
 ${}^c$ Institute of High Energy Physics, CAS, P.O. Box 918(4) Beijing 100049,  China \\
 ${}^d$ School of Physics, University of Chinese Academy of Sciences, Beijing 100049,  China \\
 ${}^e$ \, College of Information Science and Engineering,
Ocean University of China, Qingdao, Shandong 266100, P.R. China
}
\end{center}

\vspace{0.2cm}
\begin{abstract}

We compute  perturbative QCD corrections to $B \to D$ form factors at leading power in $\Lambda/m_b$,
at large hadronic recoil, from the light-cone sum rules (LCSR) with $B$-meson distribution amplitudes in HQET.
QCD factorization for the vacuum-to-$B$-meson correlation function
with an interpolating current for the $D$-meson is demonstrated explicitly at one loop with the power counting scheme
$m_c \sim {\cal O} \left  (  \sqrt{\Lambda \, m_b}   \right ) $.
The jet functions encoding information of the hard-collinear
dynamics in the above-mentioned correlation function are complicated by the appearance of an additional hard-collinear scale
$m_c$, compared to the counterparts  entering the factorization formula of
the vacuum-to-$B$-meson correction function for the construction of $B \to \pi$ from factors.
Inspecting the next-to-leading-logarithmic sum rules for the form factors of $B \to D \ell \nu$   indicates that
perturbative corrections to the hard-collinear functions are more profound than that for the hard functions,
with the default theory inputs, in the physical kinematic region.
We further compute the subleading power correction induced by the three-particle quark-gluon
distribution amplitudes of the $B$-meson at tree level employing the background gluon field approach.
The LCSR predictions for the semileptonic $B \to D \ell \nu$ form factors are
then extrapolated to the entire kinematic region with the $z$-series parametrization.
Phenomenological implications of our determinations for the form factors $f_{BD}^{+, 0}(q^2)$
are explored by investigating the (differential) branching fractions and the $R(D)$  ratio of $B \to D \ell \nu$
and by determining the CKM matrix element $|V_{cb}|$ from the total decay rate of $B \to D \mu \nu_{\mu}$.

\end{abstract}

\vfil

\end{titlepage}

\section{Introduction}
\label{sect:Intro}

Precision calculations of the semileptonic heavy-to-heavy $B \to D \ell \nu$ decays
are indispensable for a stringent test of the CKM matrix element $|V_{cb}|$ exclusively
and for a deep understanding of the strong interaction mechanism in the heavy-quark system
from both QCD and heavy-particle effective field theories.  The long-standing tension between
the exclusive and inclusive determinations of $|V_{cb}|$ \cite{Agashe:2014kda} as well as the topical
$R(D) \equiv {\cal BR}(B \to D \tau \nu_{\tau})/{\cal BR}(B \to D \ell \nu)$ anomaly \cite{Amhis:2016xyh}
necessitates further QCD calculations of $B \to D$ form factors with an increasing accuracy.
Employing the heavy-quark expansion (HQE) technique, systematic studies of $B \to D$ form factors near
zero recoil  were performed including both perturbative QCD corrections
and subleading power contributions (see \cite{Neubert:1993mb} for a review).
Very recently, the unquenched lattice-QCD calculations of $B \to D \ell \nu$ form factors
near zero recoil were reported by the FNAL/MILC Collaboration \cite{Lattice:2015rga} using the asqtad-improved fermions
for the light valence quarks and the improved Wilson (``clover") fermions for the heavy valence quarks
and by the HPQCD Collaboration \cite{Na:2015kha} independently  applying the NRQCD action for bottom and the highly improved
staggered quark action for charm quarks together with $N_f=2+1$ MILC gauge configurations.
However, extrapolating the HQE and lattice calculations of $B \to D$ form factors to the entire physical kinematic range
can be only achieved with certain parametrizations for the momentum dependence of form factors, which introduce an additional
source of theoretical uncertainties for the determination of $|V_{cb}|$.
A direct QCD computation of  the form factors of $B \to D \ell \nu$ at large recoil is therefore highly in demand
to provide complementary information on the hadronic dynamics for a better understanding of the form-factor shapes.
The major objective of this paper is to carry out perturbative QCD corrections to the two-particle contributions
to $B \to D$ form factors at large recoil from the light-cone sum rules (LCSR) with $B$-meson distribution amplitudes (DA)
following the techniques developed in \cite{Khodjamirian:2005ea,Khodjamirian:2006st,DeFazio:2005dx,DeFazio:2007hw},
in an attempt to extend the leading-order calculation of the corresponding vacuum-to-$B$-meson
correlation function performed in \cite{Faller:2008tr}.

Applying the light-cone operator-product expansion (OPE) and parton-hadron duality,
the LCSR approaches have been proven their usefulness in computing both the local and non-local
hadronic matrix elements  for the theory description of  hadronic processes with large momentum transfer.
Demonstrating QCD factorization for the vacuum-to-$B$-meson correlation function at leading power
in $\Lambda/m_b$ is evidently  the first step in the construction of sum rules for $B \to D \ell \nu$ form factors.
To this end, we need to establish the power counting scheme for the distinct momentum scales involved in
the correlation function under consideration. In contrast to the correlation function for the heavy-to-light
form factors, the appearance of an additional energy scale (the charm-quark mass) further complicates perturbative
QCD factorization for the vacuum-to-$B$-meson correlation function with an interpolating current for the $D$-meson.
Motivated by the mass hierarchy between the bottom and charm quarks numerically, we will adopt a novel power counting scheme
$m_c \sim {\cal O} \left  (  \sqrt{\Lambda \, m_b}   \right ) $, as implemented in the analysis of the inclusive semileptonic
$B \to X_c \ell \nu$ decays \cite{Boos:2005by,Boos:2005qx} with shape functions of the $B$-meson, instead of the popular
counting scheme  $m_c \sim {\cal O}(m_b)$ widely used in the heavy-quark physics
(see, for instance \cite{Gambino:2012rd,Beneke:2000ry}).  It is then apparent that the on-shell bottom-quark field,
the charm-quark field appeared in the interpolating current for the  $D$-meson  and the  external light-quark field
can be identified as hard, hard-collinear and soft modes in QCD following the convention of  \cite{Wang:2015vgv}.
Despite the fact that implementing the light-cone OPE of the above-mentioned correlation function with three distinct
momentum scales can be readily formulated in the framework of soft-collinear effective theory (SCET) including a missive
hard-collinear quark \cite{Rothstein:2003wh,Leibovich:2003jd}, we will employ an alternative strategy,
namely the method of regions \cite{Beneke:1997zp}, to compute the short-distance functions entering the QCD factorization formulae
for the considered correlation function, following closely \cite{Wang:2015vgv,Wang:2015ndk,Wang:2016qii}.

With the specified power counting scheme for the charm-quark field, the hard functions
from integrating out the dynamical fluctuations at the ${\cal O}(m_b)$ scale can be easily shown
to be  identical to the  perturbative matching coefficients of the QCD weak current
$\bar q \, \gamma_{\mu} \, (1-\gamma_5) \, b$ in SCET.
However, the jet functions involved in the factorization formulae for the  vacuum-to-$B$-meson correlation function
develop a non-trivial dependence on the charm-quark mass, yielding an interesting source of the
large-energy symmetry breaking effects for the vector and scalar $B  \to D$ form factors.
We will further verify explicitly at one loop that the jet functions for the vacuum-to-$B$-meson correlation function with
an interpolating current for the $D$-meson, in the $m_c \to 0$  limit, can be reduced to the corresponding hard-collinear functions
for the counterpart correlation function used in the construction of  the sum rules for $B \to \pi$ form factors.
QCD resummation for the parametrically large logarithms involved in the hard functions is achieved
at next-to-leading-logarithmic (NLL) accuracy with the standard renormalization-group (RG) approach.
In addition, we will evaluate the subleading power correction to the vacuum-to-$B$-meson correlation function from the three-particle
$B$-meson DA at leading order in $\alpha_s$, employing the light-cone OPE in the background field approach.

An alternative approach to compute the form factors of $B \to D \ell \nu$ from
QCD sum rules was adopted in \cite{Fu:2013wqa,Li:2009wq} with the non-perturbative strong interaction dynamics
encoded in the $D$-meson DA on the light cone. However, both calculations for the vacuum-to-$D$-meson correlation
function with an interpolating current for the $B$-meson presented in \cite{Fu:2013wqa,Li:2009wq} were carried out
at tree level without specifying the power counting scheme for the charm-quark field and without implementing
the perturbative QCD constraints for the $D$-meson DA. Yet another QCD-based approach to evaluate $B \to D$ form factors
in the framework of transverse-momentum-dependent (TMD) factorization was proposed in \cite{Li:1994zm,Kurimoto:2002sb}
with the power counting  $m_b \gg m_c \gg \Lambda$, which was further updated in \cite{Fan:2013qz} recently.
Albeit with the technical progresses on the computations of perturbative matching coefficients
\cite{Li:2010nn,Li:2012nk,Li:2012md,Li:2013xna},
TMD factorization for hard exclusive processes is still not well established  conceptually due to the lack of
a definite power counting scheme for the intrinsic transverse momentum \cite{Wang:2015qqr} and the Wilson-line structure of TMD wave functions
needs to be constructed appropriately to avoid both the rapidity and pinch singularities in the infrared subtraction \cite{Li:2014xda}.

This paper is structured as follows. In Section \ref{sect:the LCSR method} we will introduce the
vacuum-to-$B$-meson correlation function for constructing the sum rules of $B \to D$ form factors
and establish the power counting scheme for the external momenta in $\Lambda/m_b$.
The essential ingredients to compute the semileptonic $B \to D \ell \nu$ form factors from the
LCSR with $B$-meson DA, including QCD factorization for the considered correlation  function and
the hadronic dispersion relation, will be also presented by working out the tree-level sum rules
at leading power in  HQE. We will turn to demonstrate QCD factorization for the
two-particle contributions to the vacuum-to-$B$-meson correlation function at ${\cal O}(\alpha_s)$
in   Section \ref{sect:NLO two-particle correction},
where the hard coefficients and the jet functions appeared in the factorization formulae
for the above-mentioned correlation function are computed  manifestly  at one loop.
The (partial) NLL resummation improved sum rules for the two-particle contributions to  $B \to D$ form factors
will be also derived here and they constitute the main new results of this paper.
In  Section \ref{sect:three-particle correction} we further calculate the three-particle contributions
to the LCSR of $B \to D \ell \nu$ form factors at tree level, which will be shown to contribute only
at subleading power in $\Lambda/m_b$ with the power counting scheme adopted here.
Phenomenological implications of the newly derived sum rules for the decay form factors of $B \to D \ell \nu$
will be explored in  Section \ref{sect:numerical analysis} with different models for $B$-meson DA,
including determinations of the shape parameters for the vector and scalar $B \to D$ form factors
with both the $z$-series expansion \cite{Bourrely:2008za}
and the Caprini-Lellouch-Neubert (CLN) parametrization \cite{Caprini:1997mu},
the differential branching fractions of $B \to D \ell \nu$ and $R(D)$ defined in the above,
as well as the extraction of the CKM matrix element $|V_{cb}|$.
We will conclude in Section \ref{sect:Conc} with a summary of main observations and
a perspective on the future refinements. We collect some useful results of loop integrals
for evaluating the vacuum-to-$B$-meson correlation function at one loop,
present the spectral representations for the convolution integrals essential to construct
the (partial) NLL LCSR of  $B \to D$ form factors and display the lengthy coefficient functions
entering the ``effective" $B$-meson DA, absorbing the hard-collinear corrections at one loop,
in Appendices \ref{app:loop integrals}, \ref{app:spectral resp} and
\ref{app: the coefficient functions of spectral density}, respectively.

\section{The framework}
\label{sect:the LCSR method}

\begin{figure}
\begin{center}
\includegraphics[width=0.4 \columnwidth]{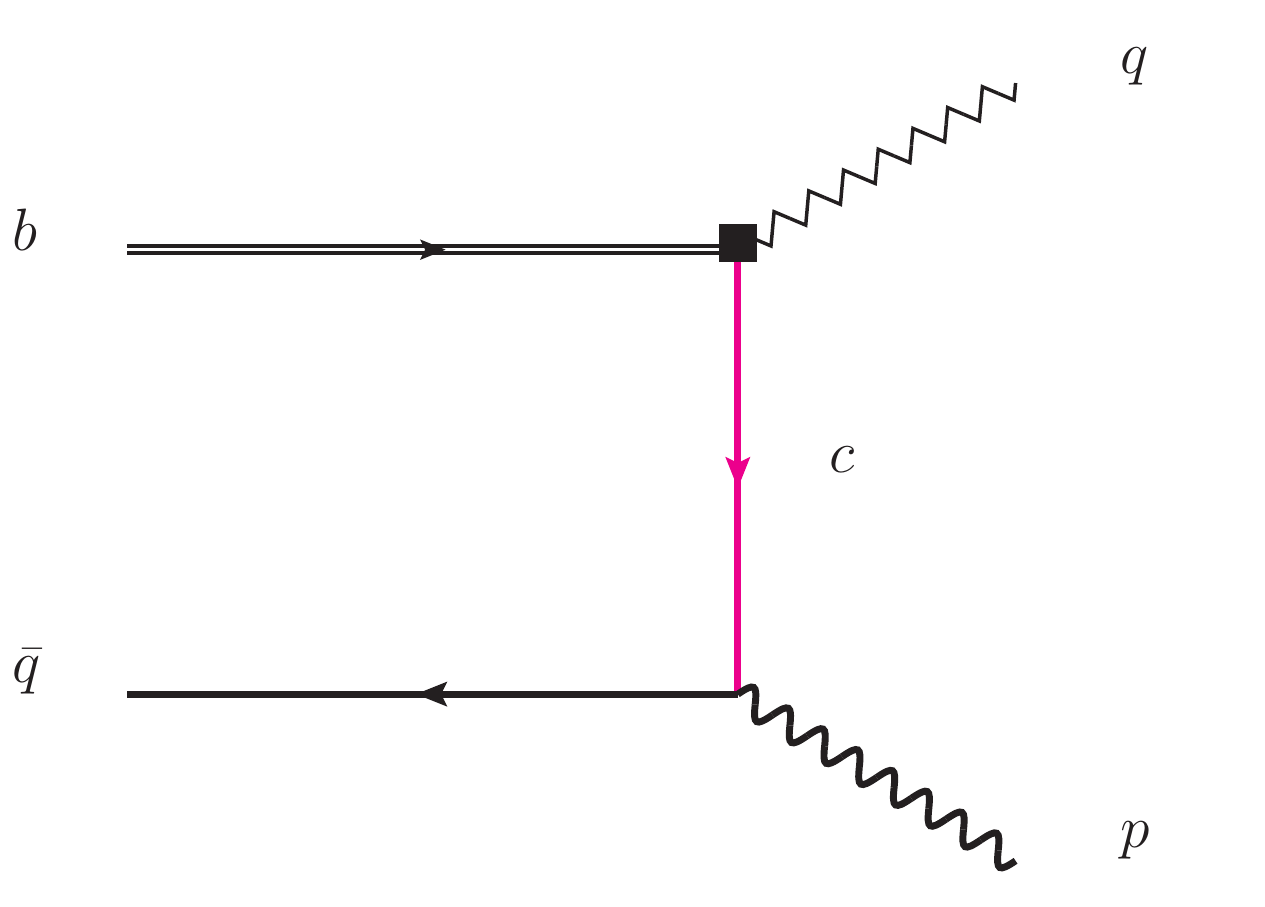}
\vspace*{0.1cm}
\caption{Diagrammatical representation of the two-particle contributions
to the vacuum-to-$B$-meson correlation function
$\Pi_{\mu}(n \cdot p,\bar n \cdot p)$ defined in (\ref{correlator: definition}) at tree level. }
\label{fig:tree-level_correlator}
\end{center}
\end{figure}

We briefly review the LCSR approach to compute $B \to D$ form factors with $B$-meson DA
following the strategy presented in \cite{Wang:2015vgv}. The vacuum-to-$B$-meson correlation
function adopted to construct the sum rules for the form factors $f_{B D}^{+}(q^2)$
and $f_{B D}^{0}(q^2)$  is defined as
\begin{eqnarray}
\Pi_{\mu}(n \cdot p,\bar n \cdot p)&=& i \, \int d^4x ~e^{i p\cdot x}
\langle 0 |T\left\{\bar{q}(x) \not n \, \gamma_5 \, c(x),
\bar{c}(0) \, \gamma_\mu  \, b(0) \right\}|\bar B(p_B) \rangle \,,
\label{correlator: definition}
\end{eqnarray}
where $p_B \equiv m_B \, v$ indicates the four-momentum of the $B$-meson and $p$
refers to the four-momentum carried by the interpolating current of the $D$-meson.
We will work in the rest frame of the $B$-meson and introduce two light-cone vectors
$n_{\mu}$ and $\bar n_{\mu}$ satisfying $n \cdot v = \bar n \cdot v = 1$ and $n \cdot \bar n=2$.
We will further employ the following power counting scheme
\begin{eqnarray}
n \cdot p =  \frac{m_B^2+m_{D}^2-q^2}{m_B} \sim  {\cal O} (m_B) \,,
\qquad | \bar n \cdot p | \sim {\cal O} (\Lambda) \,, \qquad
m_c \sim {\cal O} \left  (  \sqrt{\Lambda \, m_b}   \right )  \,,
\label{power counting scheme: charm quark}
\end{eqnarray}
at large hadronic recoil, where $q=p_B-p$ is the transfer momentum and
$\Lambda$ is a hadronic scale of order $\Lambda_{\rm QCD}$.
For the space-like interpolating momentum $p$, the leading power contribution to the correlation function
(\ref{correlator: definition}) at tree level can be achieved by evaluating the diagram in figure \ref{fig:tree-level_correlator}
 with the light-cone OPE and we obtain
\begin{eqnarray}
\Pi_{\mu, \, {\rm 2P}}(n \cdot p,\bar n \cdot p)&=& -i \, \tilde{f}_B(\mu) \, m_B \, \bar{n}_{\mu} \,
\int_0^{\infty} \, d \omega \, \frac{\phi_B^{-}(\omega, \mu)}{\bar n \cdot p - \omega - \omega_c + i \, 0}
\,\,  + {\cal O}(\alpha_s) \nonumber \\
&=&  -i \, \tilde{f}_B(\mu) \, m_B \, \bar{n}_{\mu} \,
\int_{\omega_c}^{\infty} \, d \omega^{\prime} \, \frac{\phi_B^{-}(\omega^{\prime}-\omega_c , \mu)}
{\bar n \cdot p - \omega^{\prime} + i \, 0} \,\,  + {\cal O}(\alpha_s) \,,
\label{tree-level result for the correlator}
\end{eqnarray}
where $\omega_c=m_c^2/ n \cdot p$ and the convolution integral in the second step corresponds to the spectral representation
of the considered correlation function. It is apparent that (\ref{tree-level result for the correlator}) reproduces
precisely the result for the corresponding correlation function defined with a pion interpolating current in the $m_c \to 0$ limit.

The light-cone DA of the $B$-meson in the coordinate space are defined as follows
\cite{Grozin:1996pq,Beneke:2000wa}:
\begin{eqnarray}
&& \langle  0 |\left ( \bar q \, Y_s\right )_{\beta}(t \, \bar{n}) \,
\left (Y_s^{\dag} b_v \right )_{\alpha}  (0) | \bar B(v)\rangle \nonumber \\
&& = - \frac{i \tilde f_B(\mu) \, m_B}{4}  \bigg \{ \frac{1+ \! \not v}{2} \,
\left [ 2 \, \tilde{\phi}_{B}^{+}(t, \mu) + \left ( \tilde{\phi}_{B}^{-}(t, \mu)
-\tilde{\phi}_{B}^{+}(t, \mu)  \right )  \! \not n \right ] \, \gamma_5 \bigg \}_{\alpha \beta}\,,
\label{def: 2P DA}
\end{eqnarray}
where $b_v$ indicates the effective bottom-quark field in heavy-quark effective theory (HQET),
the soft gauge link is given by
\begin{eqnarray}
Y_s(t \, \bar n)= {\rm P} \, \left \{ {\rm  Exp} \left [   i \, g_s \,
\int_{- \infty}^{t} \, dx \,  \bar n  \cdot A_{s}(x \, \bar n) \right ]  \right \} \,,
\label{def: soft gauge links}
\end{eqnarray}
and $\phi_B^{\pm}(\omega, \mu)$ are obtained from the Fourier transformation of $\tilde{\phi}_{B}^{\pm}(t, \mu)$.
The static $B$-meson decay constant $\tilde f_B(\mu)$ can be further expressed in terms of the
QCD decay constant 
\begin{eqnarray}
\tilde{f}_B(\mu)= \left \{  1 -  {\alpha_s(\mu) \, C_F \over 4 \, \pi} \,
\left [3\, \ln{\mu \over m_b} + 2  \right ] \right \}^{-1} \, f_B \,.
\label{hard matching of the fB}
\end{eqnarray}

Employing the standard definitions for the  decay constant of the $D$-meson and
for $B \to D$ transition form factors
\begin{eqnarray}
\langle D(p)|  \bar c \gamma_{\mu} b| \bar B (p_B)\rangle
&=& f_{B D}^{+}(q^2) \, \left [ 2 \, p_{\mu}  +\left ( 1 - \frac{m_B^2-m_{D}^2}{q^2} \right ) q_{\mu}  \right ]
+  f_{B D}^{0}(q^2) \, \frac{m_B^2-m_{D}^2}{q^2} q_{\mu} \,, \nonumber \\
\langle 0  |\bar q \! \not n \, \gamma_5 \, u |  D(p) \rangle &=&  i \, n \cdot p \, f_{D} \,,
\end{eqnarray}
it is straightforward to write down the hadronic dispersion relation for the correlation function
\begin{eqnarray}
\Pi_{\mu}(n \cdot p,\bar n \cdot p) &=&
\frac{i \, f_{D}  \, m_B}{2 \, (m_{D}^2/n \cdot p - \bar n \cdot p)}
\bigg \{  \bar n_{\mu} \, \left [ \frac{n \cdot p}{m_B} \, f_{B D}^{+} (q^2) + f_{B D}^{0} (q^2)  \right ]
\nonumber \\
&& \hspace{0.4 cm} +   n_{\mu} \, \frac{m_B}{n \cdot p-m_B}  \, \,
\left [ \frac{n \cdot p}{m_B} \, f_{B D}^{+} (q^2) -  f_{B D}^{0} (q^2)  \right ] \bigg \} \, \nonumber \\
&& \hspace{0.4 cm} + \int_{\omega_s}^{+\infty}  \, d \omega^{\prime} \, \frac{1}{\omega^{\prime} - \bar n \cdot p - i 0} \,
\left [ \rho^{h}(\omega^{\prime}, n \cdot p)  \, n_{\mu} \,
+ \tilde{\rho}^{h}(\omega^{\prime}, n \cdot p)  \, \bar{n}_{\mu}  \right ] \,,
\label{hadronic representation}
\end{eqnarray}
at leading power in $\Lambda/m_b$, where $\omega_s$ is the effective threshold parameter for the $D$-meson channel.
Applying the parton-hadron duality approximation for the hadronic dispersion integral and performing the Borel transformation
with respect to the variable $\bar n \cdot p$ lead to the LCSR for the two $B \to D$ form factors at tree level
\begin{eqnarray}
f_{B D, \, {\rm 2P}}^{+}(q^2) &=& \frac{\tilde f_B(\mu) \, m_B}{f_{D} \,n \cdot p} \,
{\rm Exp} \left [ {m_{D}^2 - m_c^2 \over n \cdot p \,\, \omega_M} \right ]
\int_0^{\omega_s-\omega_c} \, d \omega^{\prime} \, {\rm Exp} \left ( - \frac{\omega^{\prime}}{\omega_M} \right ) \,  \phi_B^{-}(\omega^{\prime}) + {\cal O}(\alpha_s) \,,
\nonumber \\
f_{B D, \, {\rm 2P}}^{0} (q^2) &=&  \frac{n \cdot p}{m_B} \, f_{B D, \, {\rm 2P}}^{+} (q^2) +  {\cal O}(\alpha_s) \,,
\label{B to D FFs at tree level}
\end{eqnarray}
where the upper limit of the $\omega^{\prime}$-integration  is consistent with $\omega_0(q^2, s_0^{D})$ derived in
\cite{Faller:2008tr} at leading power in $\Lambda/m_b$ keeping in mind the replacement rule $s_D \to n \cdot p \, \omega_s$.
In contrast to the sum rules for $B \to \pi$ form factors, the power counting rules of $\omega_c$,
$\omega_s$ and $\omega_M$ now read
\begin{eqnarray}
\omega_c \sim \omega_s \sim {\cal O} \left (\Lambda \right )  \,,  \qquad
\omega_s - \omega_c  \sim {\cal O} \left ( \Lambda^{3/2}  /  m_b^{1/2} \right )  \,, \qquad
\omega_M \sim {\cal O} \left ( \Lambda^{3/2}  /  m_b^{1/2} \right )  \,.
\label{power counting rule for sum rule parameters}
\end{eqnarray}
Based upon the canonical behaviour for the $B$-meson DA  $\phi_B^{-}(\omega^{\prime})$,
one can readily deduce the power counting of  $f_{B D}^{+}(q^2)$ at large hadronic recoil
as ${\cal O} \left ((\Lambda/m_b)^{3/2} \right )$ from the tree-level sum rules (\ref{B to D FFs at tree level}),
different from the scaling law $f_{B D}^{+}(q^2) \sim {\cal O}(1)$ \cite{Beneke:2000ry}
obtained with the counting scheme $m_c \sim m_b$.
In addition, the two  form factors $f_{B D}^{+,0} (q^2)$  obtained in the above  respect the large-energy symmetry relation
as discussed in \cite{Beneke:2000wa} and the symmetry breaking effects can arise from both perturbative QCD corrections
to the short-distance coefficient functions and the subleading power contributions from the higher-twist DA of the $B$-meson.

\section{Two-particle contributions to the LCSR at ${\cal O}(\alpha_s)$}
\label{sect:NLO two-particle correction}

The purpose of this section is to demonstrate QCD factorization for the two-particle contribution
to the vacuum-to-$B$-meson  correlation function (\ref{correlator: definition})
\begin{eqnarray}
\Pi_{\mu, \, {\rm 2P}}(n \cdot p,\bar n \cdot p) &=& - i \, \tilde f_B(\mu) \, m_B \,
\sum_{i= \pm} \, \int_0^{\infty} \,  \frac{d \omega}{\bar n \cdot p - \omega - \omega_c + i 0} \, \nonumber  \\
&&\hspace{-0.5 cm}  \bigg [ C_{i, n}(n \cdot p) \,   J_{i, n}(\bar n \cdot p, \omega)   \, n_{\mu}
+ C_{i, \bar n}(n \cdot p) \,   J_{i, \bar n}(\bar n \cdot p, \omega)  \, \bar n_{\mu}  \bigg ]
\,  \phi_B^{i}(\omega)  \,,
\label{master factorization formula}
\end{eqnarray}
at leading power in HQE.  We will compute the hard matching coefficients
$C_{i, k}$ ($i= \pm, \, k= n  \,\, {\rm or } \,\, \bar n$)
and the jet functions $J_{i, k}$ at next-to-leading order (NLO) in $\alpha_s$ manifestly 
and perform QCD resummation for the large logarithms in the hard functions
at NLL accuracy with the RG evolution in momentum space. Since the perturbative matching coefficients
must be independent of the external partonic states in the OPE calculation, we will choose the initial state
to be of the minimal quark and gluon degrees of freedom $| b(p_b) \, \bar q (k) \rangle$  for the convenience.

\subsection{Perturbative matching functions at NLO}

\begin{figure}
\begin{center}
\includegraphics[width=0.9 \columnwidth]{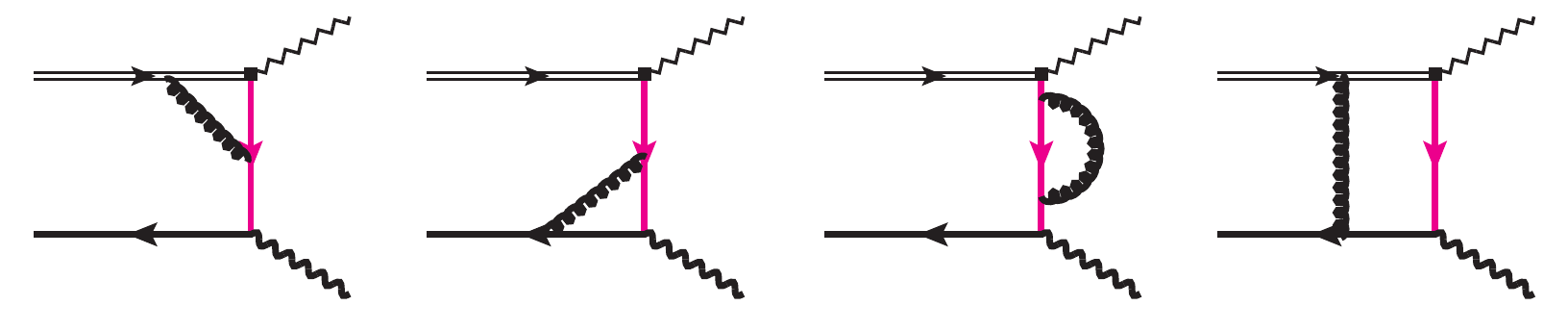} \\
(a) \hspace{3 cm} (b) \hspace{3 cm}  (c) \hspace{3 cm}   (d)
\vspace*{0.1cm}
\caption{Diagrammatical representation of the two-particle contribution to the vacuum-to-$B$-meson correlation function
$\Pi_{\mu}(n \cdot p,\bar n \cdot p)$ defined in (\ref{correlator: definition}) at ${\cal O} (\alpha_s)$. }
\label{fig:NLO_correlator}
\end{center}
\end{figure}

The NLO hard and jet functions entering the factorization formula (\ref{master factorization formula})
can be extracted by computing the leading power contributions of
the one-loop QCD diagrams displayed in figure \ref{fig:NLO_correlator} from the hard and hard-collinear regions.
It is evident that the soft contributions to the above-mentioned QCD diagrams at leading power in $\Lambda/m_b$
will be cancelled out precisely by the infrared subtraction terms following the presentation \cite{Wang:2015vgv}
and we will mainly focus on the hard and hard-collinear contributions of the one-loop diagrams
in figure \ref{fig:NLO_correlator} in the following.

\subsubsection{Weak vertex diagram}

We are now ready to compute the one-loop correction to the weak vertex diagram displayed
in figure \ref{fig:NLO_correlator} (a)
\begin{eqnarray}
\Pi_{\mu,  \, weak}^{(1)}
&=& \frac{i g_s^2 \, C_F}{(p-k)^2 -m_c^2 + i 0} \,
\int \frac{d^D \, l}{(2 \pi)^D} \,   \frac{1}{[(p-k+l)^2 -m_c^2 + i 0][(m_b v+l)^2 -m_b^2+ i 0] [l^2+i0]}  \nonumber  \\
&& \bar q(k)  \! \not n \,\gamma_5  \,\!  (\! \not p - \! \not k  + m_c) \,\, \gamma_{\rho}
\, (\! \not p - \! \not k  + \! \not l + m_c)
\, \gamma_{\mu} \, (m_b  \! \not v +  \! \not l+ m_b )\, \gamma^{\rho} \, b(v)  \,,
\label{diagram a: expression}
\end{eqnarray}
where the external bottom and light quarks are taken to be on the mass-sell with the momenta
$m_b \, v$ and $k$ (with $k^2=0$). Employing the power counting scheme for the external momenta
\begin{eqnarray}
n \cdot p \sim {\cal O}(m_b) \, \qquad \bar n \cdot p \sim {\cal O}(\Lambda)
\qquad  k \sim {\cal O}(\Lambda) \,,
\label{power counting scheme}
\end{eqnarray}
it is straightforward to identify that the leading power contributions to the scalar integral
\begin{eqnarray}
I_1 = \int \frac{d^D \, l}{(2 \pi)^D} \, \frac{1}{[(p-k+l)^2 -m_c^2 + i 0][(m_b v+l)^2 -m_b^2+ i 0] [l^2+i0]}
\end{eqnarray}
come from the hard, hard-collinear and soft regions of the loop momentum.
Evaluating the hard contribution from the weak vertex diagram yields
\begin{eqnarray}
\Pi_{\mu,  \, weak}^{(1), \, h} 
= - \frac{1}{\bar n \cdot p -\bar n \cdot k -\omega_c} \, \left [ \bar n_{\mu}  \,\,  C_{h, \bar n}^{(weak)}(n \cdot p)
+  n_{\mu}  \,\,  C_{h, n}^{(weak)}(n \cdot p) \right ] \,\,  \bar q(k)  \! \not n \,\gamma_5  \,\, b(v)  \,,
\label{diagram a: hard contribution}
\end{eqnarray}
where the two hard functions are given by
\begin{eqnarray}
 C_{h, \bar n}^{(weak)}(n \cdot p) &=& - \frac{\alpha_s \, C_F}{4 \, \pi} \, \bigg [ {1 \over \epsilon^2} +
{1 \over \epsilon} \, \left ( 2 \, \ln {\mu \over  n \cdot p} + 1  \right ) + 2 \, \ln^2 {\mu \over  n \cdot p}
+ 2 \,\ln {\mu \over  m_b}  \nonumber \\
&&  - 2 \, {\rm Li_2} \left (1 - {1 \over r} \right )
-\ln^2 r +{2-r \over r-1} \, \ln r +{\pi^2 \over 12} + 3  \bigg ] \,,  \nonumber \\
 C_{h,  n}^{(weak)}(n \cdot p) &=& - \frac{\alpha_s \, C_F}{4 \, \pi} \,
 \left [ {1 \over r-1} \, \left ( 1 +  {r \over 1-r}  \, \ln r  \right ) \right ]  \,,
\label{diagram a: hard function}
\end{eqnarray}
with $r=n \cdot p/m_b$. 

The hard-collinear contribution to the weak vertex diagram can be further computed by expanding
(\ref{diagram a: expression}), in the hard-collinear region, at leading power in $\Lambda/m_b$
\begin{eqnarray}
\Pi_{\mu,  \, weak}^{(1), \, hc}&=& \frac{i \, g_s^2 \, C_F}{\bar n \cdot p -\bar n \cdot k -\omega_c} \,
\int \frac{d^D \, l}{(2 \pi)^D} \,
\frac{2 \, n \cdot (p+l) \,\, \bar n_{\mu}}{[ n \cdot (p+l) \,
\bar n \cdot (p-k+l) + l_{\perp}^2  -m_c^2 + i 0][ n \cdot l+ i 0] [l^2+i0]}   \nonumber  \\
&&  \times   \bar d(k)  \! \not n \,\gamma_5  \,\, b(v)  \nonumber \\
&=& - \frac{1}{\bar n \cdot p -\bar n \cdot k -\omega_c} \,
\left [ \bar n_{\mu}  \,\,  J_{-, \bar n}^{(weak)}(\bar n \cdot p)
 \right ] \,\,  \bar q(k)  \! \not n \,\gamma_5  \,\, b(v)  \,.
\label{diagram a: hard-collinear contribution}
\end{eqnarray}
Applying the results of loop integrals presented in Appendix \ref{app:loop integrals} leads to
\begin{eqnarray}
J_{-, \bar n}^{(weak)}(\bar n \cdot p) &=& \frac{\alpha_s \, C_F}{4 \, \pi}
\bigg \{ {2 \over \epsilon^2} +
{2 \over \epsilon} \, \left [ \ln {\mu^2 \over  n \cdot p \, (\omega - \bar n \cdot p)}
- \ln \left (1+ r_1 \right ) + 1  \right ]  + \ln^2 {\mu^2 \over  n \cdot p \, (\omega - \bar n \cdot p)} \, \nonumber \\
&& + \, 2 \, \ln {\mu^2 \over  n \cdot p \, (\omega - \bar n \cdot p)}
- 2\, \ln (1+r_1) \, \left [ \ln {\mu^2 \over  n \cdot p \, (\omega - \bar n \cdot p)} + 1 \right ] + \ln^2 (1+r_1) \nonumber \\
&& + \, 2 \, r_1 \,  \ln \left ( {r_1 \over 1+ r_1} \right )
- \, 2 \, {\rm Li}_2  \left( {1 \over 1+ r_1}\right ) + {\pi^2 \over 6} + 4 \bigg \}  \,,
\label{diagram a: jet function}
\end{eqnarray}
with $r_1=m_c^2/ \left [n \cdot p \,\, \bar n \cdot (k -  p) \right ]$ and $\omega \equiv \bar n \cdot k$.
It is apparent that our result of $J_{-, \bar n}^{(weak)}$ reproduces the hard-collinear contribution to
the weak vertex diagram, displayed in (29) of \cite{Wang:2015vgv},
for constructing the sum rules of $B \to \pi$ form factors in the $m_c \to 0$ (i.e., $r_1 \to 0$) limit.

Proceeding in a similar manner,  we can extract the soft contribution to the weak vertex diagram
by expanding (\ref{diagram a: expression}) in the soft region
\begin{eqnarray}
\Pi_{\mu,  \, weak}^{(1), \, s}&=& \frac{i \, g_s^2 \, C_F}{\bar n \cdot p -\bar n \cdot k -\omega_c} \,
\int \frac{d^D \, l}{(2 \pi)^D} \,
\frac{ \bar n_{\mu}}{[\bar n \cdot (p-k+l) - \omega_c + i 0][v \cdot l + i 0] [l^2+i0]}     \nonumber  \\
&&  \times   \bar q(k)  \! \not n \,\gamma_5  \,\, b(v) \,,
\label{diagram a: soft contribution}
\end{eqnarray}
which is precisely the same as the soft subtraction term defined by the convolution integral of the
partonic DA of the $B$-meson  at NLO in $\alpha_s$, calculable from the Wilson-line Feynman rules,
and the tree-level short-distance function (see \cite{Wang:2015vgv} for more discussion).
We are then led to conclude that the soft QCD dynamics of the weak vertex diagram is indeed characterized
by the $B$-meson DA at leading power in $\Lambda/m_b$, independent of the renormalization scheme.

\subsubsection{$D$-meson vertex diagram}

We turn to compute the one-loop QCD correction to the $D$-meson vertex diagram displayed
 in figure \ref{fig:NLO_correlator}(b)
 \begin{eqnarray}
\Pi_{\mu,  \, D}^{(1)} &=& - \frac{i \, g_s^2 \, C_F}{n \cdot p \, (\bar n \cdot p -\omega -\omega_c)} \,
\int \frac{d^D \, l}{(2 \pi)^D} \,   \frac{1}{[(p-l)^2 -m_c^2 + i 0][(l-k)^2 + i 0] [l^2+i0]}  \nonumber  \\
&& \bar q(k)  \, \gamma_{\rho}  \, \! \not l \! \not n \,\gamma_5  \,\,  (\! \not p  - \! \not l + m_c)
\, \gamma^{\rho} \, (\! \not p -  \! \not k + m_c)\, \gamma_{\mu} \, b(v)  \,.
\label{diagram b: expression}
\end{eqnarray}
Inspecting the scalar integral and the Dirac algebra of (\ref{diagram b: expression}) with the power counting scheme
(\ref{power counting scheme}), one can verify that the leading power contributions to the $D$-meson vertex diagram
only arise from the hard-collinear and soft regions of the loop momentum.
As already discussed in \cite{Wang:2015vgv}, it turns out to be more apparent to compute the loop integrals directly
and then to expand the obtained expression up to the leading power in $\Lambda/m_b$, instead of applying the strategy
of regions.   Employing the results of loop integrals collected in Appendix \ref{app:loop integrals} and the light-cone
projector of the $B$-meson in momentum space derived in \cite{Beneke:2000wa,Bell:2013tfa} yields
 \begin{eqnarray}
\Pi_{\mu,  \, D}^{(1)} &=& \Pi_{\mu,  \, D}^{(1), \, hc} =
- \frac{i \, \tilde{f}_B(\mu) \, m_B}{\bar n \cdot p -\omega -\omega_c} \,
\bigg \{  \bar n_{\mu} \, \phi_B^{-}(\omega) \, J_{-, \bar n}^{(D)}(\bar n \cdot p)
+ n_{\mu} \, \phi_B^{+}(\omega) \, J_{+, n}^{(D)}(\bar n \cdot p)  \nonumber \\
&& - {m_c \over \bar n \cdot p}  \, \phi_B^{+}(\omega) \, \bar n_{\mu} \, J_{+, \bar n}^{(D)}(\bar n \cdot p)
- {2 \, m_c^2 \over p^2} \,
\left [ \bar n_{\mu} \, \phi_B^{-}(\omega) +  n_{\mu} \, \phi_B^{+}(\omega) \right ]
\, \delta J^{(D)}(\bar n \cdot p)  \bigg \}  \,,
\label{diagram b: result}
\end{eqnarray}
where we have introduced the following jet functions
\begin{eqnarray}
J_{-, \bar n}^{(D)}(\bar n \cdot p) &=& \frac{\alpha_s \, C_F}{4 \, \pi} \,
\bigg \{ \left [ {2 (1+r_2)  \over r_3} \, \ln \left ( {1 + r_2 + r_3 \over 1+r_2} \right )  - 1  \right ]
\left [ {1 \over \epsilon}  + \ln \left ( - {\mu^2 \over p^2} \right ) \right ]  \nonumber \\
&& - {2 (1+r_2)  \over r_3} \, \left [ {\rm Li}_2 \left ( {1 \over 1+r_2} \right ) -
{\rm Li}_2 \left ( {1+r_3 \over 1+r_2+r_3} \right )   \right ] + \ln \left ( 1 +r_2 \right )   \nonumber \\
&& - {1+r_2 \over r_3} \,\ln \left ( {1 + r_2 + r_3 \over 1+r_2} \right ) \,
\bigg [ \ln \left ( {1 + r_2 + r_3 \over 1+r_2} \right ) - \frac{2 (2+r_3)(1+r_2+r_3)}{(1+r_2)(1+r_3)} \nonumber \\
&& \hspace{0.5 cm} + \, 2 \, \ln \left ( 1+r_2 \right )  \bigg ]
+ \frac{r_2 \, \left[ r_2 (1+r_3) -2 \right ]} {1+r_3}  \, \, \ln \left ( {1+r_2 \over r_2} \right ) \,
- (4+r_2) \, \bigg \}  \,,  \\
J_{+, n}^{(D)}(\bar n \cdot p) &=& \frac{\alpha_s \, C_F}{4 \, \pi} \, \bigg \{ \frac{(1+r_3)^2-r_2^2}{r_3(1+r_3)} \,
\ln \left ( {1 + r_2 + r_3 \over 1+r_2} \right ) + \frac{r_2^2 \, (2+r_3)}{1+r_3} \,
\ln \left ( {1 + r_2  \over r_2} \right )  - r_2  \bigg \} \,,  \hspace{0.8 cm} \\
J_{+, \bar n}^{(D)}(\bar n \cdot p) &=& \frac{\alpha_s \, C_F}{4 \, \pi} \, \bigg \{ \frac{(1+r_2+r_3)^2}{r_3(1+r_3)^2}
\, \ln \left ( {1 + r_2 + r_3 \over 1+r_2} \right )  + \, {r_2 \over 1+ r_3}   \nonumber \\
&&  - \frac{r_2 \, \left [ 2(1+r_3) + r_2(2+r_3) \right ]}{(1+r_3)^2} \,\ln \left ( {1 + r_2  \over r_2} \right )
\bigg \} \,, \\
\delta J^{(D)}(\bar n \cdot p) &=& \frac{\alpha_s \, C_F}{4 \, \pi} \, \bigg \{
\frac{1+r_2+r_3}{r_3(1+r_3)} \, \ln \left (1+r_2+r_3 \right )
- {1+r_2 \over r_3} \, \ln \left ( 1+r_2 \right )  + {r_2 \over 1+r_3} \, \ln r_2  \bigg \}  \,,
\label{diagram b: jet functions}
\end{eqnarray}
with
\begin{eqnarray}
r_2 = - \frac{m_c^2}{p^2} \,, \qquad
r_3 = - \frac{\bar n \cdot k}{\bar n \cdot p}\,.
\label{def: r2 and r3}
\end{eqnarray}
Here, we have taken advantage of the fact that the soft contribution to the $D$-meson vertex diagram
vanishes in dimensional regularization (a similar observation already made in \cite{Wang:2015vgv}).

Several remarks on the resulting expressions for the QCD correction to the $D$-meson
vertex diagram presented in (\ref{diagram b: result}) are in order.
\begin{itemize}
\item{It is evident that the linear term in the charm-quark mass
corresponds to the power enhanced effect compared to the remaining contributions due to the power counting scheme
$m_c \sim {\cal O} \left  (  \sqrt{\Lambda \, m_b}   \right )$ and $\bar n \cdot p \sim {\cal O}(\Lambda)$.
This observation can be readily understood from the fact that the strange-quark mass effect
in $B \to K$ form factors is suppressed by $m_s/\Lambda$,
but not power of $m_s/m_b$,  compared to the leading power contribution in HQE \cite{Leibovich:2003jd},
and the charm-quark mass effect would naturally generate corrections to  $B \to D$ form factors  with a scaling factor
of $m_c/\Lambda \sim {\cal O} (\sqrt{m_b / \Lambda})$ in light of our power counting scheme.
(This clearly does not imply an expansion of $m_c/\Lambda$ in the QCD calculation of
the correlation function (\ref{correlator: definition}).)
To develop a better understanding of such power-enhancement mechanism, we inspect the linear charm-quark mass term
in (\ref{diagram b: expression}) before performing the loop integration
\begin{eqnarray}
\Pi_{\mu,  \, D}^{(1),\, m_c} &=& - \frac{i \, g_s^2 \, C_F}{\bar n \cdot p -\omega -\omega_c} \,
{m_c \over n \cdot p}  \,\, \bar n _{\mu} \,\, \bar q(k)  \,  \! \not {\bar n} \,\gamma_5  \, b(v) \nonumber \\
&&  \times \int \frac{d^D \, l}{(2 \pi)^D} \,
\frac{(D-2) \, (n \cdot l)^2}{[(p-l)^2 -m_c^2 + i 0][(l-k)^2 + i 0] [l^2+i0]} \,.
\end{eqnarray}
As the loop integral in the hard-collinear region induces a power-enhanced factor
${\cal O} \left (m_b / \Lambda \right )$, one can then identify the scaling behaviour of the charm-quark
mass term as  ${\cal O} (\sqrt{m_b / \Lambda})$ with respect to the tree-level contribution.
Furthermore, one  can  investigate the charm-quark mass effect directly in the context of
$B \to D$ form factors  applying the QCD factorization approach. A straightforward calculation of the
spectator interaction diagram with a hard-collinear gluon exchange between the energetic charm quark and the light
spectator quark yields
\begin{eqnarray}
\langle D(p)| \bar c \, \gamma_{\mu}  \, b | \bar B(p_B)\rangle^{\rm HSI}
\propto \frac{\alpha_s \, C_F}{N_c} \, {m_c \over n \cdot p} \,\bar n_{\mu} \,
\int_0^1 \, d u \, \frac{\phi_D(u)}{\bar u}   \, \int_0^{\infty} \, d \omega \, \frac{\phi_B^{+}(\omega)}{\omega^2}
+ ... ,
\label{B to D FF in QCDF}
\end{eqnarray}
where $\phi_D(u)$ is the twist-2 DA of the $D$-meson on the light cone.
It is evident that the second convolution integral in  (\ref{B to D FF in QCDF}) suffers from the end-point
divergence and the above-mentioned charm-quark mass effect should be identified as  the non-factorizable contribution
to $B \to D$ form factors.
}

\item{Since there is no power enhanced contribution proportional to the charm-quark mass at tree level,
the NLO jet function $J_{+, \bar n}^{(D)}(\bar n \cdot p)$ must be infrared finite to validate the factorization
formula (\ref{master factorization formula}) for the considered correlation function.}

\item {Setting $m_c \to 0$ (namely $r_2 \to 0$), our results of $J_{-, \bar n}^{(D)}(\bar n \cdot p)$
and $J_{+, n}^{(D)}(\bar n \cdot p)$ will be reduced to the corresponding hard-collinear contributions
from the pion vertex diagram as displayed in (38) of \cite{Wang:2015vgv}.}

\item{Applying the Wilson-line Feynman rules, one can easily verify that the soft contribution to the $D$-meson vertex diagram,
at leading power in $\Lambda/m_b$, is cancelled out precisely by the  corresponding soft subtraction term.}
\end{itemize}

\subsubsection{Wave function renormalization}

We proceed to compute the self-energy correction to the intermediate charm-quark propagator shown in
figure \ref{fig:NLO_correlator}(c). With the expressions of loop integrals collected in  Appendix
\ref{app:loop integrals}, we obtain
\begin{eqnarray}
\Pi_{\mu,  \, wfc}^{(1)}&=& - \frac{1}{\bar n \cdot p -\bar n \cdot k -\omega_c} \,
\left [ \bar n_{\mu}  \,\,  J_{-, \bar n}^{(wfc)}(\bar n \cdot p)
 \right ] \,\,  \bar q(k)  \! \not n \,\gamma_5  \,\, b(v)  \,,
\label{diagram c: hard-collinear contribution}
\end{eqnarray}
which is apparently free of soft and collinear divergences. The resulting jet function reads
\begin{eqnarray}
J_{-, \bar n}^{(wfc)}(\bar n \cdot p) &=& \frac{\alpha_s \, C_F}{4 \, \pi} \,
\left [  J_{-, \bar n}^{(wfc, \, 1)}(\bar n \cdot p)
 - {2 \, r_1 \over 1 + r_1} \,  J_{-, \bar n}^{(wfc, \, 2)}(\bar n \cdot p) \right ] \,, \\
 J_{-, \bar n}^{(wfc, \, 1)}(\bar n \cdot p) &=&  -\left \{ {1 \over \epsilon}
 + \ln \left (- {\mu^2 \over (p-k)^2}  \right  )  + r_1^2 \, \ln \left ( {1+ r_1 \over r_1} \right )
 - \ln \left ( 1+r_1 \right )  + 1 - r_1 \right \}  \,, \\
J_{-, \bar n}^{(wfc, \, 2)}(\bar n \cdot p) &=&  {3 \over \epsilon} + 3 \, \ln \left ( - {\mu^2 \over (p-k)^2}  \right  )
- r_1 \, (r_1 +4) \, \ln \left ( {1+ r_1 \over r_1} \right ) - 3 \, \ln \left ( 1+r_1 \right ) +  r_1 +5 \,, \hspace{0.8 cm}
\end{eqnarray}
where the ultraviolet divergence of  $J_{-, \bar n}^{(wfc, \, 2)}$ will be subtracted after the charm-quark mass renormalization
dependent on the subtraction scheme. We will employ the $\overline {\rm MS}$ renormalization scheme for the charm-loop mass, which is appropriate for a Lagrange parameter entering the short-distance matching function in the OPE calculation, and more discussion on the
renormalization schemes of the charm-quark mass can be found in \cite{Boos:2005qx}.

It is straightforward to compute the matching coefficients from the wave function renormalization of the external quarks
\begin{eqnarray}
\Pi_{\mu,  \, bwf}^{(1)} - \Phi_{b \bar q, bwf}^{(1)} \otimes T^{(0)}_{\mu}&=&
 - \frac{1}{\bar n \cdot p -\bar n \cdot k -\omega_c} \,
\left [ \bar n_{\mu}  \,\,  C_{-, \bar n}^{(bwf)}(n \cdot p)
 \right ] \,\,  \bar q(k)  \! \not n \,\gamma_5  \,\, b(v)  \,, \\
\Pi_{\mu,  \, qwf}^{(1)} - \Phi_{b \bar q, qwf}^{(1)} \otimes T^{(0)}_{\mu}&=& 0 \,,
\label{wf renormalization of external quarks}
\end{eqnarray}
where $\Phi_{b \bar q}$ indicates the partonic DA of the $B$-meson defined in (12) of \cite{Wang:2015vgv},
the tree-level hard kernel $ T^{(0)}_{\mu}$ can be readily deduced from  (\ref{tree-level result for the correlator})
and the explicit expression of $C_{-, \bar n}^{(bwf)}$ is given by
\begin{eqnarray}
C_{-, \bar n}^{(bwf)}(n \cdot p) &=& - \frac{\alpha_s \, C_F}{8 \, \pi} \,
\bigg [{3 \over \epsilon} + 3 \, \ln {\mu^2 \over m_b^2} + 4 \bigg ] \,.
\end{eqnarray}

\subsubsection{Box diagram}

The one-loop QCD correction to the box diagram displayed in figure \ref{fig:NLO_correlator}(d)
can be further computed as
\begin{eqnarray}
\Pi_{\mu,  \, box}^{(1)}
&=& - i \, g_s^2 \, C_F \,
\int \frac{d^D \, l}{(2 \pi)^D} \,   \frac{1}{[(p_b +l)^2 -m_b^2+ i 0][(p-k+l)^2 - m_c^2 + i 0] [(k-l)^2+i0][l^2+i0]}  \nonumber  \\
&& \bar q(k)  \, \gamma_{\rho}  \,  \,\,  (\! \not k - \! \not l)
\! \not n \,\gamma_5  \,\,  (\! \not p - \! \not k  + \! \not l + m_c) \, \gamma_{\mu} \,\!
\, (\! \not p_b +  \! \not l+ m_b )\, \gamma^{\rho} \, b(v)  \,.
\label{diagram d: expression}
\end{eqnarray}
Taking advantage of the power counting scheme (\ref{power counting scheme}), one can identify that
the leading power contributions to the box diagram come from the hard-collinear and soft regions
of the loop momentum. Expanding the QCD expression of  the box diagram (\ref{diagram d: expression})
in the hard-collinear region to the leading power in $\Lambda/m_b$  yields
\begin{eqnarray}
\Pi_{\mu,  \, box}^{(1), \, hc} &=& - \frac{i \, g_s^2 \, C_F }{m_b}
\,  \int \frac{d^D \, l}{(2 \pi)^D} \,\,
 \bar q(k)  \, \left [ (D-2) \, n \cdot l \, \! \not \bar n - 2 \, m_b \,  \! \not  n \right ] \, \gamma_5 \, b(v) \nonumber \\
&&  \,  \times \frac{n \cdot (p+l) \,\, \bar n_{\mu}}{[ n \cdot (p+l) \, \bar n \cdot (p-k+l) + l_{\perp}^2 - m_c^2 + i 0]
[n \cdot l \, \bar n (l-k)+ l_{\perp}^2  + i 0][l^2+i0]} \,.
\label{diagram d: hard-collinear region}
\end{eqnarray}
Employing the results of loop integrals collected in Appendix \ref{app:loop integrals}
and the momentum-space projector of the $B$-meson leads to
 \begin{eqnarray}
\Pi_{\mu,  \, box}^{(1), hc} &=& - \frac{i \, \tilde{f}_B(\mu) \, m_B}{\bar n \cdot p -\omega -\omega_c} \,
\bar n_{\mu} \, \left  \{   \phi_B^{-}(\omega) \, J_{-, \bar n}^{(box)}(\bar n \cdot p)
+ \phi_B^{+}(\omega) \, J_{+, \bar n}^{(box)}(\bar n \cdot p)  \right \}  \,,  \\
J_{-, \bar n}^{(box)} &=& \frac{\alpha_s \, C_F}{2 \, \pi} \, {(1+r_1) (1+r_3) \over r_3} \,
\bigg \{ \ln \left (1- {r_4 \over 1 + r_1} \right ) \bigg [ {1 \over \epsilon} +
\ln \left ( {\mu^2 \over n \cdot p \, (\omega - \bar n \cdot p)} \right ) \nonumber \\
&& - {1 \over 2} \, \ln \left (1- {r_4 \over 1+r_1} \right ) - \ln \left ( 1+r_1 \right )
+ 1 + {r_1 \over 1- r_4}\bigg ]  + {\rm Li}_2 \left (1- {r_1 \over 1+r_1-r_4} \right )  \nonumber \\
&& - {\rm Li}_2 \left ({1 \over 1+r_1} \right )
- {r_1 r_4 \over 1-r_4} \, \ln \left ({r_1 \over 1+r_1} \right )\bigg \} \,,  \\
J_{+, \bar n}^{(box)} &=&  \frac{\alpha_s \, C_F}{4 \, \pi} \, {(1+r_1) (1+r_3) \over r_3} \, r
\, \bigg \{  \left ( {r_1^2 \over (1-r_4)^2}  -1 \right ) \, \ln \left (1-r_4+r_1 \right )
- {r_1 \, r_4 \over 1- r_4} \nonumber \\
&& + \, (1-r_1^2) \, \ln \left (1+r_1 \right ) - {r_1^2 \, r_4 \, (2-r_4) \over (1-r_4)^2} \, \ln r_1 \bigg \} \,,
\label{diagram d: result}
\end{eqnarray}
where $r_4=r_3/(1+r_3)$ with $r_3$ defined in (\ref{def: r2 and r3}).
One can readily verify that the resulting jet functions $J_{-, \bar n}^{(box)}$
and $J_{+, \bar n}^{(box)}$ reproduce the hard-collinear contribution to the one-loop box diagram
for the vacuum-to-$B$-meson correlation function with a pion interpolating current as presented in (52)
of \cite{Wang:2015vgv}, in the $m_c \to 0$ limit.

The  soft contribution to the one-loop box diagram at leading power in $\Lambda/m_b$ can be further
extracted from (\ref{diagram d: expression}) as follows
\begin{eqnarray}
\Pi_{\mu,  \, box}^{(1), \,s}
&=& - \frac{g_s^2 \, C_F}{2}\,
\int \frac{d^D \, l}{(2 \pi)^D} \,   \frac{1}{[ v \cdot l+ i 0]
[\bar n \cdot (p-k+l) -\omega_c + i 0] [(k-l)^2+i0][l^2+i0]}  \nonumber  \\
&& \bar q(k)  \, \! \not v  \,  (\! \not k - \! \not l)  \,\,
\! \not n \,\gamma_5  \,\,  \! \not \bar n \,\, \gamma_{\mu} \, b(v)  \,,
\label{diagram d: soft region}
\end{eqnarray}
which is precisely the same as the soft subtraction term  $\Phi_{b \bar q, box}^{(1)} \otimes T^{(0)}_{\mu}$
computed with the Wilson-line Feynman rules.
We then conclude that the soft dynamics of the vacuum-to-$B$-meson correlation function under discussion
is indeed parameterized by the light-cone DA of the $B$-meson in HQET.

\subsubsection{The NLL hard and jet functions}

Now we are in a position to present the one-loop hard and jet functions entering the factorization formula
({\ref{master factorization formula}}) with resummation of the large logarithms by solving the
corresponding  RG equations in momentum space.
Putting different pieces together, we first derive the renormalized hard coefficients including
the ${\cal O}(\alpha_s)$ corrections
\begin{eqnarray}
C_{+,n} &=& C_{+,\bar n}=1 \,, \qquad
C_{-,n}=C_{h,  n}^{(weak)} = - \frac{\alpha_s \, C_F}{4 \, \pi} \,
 \left [ {1 \over r-1} \, \left ( 1 +  {r \over 1-r}  \, \ln r  \right ) \right ]  \,,  \nonumber \\
 C_{-,\bar n} &=&  1 + C_{h,  \bar n}^{(weak)} +
\left [ \Pi_{\mu,  \, bwf}^{(1)} - \Phi_{b \bar q, bwf}^{(1)} \otimes T^{(0)}_{\mu}  \right ] \nonumber \\
&=& 1 - \frac{\alpha_s \, C_F}{4 \, \pi}\,  \bigg [ 2 \, \ln^2 {\mu \over n \cdot p}
+ 5 \, \ln {\mu \over m_b} - 2 \, {\rm Li_2} \left (1 - {1 \over r} \right )  - \ln^2 r
+ {2-r \over r-1} \, \ln r  \nonumber \\
&&  +  {\pi^2 \over 12} + 5 \bigg ] \,,
\label{NLO hard functions}
\end{eqnarray}
and  the renormalized jet functions at one-loop accuracy are given by
\begin{eqnarray}
J_{+,n} &=& J_{+, n}^{(D)} - {2 \, m_c^2 \over p^2} \, \delta J^{(D)} \nonumber \\
&=& \frac{\alpha_s \, C_F}{4 \, \pi} \, \bigg \{ \frac{(1+r_2+r_3)^2}{r_3(1+r_3)} \,
\ln \left ({1 + r_2 + r_3 \over 1 + r_2} \right )
+ {r_2 \over 1+r_3} \, \left [ r_2 r_3 \,  \ln \left ({1 + r_2  \over r_2} \right )
- r_3 -1 \right ] \bigg \} \,, \hspace{1 cm} \\
J_{+,\bar n} &=& - {m_c \over \bar n \cdot p} \, J_{+, \bar n}^{(D)} +  J_{+, \bar n}^{(box)} \nonumber \\
&=& \frac{\alpha_s \, C_F}{4 \, \pi} \, \bigg \{ - {m_c \over \bar n \cdot p} \,
\bigg [ \frac{(1+r_2+r_3)^2}{r_3(1+r_3)^2} \, \ln \left ( {1 + r_2 + r_3 \over 1+r_2} \right )
+ \, {r_2 \over 1+ r_3}
\nonumber \\
&& - \frac{r_2 \, \left [ 2(1+r_3) + r_2(2+r_3) \right ]}{(1+r_3)^2} \,
\ln \left ( {1 + r_2  \over r_2} \right )    \bigg ]  + \frac{r \, (1+r_2+r_3)}{(1+r_3)^2}  \, \nonumber \\
&&  \times  \bigg [  \frac{(1+r_3)^2-r_2^2}{r_3} \, \ln \left ( {1+r_2+r_3 \over 1+r_2} \right )
 + r_2^2 (r_3+2) \, \ln \left ( {1+r_2 \over r_2} \right ) - r_2 (1+r_3) \bigg ] \bigg \}  \,, \\
 J_{-,  n} &=& 1 \,, \\
 J_{-,  \bar n} &=& 1 +  \left [ J_{-, \bar n}^{(weak)} +  J_{-, \bar n}^{(D)} - {2 \, m_c^2 \over p^2}
 \, \delta J^{(D)} +  J_{-, \bar n}^{(wfc)} + J_{-, \bar n}^{(box)} \right ] \,  \nonumber \\
&=& 1 + \frac{\alpha_s \, C_F}{4 \, \pi} \,
\bigg \{  \ln^2 \left ( {\mu^2 \over  n \cdot p \, (\omega - \bar n \cdot p)}  \right )
 - 2 \,  \left [ \ln \left ( {(1+r_2+r_3)^2 \over (1+r_2)(1+r_3)} \right ) + {3 r_2 \over 1+ r_2+r_3 } \right ] \nonumber \\
&& \times  \ln \left ( {\mu^2 \over  n \cdot p \, (\omega - \bar n \cdot p)}  \right )
+ 2\, \ln^2 \left (1+r_2+r_3 \right )
- 4\, \ln  \left (1+r_2+r_3 \right )  \, \ln \left (1+ r_3 \right ) \nonumber \\
&& + \ln^2 \left (1+ r_3 \right ) + \left [ {2 (1+r_2) \over r_3}
+ \frac{r_2 \left (r_2 + 2 (1+r_3) \right )}{(1+r_3)^2} - 1 \right ]
\, \ln  \left (1+r_2+r_3 \right ) \nonumber \\
&& + 2 \, \left [ \ln \left (1+ r_2 \right ) - {3 r_2 \over 1+r_2+r_3}\right ] \, \ln \left (1+ r_3 \right )
+  {(1+r_2)(r_3(1+r_2) -2) \over r_3} \,\ln \left (1+ r_2 \right )  \nonumber \\
&& - \ln^2 \left (1+ r_2 \right ) + r_2 \, \left [ {6 \over 1+ r_2+r_3}
- {r_2 \over (1+r_3)^2 } - {2 \over 1+r_3} - r_2 - 2 \right ] \, \ln r_2  + {\pi^2 \over 6} -1  \nonumber \\
&&  -4 \, {\rm Li}_2 \left ( {1+r_3 \over 1+r_2+r_3} \right )
+ 2\,  {\rm Li}_2 \left ( {1 \over 1+r_2} \right )
- r_2 \left [{8 \over 1 + r_2+r_3} + {1 \over 1+r_3}  + 1  \right ]  \bigg \}   \,.
\label{NLO jet functions}
\end{eqnarray}
It is apparent that the hard functions $C_{-,n}$ and $C_{-,\bar n}$ are
identical to the perturbative matching coefficients of the weak current
$\bar q \, \gamma_{\mu} \, (1-\gamma_5) \, b$  from QCD onto SCET,
as discussed in \cite{Wang:2015vgv}, due to the power counting rule of the charm-quark mass
displayed in (\ref{power counting scheme: charm quark}).
However,  the hard-collinear functions entering the factorization formula
(\ref{master factorization formula}) turn out to be significantly more involved, due to the massive
charm-quark effect, than the counterpart jet functions for the massless hard-collinear quark.

To verify the factorization-scale independence of the correlation function $\Pi_{\mu}$ at
${\cal O}(\alpha_s)$, we make use of the RG evolution equation of the charm-quark mass
\cite{Chetyrkin:1997dh,Vermaseren:1997fq}
\begin{eqnarray}
\frac{d \, \ln m_c(\mu)}{d \ln \mu} = -  \sum_{n=0} \, \left ({\alpha_s(\mu) \over 4 \pi} \right )^{n+1} \,
\gamma_m^{(n)} \,, \qquad \gamma_m^{(0)}= 6 \, C_F  \,,
\end{eqnarray}
it is then straightforward to write down
\begin{eqnarray}
&& \frac{d \,}{d \, \ln \mu} \,\Pi_{\mu, \, {\rm 2P}}(n \cdot p, \bar n \cdot p) \,
= - i \, \tilde{f}_B(\mu) \, m_B \, \frac{\alpha_s \, C_F}{4 \, \pi} \,
\int_0^{\infty} \, d \omega \,
\frac{\phi_B^{-}(\omega, \mu)}{\bar n \cdot p - \omega - \omega_c - i 0} \, \nonumber \\
&&  \times \bigg \{  {12 \, r_2 \over 1 + r_2 +r_3}
+ 4 \,  \left [ \ln \left ( {\mu^2 \over  n \cdot p \, (\omega - \bar n \cdot p)}  \right )
- \ln \left ( {(1+r_2+r_3)^2 \over (1+r_2)(1+r_3)} \right ) - {3 r_2 \over 1+ r_2+r_3 } \right ]   \nonumber \\
&& - \left [ 4 \, \ln \left ( {\mu \over n \cdot p} \right ) + 5 \right ]  \bigg \}
- i \, m_B \, \int_0^{\infty} \,
\frac{d \omega}{\bar n \cdot p - \omega - \omega_c - i 0} \, {d \over d \ln \mu} \,
\left [ \tilde{f}_B(\mu) \,  \phi_B^{-}(\omega, \mu) \right ] \,,
\label{scale independence of the correlator}
\end{eqnarray}
where the three terms in the bracket arise from the RG running of the charm-quark mass,
the jet function $ J_{-,  \bar n}$ and the hard function $ C_{-,\bar n}$, respectively.
Applying the one-loop evolution equation of the $B$-meson DA $\phi_B^{-}(\omega, \mu)$
in the absence of the light-quark mass effect 
\begin{eqnarray}
{d \over d \, \ln \mu} \, \left [  \tilde{f}_B(\mu) \,  \phi_B^{-}(\omega, \mu) \right ]
= - \frac{\alpha_s \, C_F}{4 \, \pi} \, \int_0^{\infty} \, d \omega^{\prime} \,
{\cal H}_{-}^{(1)}(\omega, \omega^{\prime}, \mu) \, \left [  \tilde{f}_B(\mu) \,  \phi_B^{-}(\omega^{\prime}, \mu) \right ]\,,
\end{eqnarray}
where the renormalization kernel ${\cal H}_{-}^{(1)}(\omega, \omega^{\prime}, \mu)$ can be found
 in \cite{Bell:2008er,DescotesGenon:2009hk},
we can readily compute the last term of (\ref{scale independence of the correlator}) as
\begin{eqnarray}
&& - i \, m_B \, \int_0^{\infty} \,
\frac{d \omega}{\bar n \cdot p - \omega - \omega_c - i 0} \, {d \over d \ln \mu} \,
\left [ \tilde{f}_B(\mu) \,  \phi_B^{-}(\omega, \mu) \right ] \nonumber \\
&& = i \, \tilde{f}_B(\mu) \, m_B \, \frac{\alpha_s \, C_F}{4 \, \pi} \,
\int_0^{\infty} \, d \omega \,
\frac{\phi_B^{-}(\omega, \mu)}{\bar n \cdot p - \omega - \omega_c - i 0} \,
 \left [ 4 \, \ln \left ( {\mu \over \omega} \right )
- 4 \, \ln \left ( {(1+r_2+r_3)^2 \over (1+r_2) r_3} \right )
- 5  \right ]  \,. \nonumber \\
\label{evolution effect of phiBminus}
\end{eqnarray}
Substituting (\ref{evolution effect of phiBminus}) into (\ref{scale independence of the correlator})
immediately leads to
\begin{eqnarray}
\frac{d \,}{d \, \ln \mu} \,\Pi_{\mu, \, {\rm 2P}}(n \cdot p, \bar n \cdot p) = {\cal O}(\alpha_s^2) \,,
\end{eqnarray}
indicating that the correlation function computed from the factorization formula
(\ref{master factorization formula}) is indeed independent of the renormalization scale
to the one-loop accuracy.

We will proceed to perform the summation of parametrically large logarithms in the perturbative expansion
of the hard matching coefficients at NLL by applying the corresponding RG equations in momentum space.
Following the arguments of \cite{Beneke:2011nf}, we will not aim at summing over logarithms
of $\mu/\mu_0$, with $\mu_0$ being a hadonic scale of ${\cal O}(\Lambda)$, from the RG evolution
of the $B$-meson DA  $\phi_B^{-}(\omega, \mu)$, since we will take the
factorization scale $\mu$ as a hard-collinear scale $\mu_{hc} \sim {\cal O} \left (\sqrt{\Lambda \, m_b} \right )$
which is  quite close to the hadronic scale $\mu_0$ numerically.
Employing the RG equations for the hard coefficient $ C_{-,\bar n}(n \cdot p, \mu)$
and the static $B$-meson decay constant $\tilde{f}_B(\mu)$
\begin{eqnarray}
\frac{d C_{-,\bar n}(n \cdot p, \mu)}{d \ln \mu} \,
&=& \left [ - \Gamma_{\rm {cusp}}(\alpha_s) \, \ln{\mu \over n \cdot p}
+ \gamma(\alpha_s)  \right ] \, C_{-,\bar n}(n \cdot p, \mu) \,,  \nonumber \\
\frac{d \tilde{f}_B(\mu)}{d \ln \mu} \,
&=&  \tilde{\gamma}(\alpha_s) \, \tilde{f}_B(\mu) \,,
\end{eqnarray}
where the various anomalous dimensions can be inferred from \cite{Beneke:2011nf}.
To achieve the resummation improved factorization formula for the considered correlation function
at  NLL accuracy,  the cusp anomalous dimension  $\Gamma_{\rm {cusp}}(\alpha_s)$ needs to be expanded at the
three-loop accuracy, while the soft anomalous dimensions $\gamma(\alpha_s)$ and
$\tilde{\gamma}(\alpha_s)$ need to be expanded up to two loops.
The NLL resummation improved expressions for  $C_{-,\bar n}$ and $\tilde{f}_B$ can  be further computed as
\begin{eqnarray}
C_{-,\bar n}(n \cdot p, \mu) &=& U_1(n \cdot p, \mu_{h1}, \mu) \, C_{-,\bar n}(n \cdot p, \mu_{h1}) \,, \nonumber \\
\tilde{f}_B(\mu) &=&  U_2(n \cdot p, \mu_{h2}, \mu) \,\tilde{f}_B(\mu_{h2}) \,,
\end{eqnarray}
where the explicit expressions of the evolution functions $U_1$ and $U_2$ can be found in \cite{Wang:2016qii}.
It is then a straightforward task to deduce the (partial) NLL resummation  improved factorization formula for
the correlation function $\Pi_{\mu}$
\begin{eqnarray}
&& \Pi_{\mu, \, {\rm 2P}}(n \cdot p,\bar n \cdot p)=  - i \, \left [ U_2(n \cdot p, \mu_{h2}, \mu) \,\tilde{f}_B(\mu_{h2})  \right ] \, m_B \,
\int_0^{\infty} \,  \frac{d \omega}{\bar n \cdot p - \omega - \omega_c + i 0} \, \nonumber  \\
&&  \times \, \bigg \{  \bigg [ C_{+, n}(n \cdot p, \mu) \,   J_{+, n}(\bar n \cdot p, \omega, \mu)   \, n_{\mu}
+ C_{+, \bar n}(n \cdot p, \mu) \,   J_{+, \bar n}(\bar n \cdot p, \omega, \mu)  \, \bar n_{\mu}  \bigg ] \,
\phi_B^{+}(\omega, \mu)  \nonumber \\
&& \hspace{0.3 cm} + \bigg [ C_{-, n}(n \cdot p, \mu) \,   J_{-, n}(\bar n \cdot p, \omega, \mu)   \, n_{\mu} +
U_1(n \cdot p, \mu_{h1}, \mu) \, C_{-,\bar n}(n \cdot p, \mu_{h1}) \,
J_{-, \bar n}(\bar n \cdot p, \omega, \mu)  \, \bar n_{\mu}  \bigg ] \,\nonumber \\
&&   \hspace{0.6 cm} \times  \, \phi_B^{-}(\omega, \mu) \bigg \} \,,
\label{resummation improved factorization formula}
\end{eqnarray}
where $\mu_{h1}$ and  $\mu_{h2}$ are the hard scales of ${\cal O}(m_b)$ and the factorization scale $\mu$ needs to be
taken at a hard-collinear scale ${\cal O}(\sqrt{m_b \, \Lambda})$.


\subsection{The NLL LCSR for $B \to D$ form factors}

We are now ready to derive the (partial) NLL resummation improved sum rules for the vector and scalar $B \to D$ form factors.
To this end, it is mandatory to work out the spectral representation of the factorization formula
for $\Pi_{\mu}$ obtained in the above, which turns out to be a nontrivial task compared to the construction of
the $B$-meson LCSR for $B \to \pi$ form factors \cite{Wang:2015vgv}.
Applying the dispersion representations of convolution integrals collected in Appendix \ref{app:spectral resp}
yields
\begin{eqnarray}
&& \Pi_{\mu, \, {\rm 2P}}(n \cdot p,\bar n \cdot p) = - i \, \left [ U_2(n \cdot p, \mu_{h2}, \mu) \,\tilde{f}_B(\mu_{h2})  \right ] \, m_B \,
\int_0^{\infty} \,  \frac{d \omega^{\prime}}{\omega^{\prime} - \bar n \cdot p  - i 0} \, \nonumber  \\
&&  \times \, \bigg \{   C_{+, n}(n \cdot p, \mu) \,   \Phi_{+, n}^{\rm eff}(\omega^{\prime}, \mu)   \, n_{\mu}
+ C_{+, \bar n}(n \cdot p, \mu) \,   \Phi_{+, \bar n}^{\rm eff}(\omega^{\prime}, \mu)\, \bar n_{\mu}   \nonumber \\
&& \hspace{0.5 cm} + C_{-, n}(n \cdot p, \mu) \,   \Phi_{-, n}^{\rm eff}(\omega^{\prime}, \mu)  \, n_{\mu}  +
U_1(n \cdot p, \mu_{h1}, \mu) \, C_{-,\bar n}(n \cdot p, \mu_{h1}) \,
\Phi_{-, \bar n}^{\rm eff}(\omega^{\prime}, \mu)  \, \bar n_{\mu}  \bigg \} \,,
\label{spectral representation of the factorization formula}
\end{eqnarray}
where we have introduced the ``effective" DA $\Phi_{i, k}^{\rm eff}(\omega^{\prime}, \mu)$
($i= \pm, \, k= n  \,\, {\rm or } \,\, \bar n$) absorbing the hard-collinear QCD corrections to the correlation function
\begin{eqnarray}
\Phi_{+, n}^{\rm eff}(\omega^{\prime}, \mu) &=& \frac{\alpha_s \, C_F}{4 \, \pi} \,
\bigg \{ {\omega_c \over \omega^{\prime}} \, \left [ {\omega^{\prime} - \omega_c \over \omega^{\prime}}  \,
\ln \left |{\omega_c - \omega^{\prime} \over \omega_c} \right | - 1  \right ] \,
 \phi_B^{+}(\omega^{\prime} - \omega_c) \, \theta(\omega^{\prime} - \omega_c)  \nonumber \\
&& - \left [ \omega_c \, \ln \left |{\omega_c - \omega^{\prime} \over \omega_c} \right | \,
\delta^{\prime}(\omega^{\prime}) - \delta(\omega^{\prime}) \right ] \,
\int_{0}^{\infty} \, d \omega \, {\omega_c \over \omega + \omega_c}  \, \phi_B^{+}(\omega) \,  \nonumber \\
&& + \, \theta(\omega^{\prime} - \omega_c)  \, \int_0^{\infty}  \, d \omega \,
\bigg [ {\omega \, \omega_c \over (\omega+\omega_c)^2 } \, \left ({\cal P} \, {1 \over \omega^{\prime} -\omega -\omega_c}
- {1 \over \omega^{\prime}}  \right ) + {\omega_c \over \omega} \, {1 \over \omega^{\prime}} \nonumber \\
&& + {\omega_c^2 \over \omega+\omega_c} \,{1 \over {\omega^{\prime}}^2}
- {\theta(\omega+\omega_c-\omega^{\prime})  \over \omega} \,
+ {\omega_c \over \omega} \, {\theta(\omega^{\prime} - \omega - \omega_c)  \over \omega-\omega^{\prime}}
\,   \bigg ]   \, \phi_B^{+}(\omega) \, \bigg \} \,, \\
\Phi_{+, \bar n}^{\rm eff}(\omega^{\prime}, \mu) &=&  \frac{\alpha_s \, C_F}{4 \, \pi} \,
\bigg \{ m_c \, \bigg [ \left ( {1 \over \omega^{\prime} - \omega_c} - {1 \over \omega^{\prime}}
- { \omega_c - \omega^{\prime} \over {\omega^{\prime}}^2} \,
\ln \left |{\omega_c - \omega^{\prime} \over \omega_c} \right | \right )   \,
\phi_B^{+}(\omega^{\prime} - \omega_c) \, \theta(\omega^{\prime} - \omega_c)  \nonumber \\
&& + {\omega_c \over \omega^{\prime}} \, \theta(\omega^{\prime} - \omega_c)  \,
\int_0^{\infty} \, d \omega \, {d \over d \omega}{\phi_B^{+}(\omega) \over \omega}
- \left ( \delta(\omega^{\prime}) - \omega_c \, \ln \left |{\omega_c - \omega^{\prime} \over \omega_c} \right | \,
\delta^{\prime}(\omega^{\prime})  \right )  \, \nonumber \\
&& \times \int_{0}^{\infty} \, d \omega \, {\omega_c \over \omega ( \omega + \omega_c)} \, \phi_B^{+}(\omega)
- \int_{0}^{\infty} \, d \omega \,  \frac{\theta(\omega^{\prime} - \omega - \omega_c)}{\omega^{\prime} - \omega}
\, \left (1 + \omega_c \, {d \over d \omega} \right ) \, {\phi_B^{+}(\omega) \over \omega}  \nonumber \\
&& +  \, \theta(\omega^{\prime} - \omega_c)  \, \int_{0}^{\infty} \, d \omega \,
\bigg (  {\omega \over (\omega+\omega_c)^2} \,
{\cal P} \, {1 \over \omega^{\prime} -\omega -\omega_c}
+  {\omega_c (2 \omega+\omega_c) \over \omega (\omega+\omega_c)^2}  \, {1 \over \omega^{\prime}}  \nonumber \\
&& - {\omega_c^2 \over \omega (\omega+\omega_c)} \, {1 \over {\omega^{\prime}}^2} \bigg ) \, \phi_B^{+}(\omega)  \bigg ]  \nonumber \\
&& + \, r \, \bigg [  \theta(\omega^{\prime} - \omega_c)  \, {\omega_c^2 \over {\omega^{\prime}}^2} \,
\int_{0}^{\infty} \, d \omega \,  \left ( 1 - \omega^{\prime} {d \over d \omega}  \right ) \,
{\phi_B^{+}(\omega) \over \omega}  \nonumber \\
&& -  \theta(\omega^{\prime} - \omega_c)  \, \int_{0}^{\infty} \, d \omega \,
 \theta(\omega + \omega_c - \omega^{\prime})  \,\left ( 1 - \omega_c {d \over d \omega}  \right )  \,
{\phi_B^{+}(\omega) \over \omega} \nonumber \\
&& +   \int_{0}^{\infty} \, d \omega \, \theta(\omega^{\prime} - \omega - \omega_c) \,
{\omega_c^2  \over \omega^{\prime}-\omega } \, {d \over  d \omega} \, { \phi_B^{+}(\omega) \over \omega} \nonumber \\
&& + \omega_c \, \left ( \delta(\omega^{\prime}) - \omega_c \, \ln \left |{\omega_c - \omega^{\prime} \over \omega_c} \right | \,
\delta^{\prime}(\omega^{\prime})  \right )  \, \int_{0}^{\infty} \, d \omega \,
{\phi_B^{+}(\omega) \over \omega }\bigg ] \bigg \}  \,,  \\
\Phi_{-, n}^{\rm eff}(\omega^{\prime}, \mu) &=&  - \phi_B^{-}(\omega^{\prime} - \omega_c) \,
 \theta(\omega^{\prime} - \omega_c) \,, \\
 \Phi_{-, \bar n}^{\rm eff}(\omega^{\prime}, \mu) &=&  - \phi_B^{-}(\omega^{\prime} - \omega_c) \,
 \theta(\omega^{\prime} - \omega_c) +
 \frac{\alpha_s \, C_F}{4 \, \pi} \, \bigg \{   \phi_B^{-}(\omega^{\prime} - \omega_c) \,
\theta(\omega^{\prime} - \omega_c) \, \rho_{-, \bar n}^{(1)}(\omega^{\prime})  \nonumber \\
&& + \left [ {d \over d \omega^{\prime}} \phi_B^{-}(\omega^{\prime}-\omega_c) \right ] \,
\theta(\omega^{\prime}-\omega_c) \, \rho_{-, \bar n}^{(2)}(\omega^{\prime})
+ \phi_B^{-}(\omega^{\prime}) \, \rho_{-, \bar n}^{(3)}(\omega^{\prime})
+ \phi_B^{-}(0) \, \rho_{-, \bar n}^{(4)}(\omega^{\prime})  \nonumber \\
&& + \, \int_{0}^{\infty} \, d \omega \,   \phi_B^{-}(\omega) \, \rho_{-, \bar n}^{(5)}(\omega, \omega^{\prime})
+ \, \int_{0}^{\infty} \, d \omega \,  \left [ {d \over d \omega}  \, \phi_B^{-}(\omega)  \right ] \,
 \rho_{-, \bar n}^{(6)}(\omega, \omega^{\prime}) \bigg \}  \nonumber \\
&& + \, \int_{0}^{\infty} \, d \omega \,  \left [ {d \over d \omega}  \, {\phi_B^{-}(\omega) \over \omega } \right ] \,
 \rho_{-, \bar n}^{(7)}(\omega, \omega^{\prime}) \bigg \}   \,,
\label{effective B-meson DA}
\end{eqnarray}
with ${\cal P}$ indicating the principle-value prescription.
The tedious expressions of the coefficient functions $ \rho_{-, \bar n}^{(i)}$ in (\ref{effective B-meson DA})
are collected in Appendix \ref{app: the coefficient functions of spectral density}.
Applying the standard strategy to construct the sum rules
for hadronic transition form factors  by matching (\ref{spectral representation of the factorization formula})
and (\ref{hadronic representation}) with the aid of the parton-hadron duality approximation and the Borel transformation
leads to
\begin{eqnarray}
&& f_D \, \, {\rm Exp} \left[- {m_{D}^2 \over n \cdot p \,\, \omega_M} \right] \,
\left \{ {n \cdot p \over m_B} \, f_{B D, \, \rm 2P}^{+}(q^2) \,, \,\,  f_{B D, \, \rm 2P}^{0}(q^2)  \right  \}  \nonumber \\
&& = - \left [ U_2(n \cdot p, \mu_{h2}, \mu) \,\tilde{f}_B(\mu_{h2})  \right ] \,
\int_{0}^{\omega_s} \, d \omega^{\prime} \,{\rm Exp} \left [- {\omega^{\prime} \over \omega_M} \right ] \,
 \nonumber \\
&& \hspace{0.5 cm} \times \bigg \{ C_{+, \bar n}(n \cdot p, \mu) \,   \Phi_{+, \bar n}^{\rm eff}(\omega^{\prime}, \mu) \,
+ U_1(n \cdot p, \mu_{h1}, \mu) \, C_{-,\bar n}(n \cdot p, \mu_{h1}) \,
\Phi_{-, \bar n}^{\rm eff}(\omega^{\prime}, \mu)   \nonumber \\
&& \hspace{1 cm} \pm {n \cdot p - m_B \over m_B} \, \left [  C_{+, n}(n \cdot p, \mu) \,   \Phi_{+, n}^{\rm eff}(\omega^{\prime}, \mu)
 + C_{-, n}(n \cdot p, \mu) \,   \Phi_{-, n}^{\rm eff}(\omega^{\prime}, \mu) \right ]  \bigg \}  \,,
\label {NLL reummation improved SR for B to D FFs}
\end{eqnarray}
which serves as  the master formula for the two-particle contributions to  $B \to D$ form factors obtained in this paper.
It is evident that the symmetry-breaking corrections to the form-factor relation  (\ref{B to D FFs at tree level})
arise from both hard and hard-collinear fluctuations as displayed in the last line of
(\ref{NLL reummation improved SR for B to D FFs}), and the power enhanced contribution due to
the charm-quark mass effect preserves the large-recoil symmetry relation.
The symmetry relations of $B \to D$ form factors at large hadronic recoil could be also obtained in SCET with massive
collinear quark fields by generalizing the discussion for  heavy-to-light form factors \cite{Beneke:2003pa,Beneke:2005gs,Bauer:2000yr,Bauer:2002aj}
and we leave a systematic investigation of the SCET formulation of $B \to D$ form factors for future work.

\section{Three-particle contributions to the LCSR at tree level}
\label{sect:three-particle correction}

The objective of this section is to compute the tree-level contribution to
the sum rules of $B \to D$ form factors from the three-particle $B$-meson DA.
To this end, we first need to demonstrate QCD factorization for the three-particle contributions
to the correlation function (\ref{correlator: definition}) with space-like interpolating momentum,
which can be achieved by evaluating the diagram displayed in figure 3 with the aid of the light-cone OPE.

\begin{figure}
\begin{center}
\includegraphics[width=0.4 \columnwidth]{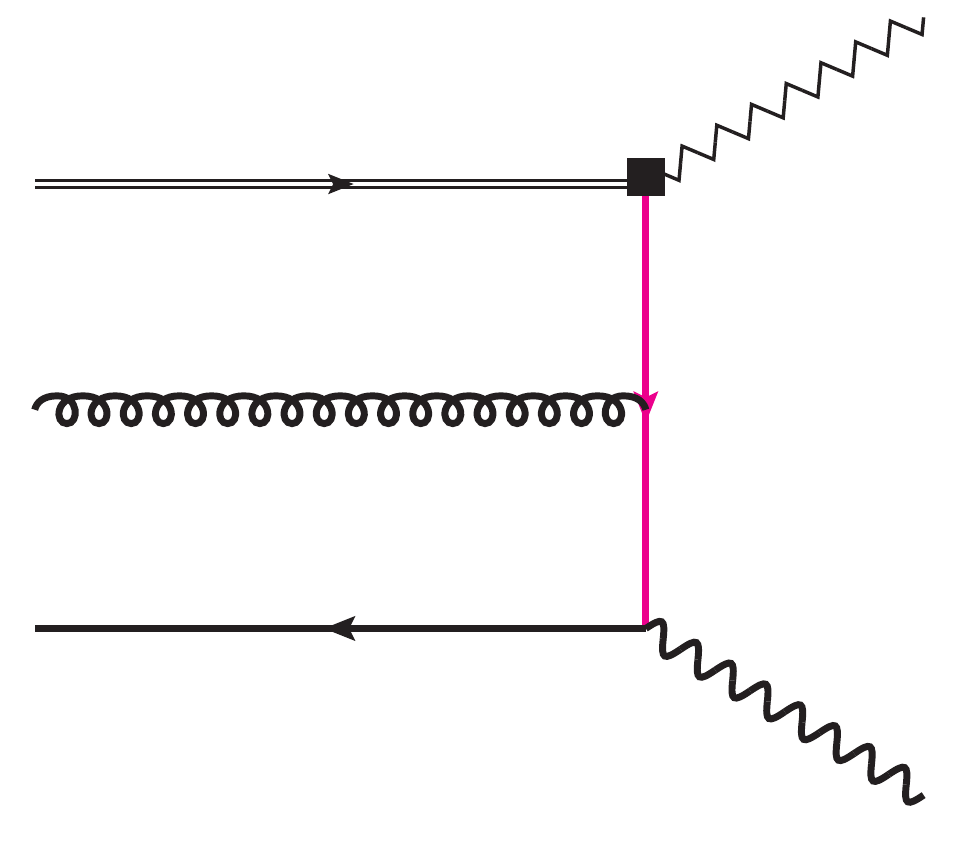}
\vspace*{0.1cm}
\caption{Diagrammatical representation of the three-particle contributions
to the vacuum-to-$B$-meson correlation function
$\Pi_{\mu}(n \cdot p,\bar n \cdot p)$ defined in (\ref{correlator: definition}) at tree level. }
\label{fig:tree-level_correlator for 3P}
\end{center}
\end{figure}

Employing the massive quark propagator near the light cone from the
background gluon field  method \cite{Balitsky:1987bk,Khodjamirian:2010vf,Khodjamirian:2012rm}
\begin{eqnarray}
S(x, 0, m_c) &=& \langle 0 | {\rm T} \, \{\bar c (x), c(0) \} | 0\rangle  \nonumber \\
& \supset &  i \, \int_0^{\infty} \,\, {d^4 k \over (2 \pi)^4} \, e^{- i \, k \cdot x} \,
\int_0^1 \, d u \, \left  [ {u \, x_{\mu} \, \gamma_{\nu} \over k^2 - m_c^2}
 - \frac{(\not \! k + m_c) \, \sigma_{\mu \nu}}{2 \, (k^2 - m_c^2)^2}  \right ]
\, G^{\mu \nu}(u \, x) \,,
\end{eqnarray}
with $G^{\mu \nu}=g_s \, T^a \, G_{\mu \nu}^a$, it is then straightforward to derive the
factorization formula
\begin{eqnarray}
\Pi_{\mu, \, {\rm 3P}} (n \cdot p,\bar n \cdot p) &=&
 - \, {i \,  \tilde{f}_B(\mu) \, m_B \over n \cdot p} \,
\int_0^{\infty} \, d \omega \, \int_0^{\infty} \, d \xi \,
\int_0^{1} \, d u \, \bigg \{ \frac{\tilde{\rho}_{2,  n}(u, \omega, \xi) }
{(\bar n \cdot p - \omega - u\, \xi - \omega_c)^2 } \,  n_{\mu}  \nonumber \\
&&
+ \left [ \frac{\tilde{\rho}_{2, \bar n}(u, \omega, \xi) }
{(\bar n \cdot p - \omega - u\, \xi - \omega_c)^2 }
+ \frac{\tilde{\rho}_{3, \bar n}(u, \omega, \xi) }
{(\bar n \cdot p - \omega - u\, \xi - \omega_c)^3 } \right ] \, {\bar n}_{\mu}
\bigg \}  \,,
\end{eqnarray}
where the coefficient functions $\tilde{\rho}_{i,  k}(u, \omega, \xi)$ are given by
\begin{eqnarray}
\tilde{\rho}_{2,  n}(u, \omega, \xi) &=& 2 \, (u-1) \,
\left[ \psi_V(\omega, \xi)  +  \psi_A(\omega, \xi) \right  ] \,, \nonumber \\
\tilde{\rho}_{2,  \bar n}(u, \omega, \xi) &=& - \psi_V(\omega, \xi)  +
\, (2 \, u-1) \, \psi_A(\omega, \xi) \,, \nonumber \\
\tilde{\rho}_{3,  \bar n}(u, \omega, \xi) &=& 2 \, ( 2 \, u-1) \,
\left [ \bar{X}_A(\omega, \xi) - 2 \,  \bar{Y}_A(\omega, \xi) \right ] \,.
\label{3P contribution: master formula}
\end{eqnarray}
The appeared three-particle DA of the $B$-meson can be defined by the position-space
matrix element on the light cone \cite{Kawamura:2001jm,Geyer:2005fb}
\begin{eqnarray}
&& \langle 0 |\bar q_{\alpha}(x) \,\, G_{\lambda \, \rho}(u \, x)
 \,\, b_{v \beta}(0)| \bar B (v)\rangle  \big |_{x^2=0} \nonumber \\
&& = {\tilde{f}_B(\mu) \, m_B \over 4} \, \int_0^{\infty} d \omega \, \int_0^{\infty} d \xi \,\,
e^{-i(\omega+ u\, \xi) \, v \cdot x} \,\,
\bigg [ \left (1 + \! \not v \right ) \,
\bigg \{  (v_{\lambda} \, \gamma_{\rho} - v_{\rho} \, \gamma_{\lambda} ) \,
\left [\Psi_A(\omega, \xi) - \Psi_V(\omega, \xi) \right ]  \nonumber \\
&& \hspace{0.6 cm} - i \, \sigma_{\lambda \rho} \, \Psi_V(\omega, \xi)
- \frac{x_{\lambda} v_{\rho} - x_{\rho} v_{\lambda} }{v \cdot x} \, X_A(\omega, \xi)
+ \frac{x_{\lambda} \gamma_{\rho} - x_{\rho} \gamma_{\lambda} }{v \cdot x} \, Y_A(\omega, \xi)
\bigg \} \, \gamma_5 \bigg ]_{\beta \alpha} \,,
\label{def: 3P DA}
\end{eqnarray}
where we have neglected the soft Wilson lines to maintain the gauge invariance of the
string operator on the left-hand side.  The following conventions
\begin{eqnarray}
\bar X_A(\omega, \xi) = \int_0^{\omega} \, d \eta \, X_A(\eta, \xi)  \,, \qquad
\bar Y_A(\omega, \xi) = \int_0^{\omega} \, d \eta \, Y_A(\eta, \xi)   \,,
\end{eqnarray}
are further introduced in (\ref{3P contribution: master formula}) for convenience.
Our current understanding of  the model independent properties of the three-particle quark-gluon $B$-meson DA
is limited to the twist-3 DA $\Phi_3(\omega, \xi) \equiv \Psi_A(\omega, \xi) - \Psi_V(\omega, \xi)$
which was shown to be completely integrable in the large $N_c$ limit and to be solvable exactly
at one loop \cite{Braun:2015pha}. Investigating perturbative QCD constraints on the higher-twist
$B$-meson DA from the OPE analysis in the partonic picture and from the renormalization  evolution equations
in momentum (or ``dual") space, along the lines of \cite{Lee:2005gza,Feldmann:2014ika}, would be interesting
for future work.

Expressing the factorization formula for the three-particle contributions to the correlation function
$\Pi_{\mu, \, {\rm 3P}}$ in a dispersion form with respect to the variable $\bar n \cdot p$ and
implementing the continuum subtraction with the aid of the parton-hadron duality relation
as well as the Borel transformation lead to the following sum rules
\begin{eqnarray}
&& f_D \, \, {\rm Exp} \left[- {m_{D}^2 \over n \cdot p \,\, \omega_M} \right] \,
\left \{ {n \cdot p \over m_B} \, f_{B D, \, \rm 3P}^{+}(q^2) \,, \,\,  f_{B D, \, \rm 3P}^{0}(q^2)  \right  \}
\nonumber \\
&& = \,  - { \tilde{f}_B(\mu) \over  n \cdot p}   \,
\left [ {\cal F}_{{\rm 3P}, \bar n}(\omega_s, \omega_M) \,  \mp \, {m_B - n \cdot p \over m_B} \,
{\cal F}_{{\rm 3P}, n}(\omega_s, \omega_M)  \right ] \,,
\label{3-particle contribution to the LCSR}
\end{eqnarray}
where the explicit expressions of the coefficient functions ${\cal F}_{{\rm 3P}, \bar n}$
and ${\cal F}_{{\rm 3P}, n}$ are
\begin{eqnarray}
{\cal F}_{{\rm 3P}, \bar n}(\omega_s, \omega_M)
&=& \int_0^{\omega_s-\omega_c} \, d \omega  \, \int_{\omega_s - \omega_c -\omega}^{\infty} \,
{d \xi \over \xi} \,\, {\rm Exp} \left ( - {\omega_s \over \omega_M} \right )  \, \nonumber \\
&& \times \, \bigg [ \tilde{\rho}_{2,  \bar n}(u, \omega, \xi)
- {1 \over 2} \, \left ( {d \over d \omega}  +  {1 \over \omega_M}\right )
\tilde{\rho}_{3,  \bar n}(u, \omega, \xi) \bigg ] \, \bigg |_{u= \frac{\omega_s - \omega_c -\omega} {\xi}}  \nonumber \\
&& + \int _{\omega_c}^{\omega_s} \, d \omega^{\prime}  \, \int_0^{\omega^{\prime}-\omega_c} \,
d \omega  \, \int_{\omega^{\prime} - \omega_c -\omega}^{\infty} \,
{d \xi \over \xi} \, {1 \over \omega_M} \, {\rm Exp} \left ( - {\omega^{\prime} \over \omega_M} \right )  \, \nonumber \\
&& \times \, \bigg [ \tilde{\rho}_{2,  \bar n}(u, \omega, \xi)
- {1 \over 2 \, \omega_M} \, \tilde{\rho}_{3,  \bar n}(u, \omega, \xi) \bigg ]
\, \bigg |_{u= \frac{\omega^{\prime} - \omega_c -\omega} {\xi}}   \,,  \\
{\cal F}_{{\rm 3P}, n}(\omega_s, \omega_M)
&=&  \int_0^{\omega_s-\omega_c} \, d \omega  \, \int_{\omega_s - \omega_c -\omega}^{\infty} \,
{d \xi \over \xi} \,\, {\rm Exp} \left ( - {\omega_s \over \omega_M} \right ) \,
\tilde{\rho}_{2,   n}(u, \omega, \xi)  \, \bigg |_{u= \frac{\omega_s - \omega_c -\omega} {\xi}}  \, \nonumber \\
&& + \int _{\omega_c}^{\omega_s} \, d \omega^{\prime}  \, \int_0^{\omega^{\prime}-\omega_c} \,
d \omega  \, \int_{\omega^{\prime} - \omega_c -\omega}^{\infty} \,
{d \xi \over \xi} \, {1 \over \omega_M} \, {\rm Exp} \left ( - {\omega^{\prime} \over \omega_M} \right )  \,
\tilde{\rho}_{2,   n}(u, \omega, \xi)  \, \bigg |_{u= \frac{\omega^{\prime} - \omega_c -\omega} {\xi}}  \,. \nonumber \\
\end{eqnarray}
It is evident from (\ref{3-particle contribution to the LCSR}) that the large-recoil symmetry breaking effect
between the vector and scalar $B \to D$ form factors can be  induced by the higher-twist contributions
from the three-particle $B$-meson DA, already at tree level.

We are now ready to derive the power counting behaviour of the three-particle correction to $B \to D$ form factors at tree level.
Applying the canonical behaviour of the three-particle $B$-meson DA  in the end-point region \cite{Khodjamirian:2006st}
\begin{eqnarray}
\Psi_V(\omega, \xi) \sim \Psi_A(\omega, \xi) \sim \xi^2 \,,  \qquad
X_A(\omega, \xi) \sim \xi (2 \omega - \xi) \,, \qquad Y_A(\omega, \xi) \sim \xi \,,
\end{eqnarray}
we can verify that the three-particle contribution to the sum rules of $B \to D$ form factors
(\ref{3-particle contribution to the LCSR}) is counted as  ${\cal O} \left ((\Lambda/m_b)^{5/2} \right )$ in the
heavy quark limit and is indeed suppressed by a factor of $\Lambda/m_b$ compared to the two-particle
contribution at tree level as presented in (\ref{B to D FFs at tree level}).
However, it needs to point out that  higher twist effects from the three-particle gluon-quark DA
can give rise to the leading power contributions to $B \to D$ form factors at ${\cal O} (\alpha_s)$
on account of the reasonings of \cite{Beneke:2003pa,Ball:2003bf} and a transparent demonstration of
this observation by computing the three-particle contribution to the sum rules at NLO in $\alpha_s$
directly is in high demand on both conceptual and phenomenological aspects.

The final expressions for the form factors of $B \to D \ell \nu$ can be obtained by adding up the two-particle
and the three-particle contributions together
\begin{eqnarray}
f_{B D}^{+}(q^2) = f_{B D, \, \rm 2P}^{+}(q^2)  + f_{B D, \, \rm 3P}^{+}(q^2)  \,, \qquad
f_{B D}^{0}(q^2) = f_{B D, \, \rm 2P}^{0}(q^2)  + f_{B D, \, \rm 3P}^{0}(q^2)  \,,
\label{the final sum rules}
\end{eqnarray}
where the detailed expressions of $f_{B D, \, \rm 2P}^{+, 0}(q^2)$ and $f_{B D, \, \rm 3P}^{+, 0}(q^2)$
 are presented in  (\ref{NLL reummation improved SR for B to D FFs}) and
(\ref{3-particle contribution to the LCSR}), respectively.

\section{Numerical analysis}
\label{sect:numerical analysis}

Having at our disposal the resummation improved sum rules for $B \to D$ form factors,
we are in a position to explore the phenomenological consequences of perturbative corrections
to the two-particle contributions and  higher-twist effects from the three-particle
quark-gluon $B$-meson DA at tree level.
Employing the obtained theory predictions for $B \to D$ form factors, we will further present
our results for the semileptonic $B \to D \ell \nu$ ($\ell = \mu, \, \tau$) decay observables,
including the  invariant-mass distributions of the lepton pair, the topical $R(D)$ ratio,
and the CKM matrix element $|V_{cb}|$.

\subsection{Theory input parameters}
\label{subsection: theory inputs}

We will first specify the theory inputs entering the LCSR for $B \to D $ form factors
shown in  (\ref{NLL reummation improved SR for B to D FFs}) and (\ref{3-particle contribution to the LCSR}).
The fundamental nonperturbative  quantities describing the soft strong interaction dynamics encoded
in the correlation function (\ref{correlator: definition}) are the $B$-meson light-cone DA
defined in  (\ref{def: 2P DA}) and (\ref{def: 3P DA}).  A detailed account of the current understanding
towards  perturbative and nonperturbative aspects of the  two-particle $B$-meson DA
was already presented in \cite{Wang:2015vgv}, following which we will consider two models of the leading twist
DA $\phi_B^{+}(\omega, \mu_0)$ proposed in \cite{Grozin:1996pq,Braun:2003wx}
\begin{eqnarray}
&&  \phi_{B,\rm I}^+(\omega,\mu_0) = \frac{\omega}{\omega_0^2} \, e^{-\omega/\omega_0} \,,
\label{the first model of the B-meson DA} \\
&&  \phi_{B,\rm II}^+(\omega,\mu_0)= \frac{1}{4 \pi \,\omega_0} \, {k \over k^2+1} \,
\left[ {1 \over k^2+1} - \frac{2 (\sigma_{1}(\mu_0) -1)}{\pi^2}  \, \ln k \right ] \,, \hspace{0.5 cm}
k= \frac{\omega}{1 \,\, \rm GeV} \,,
\label{the second model of the B-meson DA}
\end{eqnarray}
where the shape parameter $\omega_0$ can be converted to the inverse moment of the leading-twist $B$-meson DA $\lambda_B(\mu_0)$.
The renormalization scale evolution of $\lambda_B(\mu)$ at one loop can be derived from the Lange-Neubert equation
of $\phi_B^{+}(\omega, \mu)$ \cite{Lange:2003ff}
\begin{eqnarray}
\frac{\lambda_B(\mu_0)}{\lambda_B(\mu)} &=&
1 + {\alpha_s(\mu_0) \, C_F \over 4 \, \pi} \, \ln {\mu \over \mu_0} \,
\left [2 - 2\, \ln {\mu \over \mu_0} - 4 \, \sigma_{1}(\mu_0) \right ] + {\cal O}(\alpha_s^2)\,.
\label{lambdab evolution}
\end{eqnarray}
The inverse-logarithmic moment  at a hadronic scale $\mu_0= 1 \,\, {\rm GeV}$
will be taken as  $\sigma_1(\mu_0)= 1.4 \pm 0.4$,  determined from a QCD sum rule analysis \cite{Braun:2003wx}.
The higher-twist two-particle DA  of the  $B$-meson   $\phi_{B}^{-}(\omega,\mu_0)$ will be determined from the QCD
equations of motion in the heavy quark limit \cite{Kawamura:2001jm}
\begin{eqnarray}
\omega \, \phi_B^-(\omega) - \int_0^\omega d\eta \left[ \phi_B^-(\eta)-\phi_B^+(\eta) \right ] 
&=& 2 \, \int_0^\omega d\eta \int_{\omega-\eta}^\infty \frac{d\xi}{\xi} \, \frac{\partial}{\partial \xi}
\left [  \Psi_A(\eta,\xi)-\Psi_V(\eta,\xi)\right ] \,,
\end{eqnarray}
which was also demonstrated to be valid  from a non-relativistic toy model manifestly at NLO in
the strong coupling constant \cite{Bell:2008er}.
With regarding to the three-particle quark-gluon DA of the $B$-meson, we will employ the exponential model
inspired from the canonical behaviour predicted by the HQET sum rules at tree level \cite{Khodjamirian:2006st}
\begin{eqnarray}
\Psi_V(\omega, \xi, \mu_0) &=& \Psi_A(\omega, \xi, \mu_0)
=\frac{\lambda_E^2}{6 \, \omega_0^4} \, \xi^2 \, e^{-(\omega+\xi)/\omega_0} \,, \nonumber \\
X_A(\omega, \xi, \mu_0) &=& \frac{\lambda_E^2}{6 \, \omega_0^4} \, \xi \, (2 \, \omega - \xi) \,
e^{-(\omega+\xi)/\omega_0} \,, \nonumber \\
Y_A(\omega, \xi, \mu_0) &=& - \frac{\lambda_E^2}{24 \, \omega_0^4} \, \xi \,
(7 \, \omega_0 - 13 \, \omega + 3 \, \xi) \, e^{-(\omega+\xi)/\omega_0} \,,
\end{eqnarray}
where the normalization constant  defined by the matrix element of the chromoelectric operator
was estimated as $\lambda_E^2(\mu_0)=(0.03 \pm 0.02 ) \, {\rm GeV^2}$ \cite{Nishikawa:2011qk},
from the two-point QCD sum rules including  higher-order perturbative and nonperturbative corrections.
We already implemented the approximation $\lambda_E=\lambda_H$ in the above expressions,
following \cite{Khodjamirian:2006st}, which is supported by the nonperturbative QCD
 calculations numerically \cite{Grozin:1996pq,Nishikawa:2011qk}.

With the matching relation (\ref{hard matching of the fB}) the static $B$-meson  decay constant
$\tilde{f}_B(\mu)$ can be traded into the QCD decay constant $f_B$, whose values will be taken from
lattice QCD simulations $f_B= (192.0 \pm 4.3) \, {\rm MeV} $ \cite{Aoki:2016frl} with $N_f= 2+1$.
Likewise, we will adopt the intervals for the $D$-meson decay constant in QCD from
lattice  simulations $f_D= (209.2 \pm 3.3) \, {\rm MeV} $ \cite{Aoki:2016frl} with $N_f= 2+1$.
In addition, we will employ the $\overline {\rm MS}$ bottom quark mass
$\overline{m_b}(\overline{m_b})= (4.193^{+0.022}_{-0.035}) \, {\rm GeV}$ \cite{Beneke:2014pta}
from non-relativistic sum rules and  the  $\overline {\rm MS}$ charm  quark mass
$\overline{m_c}(\overline{m_c})= (1.288 \pm 0.020) \, {\rm GeV}$ \cite{Dehnadi:2015fra}
from relativistic sum rules,  employing the  quark vector correlation function
computed at ${\cal O}(\alpha_s^3)$.
Following \cite{Beneke:2011nf,Wang:2015vgv}, the hard scales $\mu_{h1}$ and $\mu_{h2}$
are taken to be equal and will be varied in the interval $[m_b/2 \,, \, 2 \, m_b]$ around the
default value $m_b$, and the factorization scale is chosen as $1.0 \, {\rm GeV} \leq \mu \leq 2.0 \, {\rm GeV}$
with the default value $\mu=1.5 \, {\rm GeV}$.

The determination of the sum rule parameters $\omega_M$ and $\omega_s$ can be achieved by implementing the
standard procedure described in \cite{Wang:2015vgv} and applying the same strategies  leads to
\begin{eqnarray}
M^2 \equiv n \cdot p \,\, \omega_M = (4.5 \pm 1.0) \, {\rm GeV^2} \,, \qquad
s_0  \equiv n \cdot p \,\, \omega_s = (6.0 \pm 0.5) \, {\rm GeV^2}  \,,
\label{choices of sum rule parameters}
\end{eqnarray}
which are  in agreement with the intervals adopted in \cite{Faller:2008tr,Khodjamirian:2009ys}.

\subsection{Predictions for $B \to D$ form factors}

We will turn to discuss the choice of the inverse moment $\lambda_B (1 \, {\rm GeV})$ which
serves as an important source for theory uncertainties,
prior to  presenting the sum rule predictions for the form factors of $B \to D \ell \nu$.
Albeit with the intensive investigations of  determining $\lambda_B$ theoretically,
the current constraints on $\lambda_B$ are still far from satisfactory, due to the emerged tension
between the NLO sum rule predictions and the implications of  hadronic $B$-meson decay data from QCD
factorization. In particular, the subleading power contributions in HQE
can give rise to sizeable impact on the determination of $\lambda_B$ numerically.
This has been demonstrated explicitly by extracting $\lambda_B$ from the partial branching fractions
of $B \to \gamma \ell \nu$ with the power suppressed effects estimated
from the dispersion approach \cite{Wang:2016qii,Wang:2016beq}.
Following the above argument, it is very plausible that the unaccounted
subleading power contributions to the LCSR for $B \to \pi$ form factors can yield significant corrections
to the fitted values of $\lambda_B (\mu_0)$ \cite{Wang:2015vgv} in addition to the systematic uncertainty induced by the
parton-hadron duality approximation.
In order to be insensitive to the unconsidered effects in the sum rule determinations \cite{Wang:2015vgv},
we will perform an independent determination of $\lambda_B (\mu_0)$ by matching the $B$-meson LCSR of
the vector $B \to D$ form factor at $q^2=0$ to the lattice-QCD calculation with an extrapolation to the large recoil
region using the $z$-series parametrization \footnote{Our major objective is to predict the $q^2$ shapes of
$B \to D$ form factors from the LCSR with the $B$-meson DA, with the normalization $f_{BD}^{+}(0)$ taken as an input.
We have verified numerically that the form-factor shapes at large recoil are insensitive to the actual value
of $\omega_0(1 \, {\rm GeV})$ within a ``reasonable" interval.}.
Taking $f_{BD}^{+}(0)=0.672 \pm 0.027$ \cite{Lattice:2015rga} as an input and implementing
 the above-mentioned matching procedure lead to
 \begin{eqnarray}
\omega_0(1 \, {\rm GeV}) &=& 570^{+38}_{-35} \, {\rm MeV} \,, \qquad {\rm (Model-I)} \,  \nonumber  \\
\omega_0(1 \, {\rm GeV}) &=& 555^{+24}_{-20} \, {\rm MeV} \,, \qquad {\rm (Model-II)} \,
\label{fitted values of omega0}
\end{eqnarray}
which differ from the intervals of $\lambda_B (\mu_0)$ obtained from matching two different versions
of sum rules for the vector $B \to \pi$ form factor \cite{Wang:2015vgv}, however, are comparable to the
values determined with distinct QCD approaches \cite{Braun:2003wx,Ball:2003fq}.
In the following we will take $\phi_{B,\rm I}^+(\omega,\mu_0)$ as our default model for the illustration purpose
and the systematic uncertainty induced by the model dependence of $\phi_{B}^{+}(\omega,\mu_0)$ will be taken into
account in the final predictions for the form factors of $B \to D \ell \nu$.

\begin{figure}
\begin{center}
\includegraphics[width=0.4 \columnwidth]{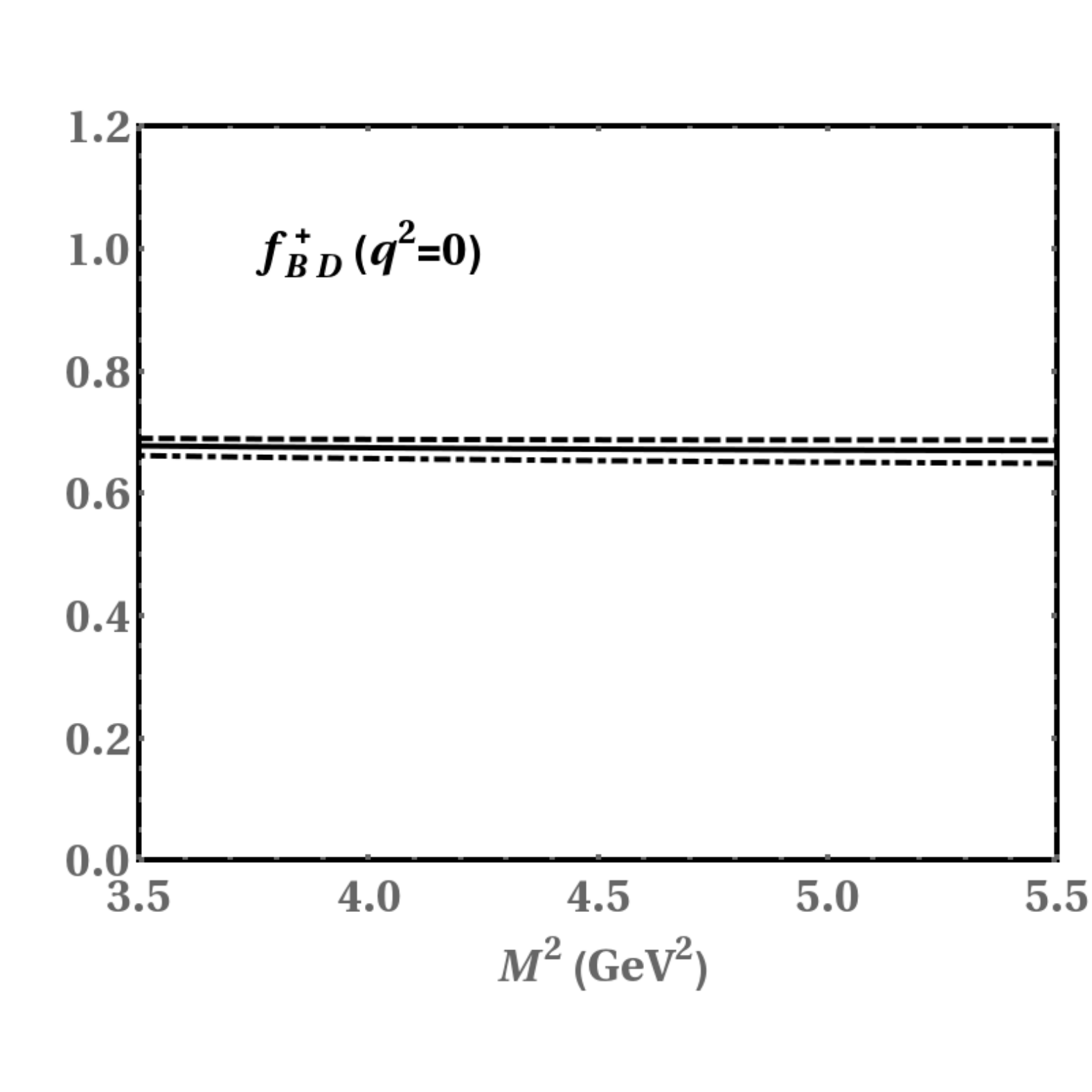} \hspace{1 cm}
\includegraphics[width=0.4 \columnwidth]{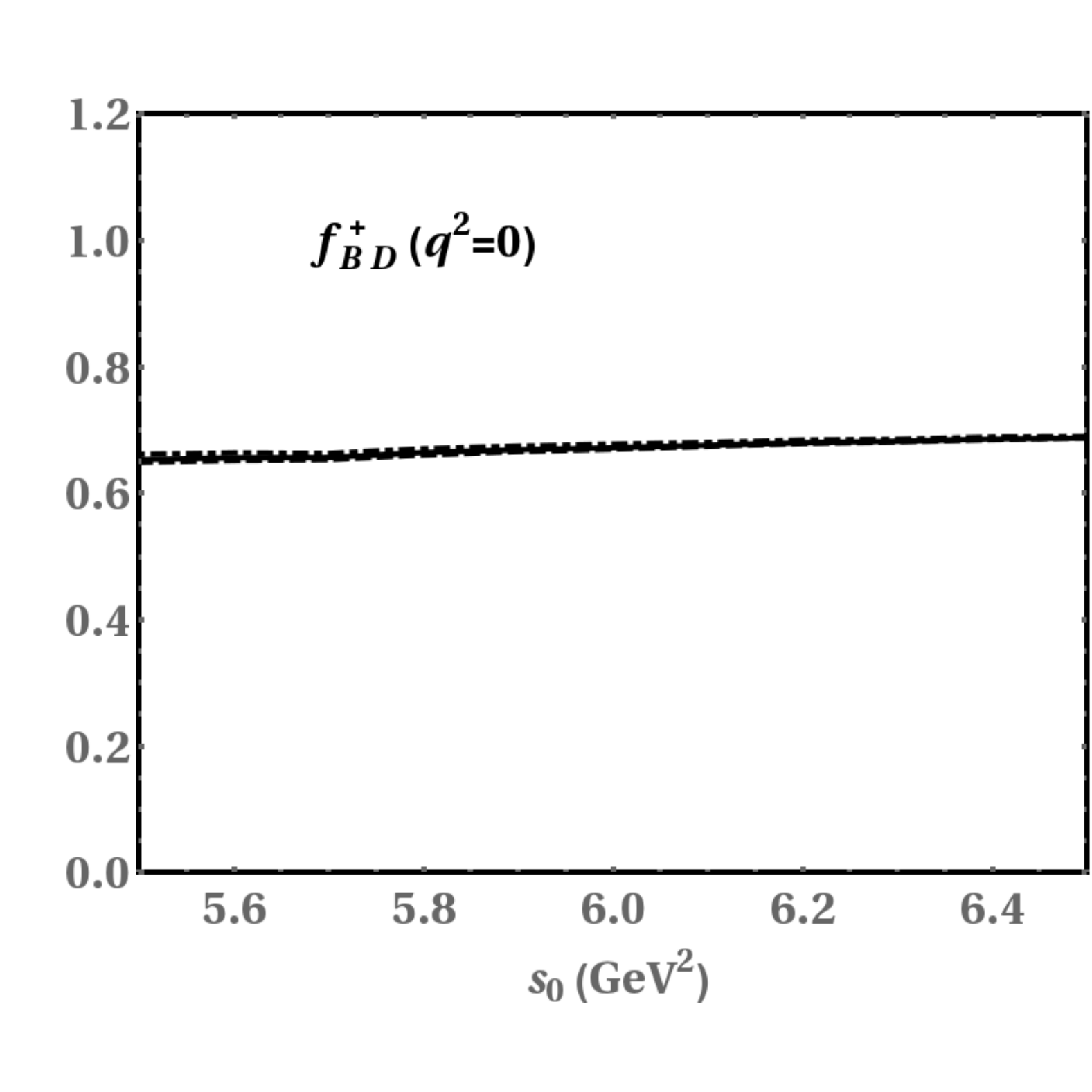} \\
\includegraphics[width=0.4 \columnwidth]{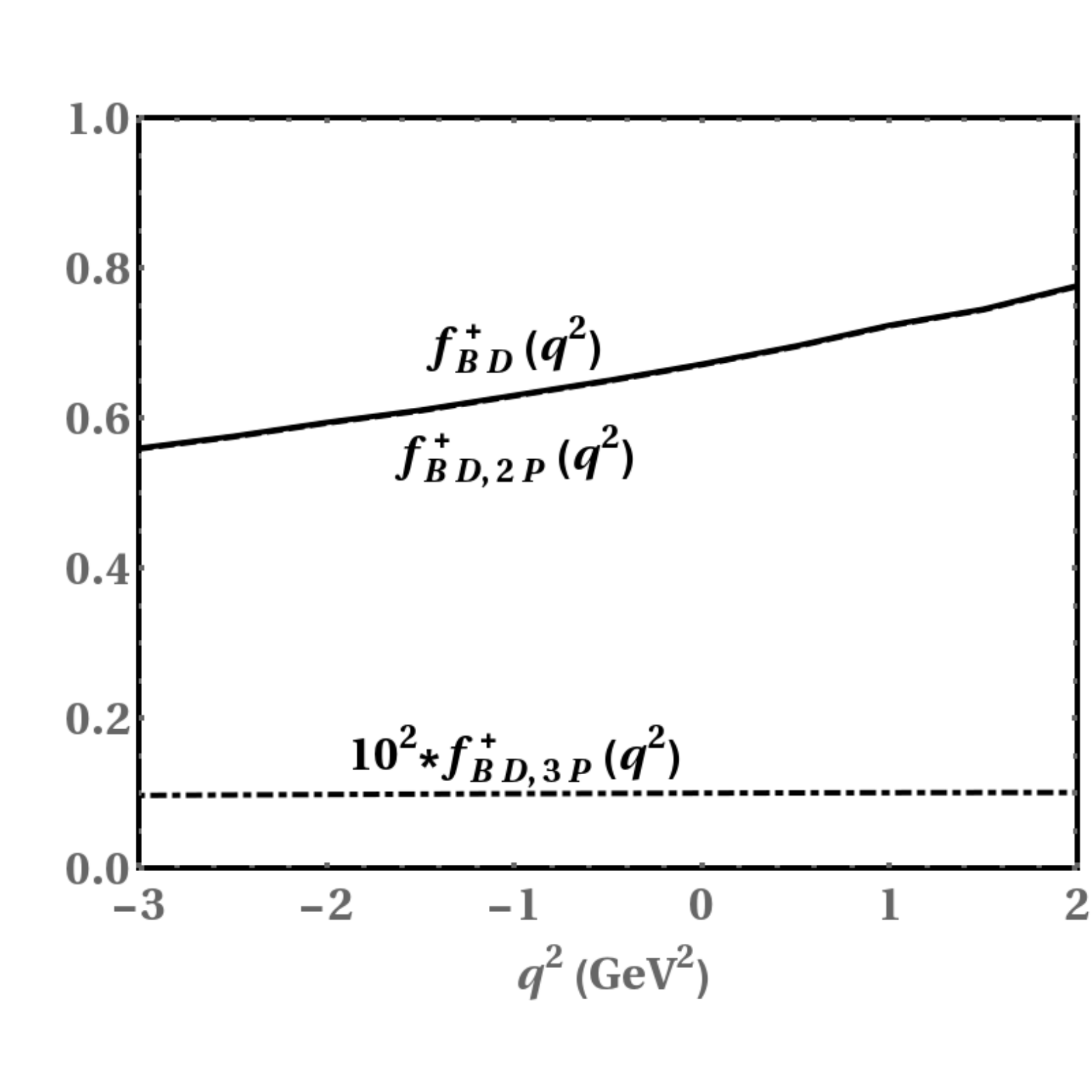}
\vspace*{0.1cm}
\caption{{\bf Top}: Dependence of the vector form factor $f_{B D}^{+}(0)$ on the Borel mass $M^2$
(left panel) and on the duality-threshold parameter $s_0$ (right  panel) at $\lambda_B(\mu_0) = 570 \, {\rm MeV}$.
The solid, dashed and dot-dashed curves are predicted from the LCSR (\ref{the final sum rules})
with $s_0=6.0 \, {\rm GeV^2}$,  $6.5 \, {\rm GeV^2}$  and  $5.5 \, {\rm GeV^2}$ (left panel),
and with $M^2=4.5 \, {\rm GeV^2}$,  $5.5 \, {\rm GeV^2}$  and  $3.5 \, {\rm GeV^2}$ (right  panel).
{\bf Bottom}: Breakdown of the form factor $f_{B D}^{+}(0)$ from the two-particle and from the three-particle contributions.
The two-particle contribution to $f_{B D}^{+}(0)$ indicated by the dashed curve is almost
indistinguishable from the total result represented by the solid curve, due to the negligible
effect from the three-particle $B$-meson DA as shown by the dot-dashed curve. }
\label{fig:SR parameter dependence of FFs and 3P effect}
\end{center}
\end{figure}

Now we are ready to explore the numerical features of the LCSR predictions for $B \to D$
form factors including the (partial) NLL resummation for the two-particle contributions and the subleading power corrections
from the three-particle quark-gluon DA.
To demonstrate the reliability of the sum rule calculations, we first display
the dependence of $f_{B D}^{+}(0)$ on the sum rule parameters $M^2$ and $s_0$ in figure
\ref{fig:SR parameter dependence of FFs and 3P effect}.
It is apparent that the variations of the Borel parameter and the effective threshold
within the intervals (\ref{choices of sum rule parameters})  only
bring about  negligible influence on  the sum rule predictions for the vector form factor
of $B \to D \ell \nu$.  We further present the separate contributions to  $f_{B D}^{+}(0)$
from the two-particle and the three-particle contributions in figure
\ref{fig:SR parameter dependence of FFs and 3P effect} in an attempt to understand the
subleading power corrections from the higher Fock states.
It can be readily observed that the tree-level contributions from the three-particle $B$-meson DA
turn out to be of minor importance numerically, approximately ${\cal O} (1 \%)$, compared to the two-particle
contributions. However,  it is worthwhile to mention that the smallness of the three-particle contribution
at leading order (LO) in perturbative expansion does not imply the insignificant impact of the subleading
power contributions  in the $B \to D \ell \nu$ decay amplitude in general, due to the yet
unaccounted power suppressed effects  induced by the  off light-cone corrections to the nonlocal  matrix element
defining the two-particle $B$-meson DA,  by the subleading power corrections to the perturbative  coefficient
functions entering the factorization formula (\ref{master factorization formula}) and by the additional
contributions generated by  new momentum regions (or equivalently, new field modes in the language of SCET)
when applying the method of regions to the evaluation of loop integrals involved in perturbative corrections
to the QCD amplitude.

\begin{figure}
\begin{center}
\includegraphics[width=0.4 \columnwidth]{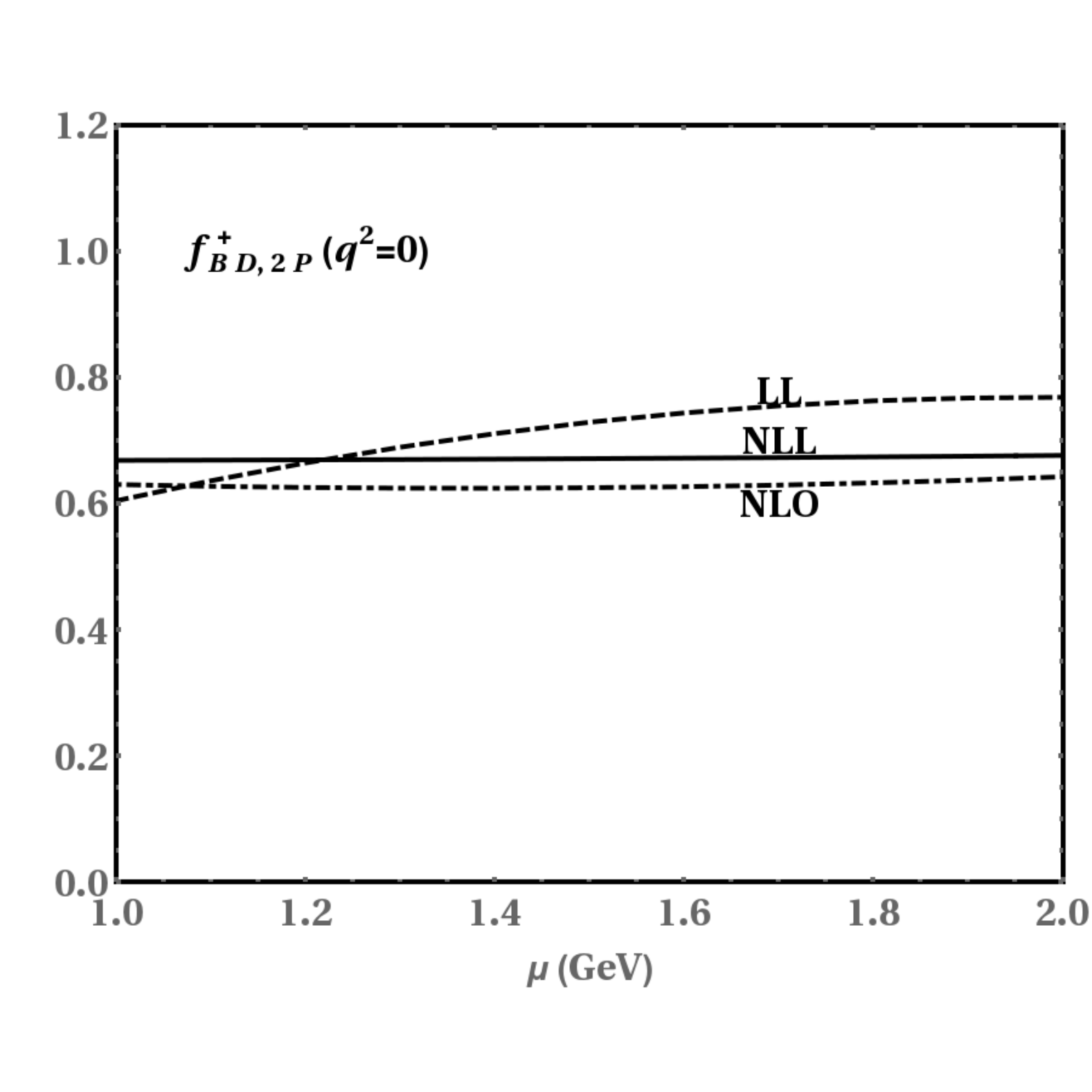} \hspace{1 cm}
\includegraphics[width=0.4 \columnwidth]{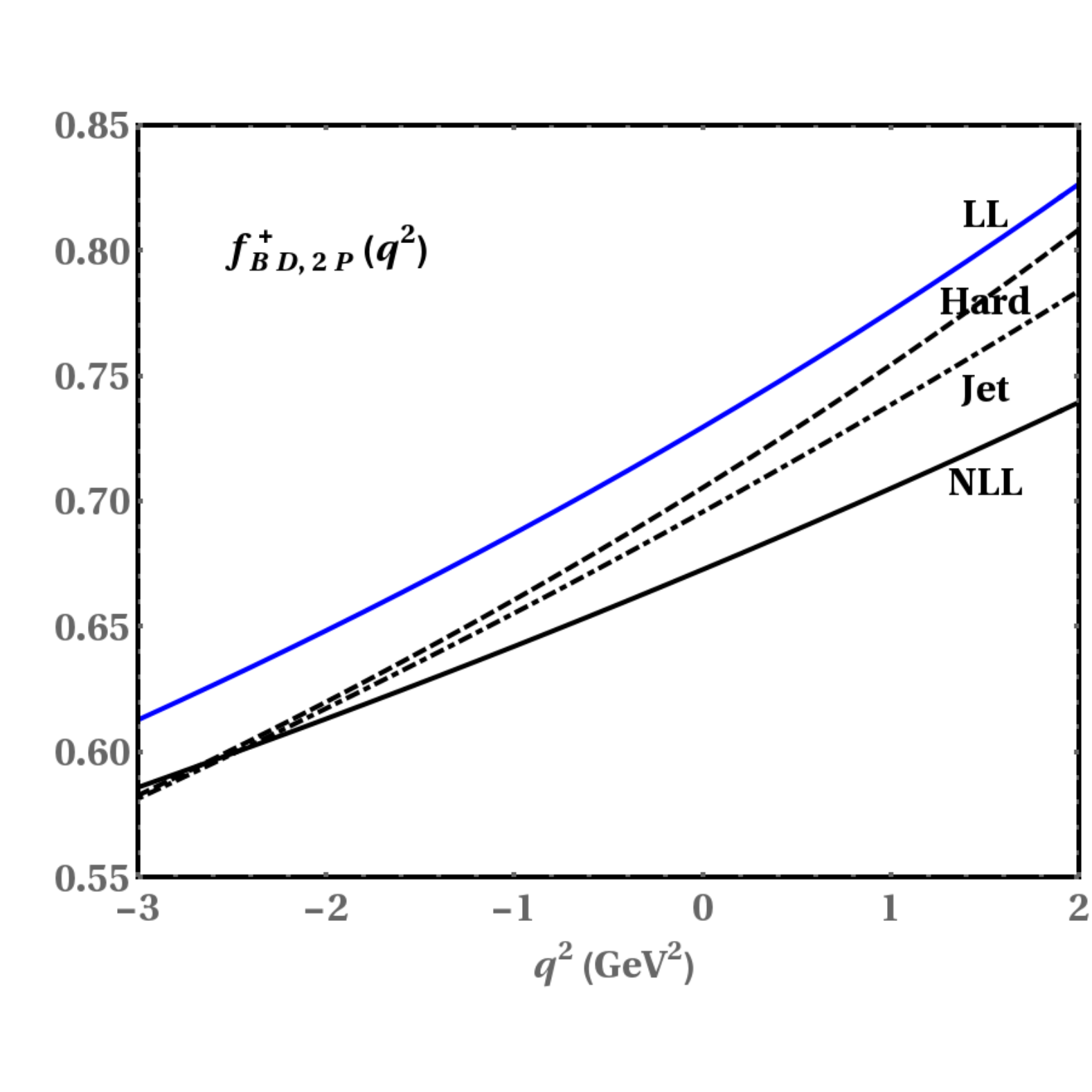}
\vspace*{0.1cm}
\caption{{\bf Left}: Factorization scale dependence of the two-particle contributions to the form factor
$f_{B D}^{+}(0)$ computed from the sum rules (\ref{NLL reummation improved SR for B to D FFs}).
The curves labelled with ``LL", ``NLO" and ``NLL" are obtained from the resulting predictions
at leading-logarithmic (LL), NLO and (partial) NLL accuracy.
{\bf Right}: Breakdown of the two-particle contributions to $f_{B D}^{+}(0)$
from the LL effect, from the NLL hard correction (``Hard") and from the NLO hard-collinear correction (``Jet"). }
\label{fig: NLO correction and resummation effect}
\end{center}
\end{figure}

\begin{figure}
\begin{center}
\includegraphics[width=0.4 \columnwidth]{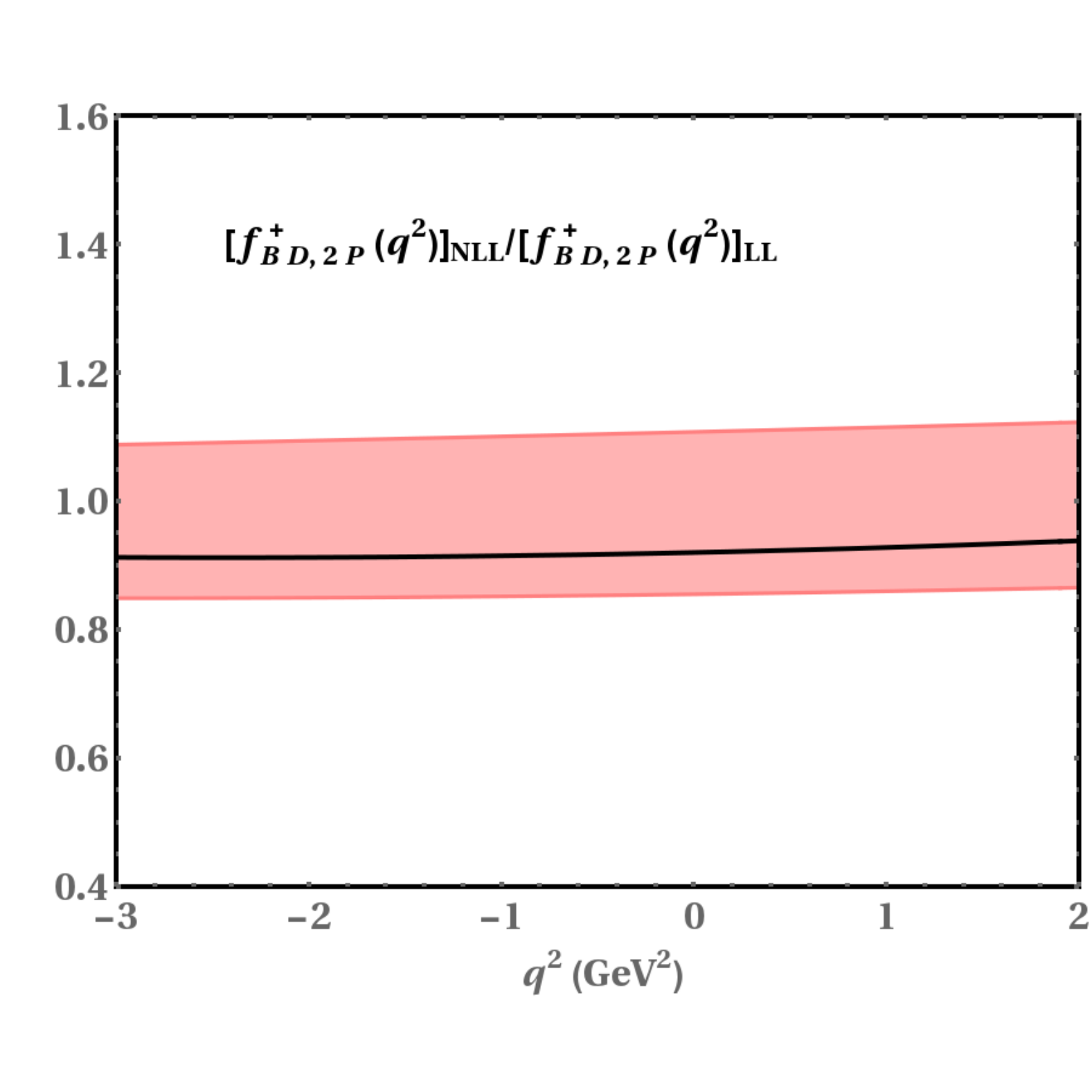} \hspace{1 cm}
\includegraphics[width=0.4 \columnwidth]{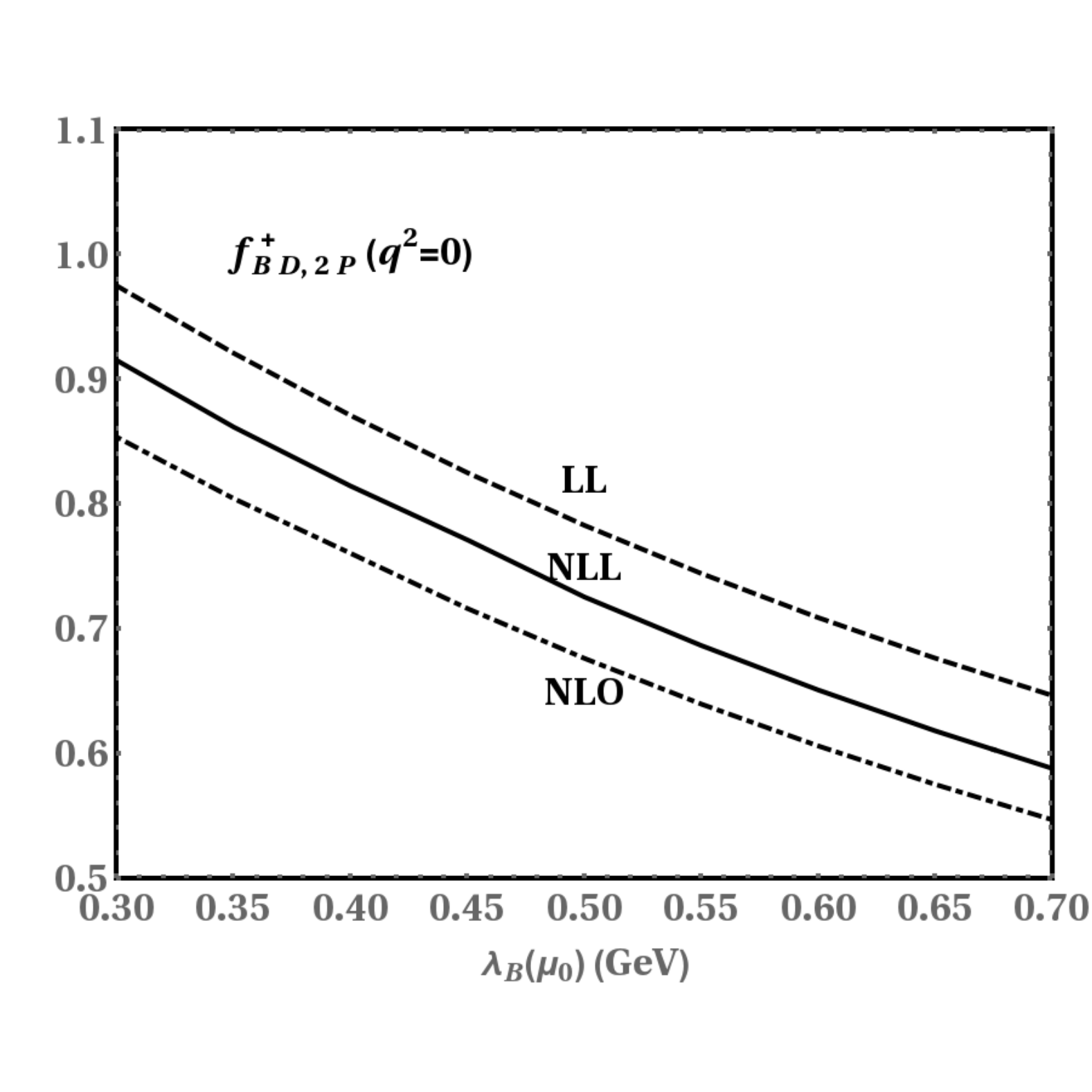}
\vspace*{0.1cm}
\caption{{\bf Left}: The $q^2$ dependence of the ratio
$[f_{B D, 2P}^{+}(q^2)]_{\rm NLL}/[f_{B D, 2P}^{+}(q^2)]_{\rm LL}$
with  uncertainties  from the variations of both the hard and hard-collinear scales.
{\bf Right}: Dependence of the two-particle contributions to $f_{BD}^{+}(q^2=0)$,
at LL, NLO and (partial) NLL accuracy, on the inverse moment of the leading twist
$B$-meson DA $\lambda_B(\mu_0)$. }
\label{fig: NLL correction and omega0 dependence}
\end{center}
\end{figure}

We proceed to investigate the impact of perturbative corrections to the short-distance functions
and resummation effects for the parametrically large logarithms on predicting
the form factors of $B \to D \ell \nu$.
It is evident from figure \ref{fig: NLO correction and resummation effect} that NLO QCD corrections
to the perturbative matching coefficients can reduce the tree-level prediction for $f_{B D}^{+}(0)$
by approximately $10 \%$ at $\mu =1.5 \, {\rm GeV}$ and the (partial) NLL resummmation effect can enhance the
NLO QCD calculation by an amount of $3 \%$ numerically at the same value of $\mu$.
Both the NLO and NLL predictions exhibit  weaker dependencies on the factorization scale
when compared to the LO result.
Inspecting the different origins of perturbative QCD corrections displayed in figure
\ref{fig: NLO correction and resummation effect} shows that the one-loop hard-collinear correction turns
out to be more pronounced than the corresponding hard correction, with the inverse moment
$\lambda_B(\mu_0) = 570 \, {\rm MeV}$, at $q^2 \geq 0$, highlighting the significance of computing the
NLO jet functions accomplished in this paper.
We further plot the theory predictions for the ratio
$[f_{B D, 2P}^{+}(q^2)]_{\rm NLL}/[f_{B D, 2P}^{+}(q^2)]_{\rm LL}$ in figure
\ref{fig: NLL correction and omega0 dependence}
with  uncertainties from the variations of both the hard and hard-collinear scales
within the intervals  given above.
To develop a better understanding of the sensitivity of $B \to D$ form factors on
the inverse moment $\lambda_B$, we also present the LO, NLO and (partial) NLL sum rule predictions
for the leading power contributions to $f_{B D}^{+}(q^2)$,
in figure \ref{fig: NLL correction and omega0 dependence},
in a wide range $300 \, {\rm MeV}  \leq \lambda_B(\mu_0) \leq 700 \, {\rm MeV}$.
We can readily observe that the sum rule predictions for  $f_{B D}^{+}(q^2)$
increase steadily with the reduction of $\lambda_B$, in analogy to the observation
for the radiative leptonic $B \to \gamma \ell \nu$ decay  in \cite{Wang:2016qii}.

As already explained in \cite{Faller:2008tr},  the light-cone OPE for the vacuum-to-$B$-meson
correlation function (\ref{correlator: definition}) can be only justified near the maximal recoil
to fulfill the power counting rules  $n \cdot p  \gg m_c \gg \Lambda$ and
$ m_c  \sim {\cal O}(\sqrt{n \cdot p \, \Lambda})$.
In addition,   QCD factorization for the correlation function (\ref{correlator: definition}) is fully
applicable at space-like momentum transfer on the basis of the power counting analysis.
It is then distinctly that the $B$-meson LCSR for the form factors of $B \to D \ell \nu$
derived in (\ref{the final sum rules}) can be trusted at
$q_{min}^2 \leq q^2 \leq q_{max}^2 = 2 \,\, {\rm GeV}^2$,
where a moderate value $q_{min}^2 = -3 \,\, {\rm GeV}^2$ will be employed (see also \cite{Khodjamirian:2009ys})
in the following analysis  for the sake of adopting the same intervals of the sum rules parameters
shown in (\ref{choices of sum rule parameters}).
In order to access the information of $B \to D$ form factors in the whole kinematic region,
we need to extrapolate the  LCSR predictions obtained above toward large momentum transfer
with a certain parametrization for the form factors.
Following the arguments of \cite{Bourrely:2008za,Khodjamirian:2009ys}, we will take advantage of
the $z$-series parametrization, in line with the analytical properties and perturbative QCD scaling
behaviours of $B \to D$ form factors, which can be readily introduced by mapping the cut complex $q^2$-plane onto the unit
disk $|z(q^2, t_0) \leq 1|$ according to the conformal transformation \cite{Bourrely:1980gp}
\begin{eqnarray}
z(q^2, t_0) = \frac{\sqrt{t_{+}-q^2}-\sqrt{t_{+}-t_0}}
{\sqrt{t_{+}-q^2}+\sqrt{t_{+}-t_0}}\,.
\label{definition of z parameter}
\end{eqnarray}
Here, the parameter $t_{+}=(m_B+m_D)^2$ is determined by the onset of the branching cut in the $| B \, D \rangle$ channel,
and $t_0 < t_{+}$ is an auxiliary parameter determining the value of $q^2$ mapped to the origin in the $z$-plane.
An optimal choice of $t_{0}=(m_B-m_D)^2$ will be implemented in the numerical computation to achieve a narrow interval
of $|z|$.

Taking into account the near-threshold behaviour from angular momentum conservation leads to
the suggested parametrization \cite{Bourrely:2008za,Khodjamirian:2009ys}
\begin{eqnarray}
f^+_{B D}(q^2) &=& \frac{f^+_{B D}(0)}{1-q^2/m_{B_c^{*}}^2}
\Bigg\{1+\sum \limits_{k=1}^{N-1}b_k\,\Bigg(z(q^2,t_0)^k- z(0,t_0)^k \nonumber\\
&&  \hspace{2.5 cm} - \, (-1)^{N-k} \frac{k}{N}\bigg[z(q^2,t_0)^N- z(0,t_0)^N \bigg] \Bigg) \Bigg\} \,
\label{z parametrization of the vector FF}
\end{eqnarray}
for the vector form factor, where $m_{B_c^{*}}= \left ( 6.330 \pm 0.009 \right ) \, {\rm GeV}$ \cite{Gregory:2009hq},
and the $z$-series expansion will be truncated at $N=2$ in the practical matching procedure.
Along the same vein, we will employ the following parametrization
\begin{eqnarray}
f^0_{B D}(q^2) = \frac{f^0_{B D}(0)}{1-q^2/m_{B_c^{(0)}}^2} \Bigg \{ 1+\sum  \limits_{k=1}^{N} \tilde{b}_k\,
\Bigg ( z(q^2,t_0)^k- z(0,t_0)^k \Bigg ) \Bigg \} \,
\label{z parametrization of the scalar FF}
\end{eqnarray}
for the scalar form factor, where $m_{B_c^{(0)}}= \left ( 6.420 \pm 0.009 \right ) \, {\rm GeV}$ \cite{Na:2015kha},
and we will only keep the terms up to ${\cal O}(z)$ in the $z$-expansion.
An alternative parametrization of $B \to D$ form factors proposed in \cite{Boyd:1995cf}
(see also \cite{Bigi:2016mdz} for a recent discussion) including the outer function
and the Blaschke factor will not be considered here following the reasonings of \cite{Bourrely:2008za}.

\begin{figure}
\begin{center}
\includegraphics[width=0.4 \columnwidth]{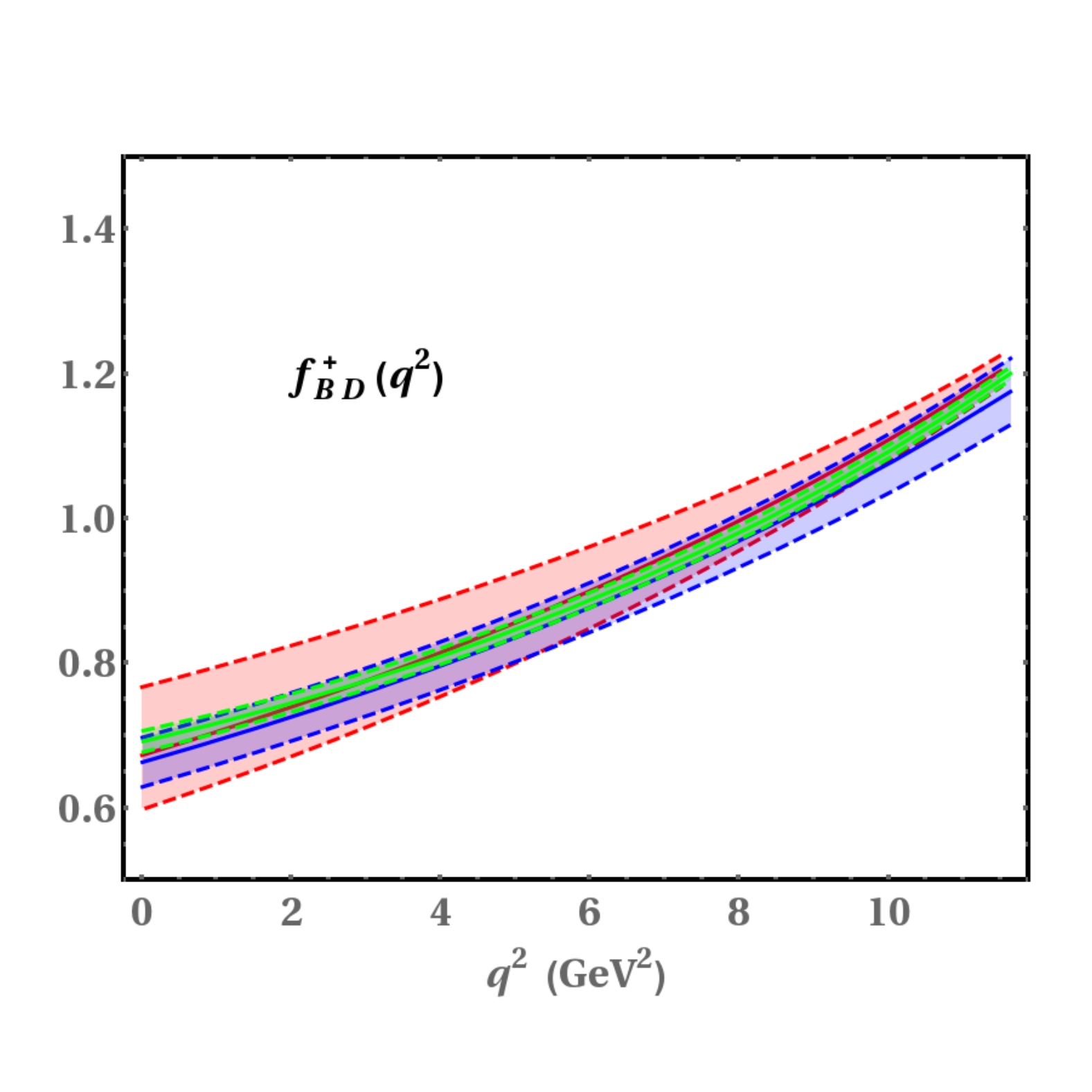} \hspace{1 cm}
\includegraphics[width=0.4 \columnwidth]{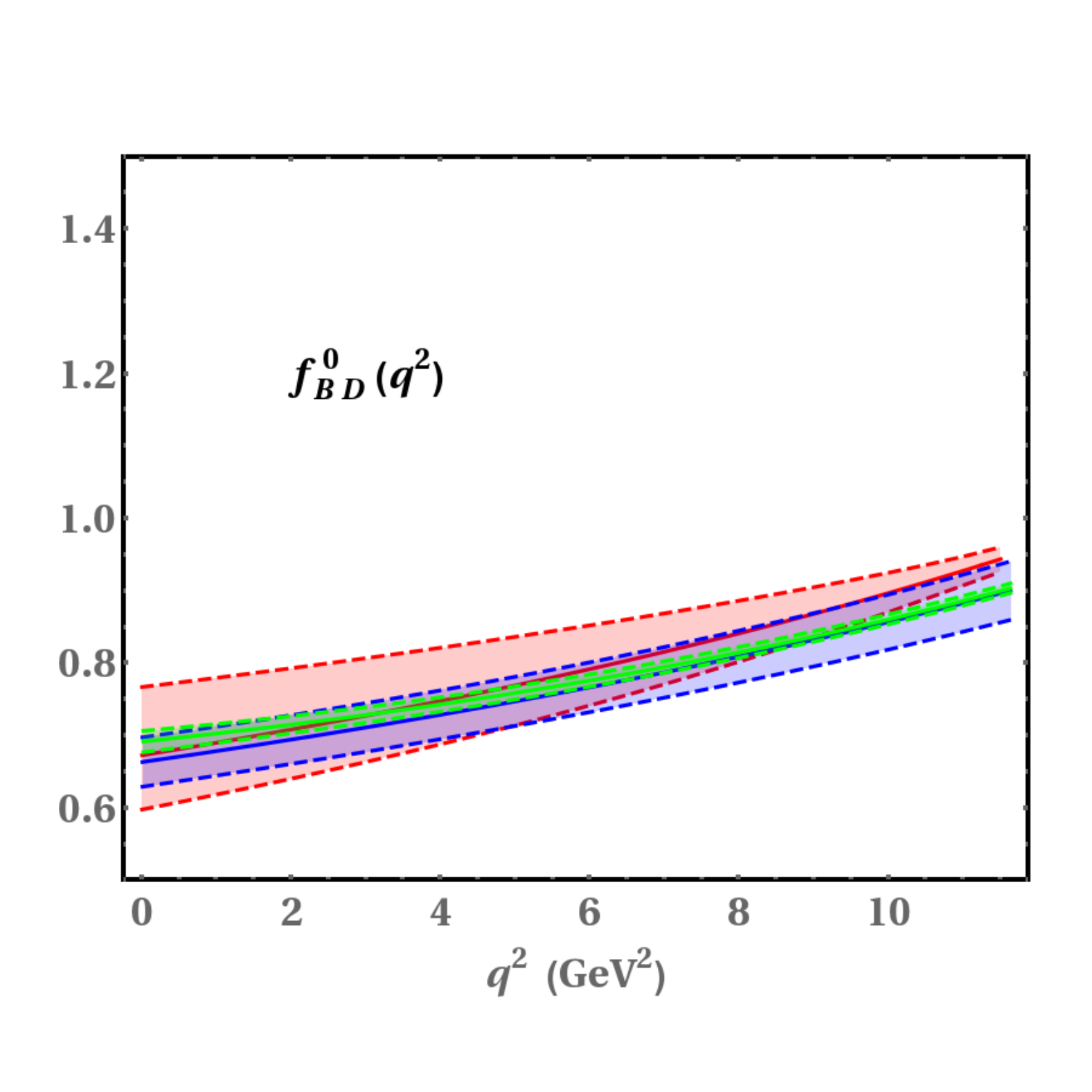}
\vspace*{0.1cm}
\caption{The transfer momentum  dependence of the form factor $f_{BD}^{+}(q^2)$ (left panel)
and of the form factor $f_{BD}^{0}(q^2)$ (right panel) predicted from the LCSR calculations,
including the two-particle contributions at (partial) NLL accuracy and the tree-level
three-particle contributions, with an extrapolation toward  large $q^2$ applying the $z$-series parametrization.
The pink, blue and green curves correspond to  theory predictions from this work, from the lattice QCD
calculations by the HPQCD Collaboration \cite{Na:2015kha}, and from a combined fit of the BarBar and Belle
data as well as the HPQCD and  FNAL/MILC calculations \cite{DeTar:2015orc}. Theory uncertainties for all the calculations
are indicated by the shaded regions. }
\label{fig: the transfer momentum dependence of form factors}
\end{center}
\end{figure}

It is now a straightforward task to implement the matching procedure described above with the aid of
the LCSR predictions at  $q_{min}^2 \leq q^2 \leq q_{max}^2$ and the $z$-series parametrization for the
determination of the momentum-transfer dependence in the entire kinematic region.
To achieve a better accuracy for the resulting shape parameters $b_1$ and $\tilde{b}_1$, we will further employ
the synthetic data points for the form factors $f_{B D}^{+, 0}(q^2)$ at $q^2= 8.47 \, {\rm GeV^2}$,
$10.05 \, {\rm GeV^2}$ and $11.63 \, {\rm GeV^2}$ from the FNAL/MILC Collaboration \cite{Lattice:2015rga}
in the numerical fitting. We present the yielding predictions for the $q^2$ dependence of $f_{B D}^{+, 0}(q^2)$
in the physical kinematic range $0 \leq q^2 \leq (m_B-m_D)^2$ with theory uncertainties in figure
\ref{fig: the transfer momentum dependence of form factors}
where recent determinations from the lattice QCD simulation combined with a similar $z$-parametrization
by the HPQCD Collaboration \cite{Na:2015kha} and from a joint fit of the BaBar and Belle data combined
with the lattice calculations \cite{DeTar:2015orc} are also shown for a comparison.
It is apparent that our predictions for the $B \to D \ell \nu$ form factors are in good agreement with
those displayed in  \cite{Na:2015kha,DeTar:2015orc} within the theory uncertainties.
We further collect the obtained results for the  shape parameters, with numerically important uncertainties,
entering the $z$-series expansion in Table \ref{table: fitted results for shape parameters},
where the predictions for form factors at $q^2=0$ are also presented for completeness.
It turns out that the dominant theory uncertainties for the shape parameters arise from the variations
of the hard scales $\mu_{h1(2)}$ and from the errors in the determination of $\omega_0$.
The relatively more precise predictions for the form factors of $B \to D \ell \nu$  at high $q^2$ are mainly
due to the  high precision lattice date points from  the FNAL/MILC Collaboration \cite{Lattice:2015rga},
whose accuracy is even significantly better than that computed by the HPQCD Collaboration \cite{Na:2015kha}.
Moreover, the LCSR predictions for $f_{B D}^{+, 0}(q^2)$ appear to grow faster with the increase of the momentum
transfer squared compared to those obtained from the lattice calculations \cite{Na:2015kha}.

\begin{table}[t!bph]
\begin{center}
\begin{tabular}{|c|c|c|c|c|c|c|c|c|}
  \hline
  \hline
&&&&&&&& \\[-3.5mm]
 Parameter & Central \,\, value & $\omega_0$ & $\sigma_1$ & $\mu$ & $\mu_{h1(2)}$ &
$M^2$ & $s_0$ &  $\phi_B^{\pm}(\omega)$ \\
  \hline
    &&&&&&&& \\[-1mm]
  $f^+_{B D}(0)$ &$0.673$ & $^{-0.027}_{+0.026}$ & $^{-0.01}_{+0.01}$ &$^{+0.015}_{-0.031}$&
$^{+0.058}_{-0.035}$&
$^{-0.003}_{+0.005}$ & $^{+0.015}_{-0.019}$ & - \\
  &&&&&&&& \\[-2mm]
\hline
&&&&&&&& \\[-3mm]
  $b_1$ & $-4.20$ & $^{-0.80}_{+0.74}$ & $^{-0.31}_{+0.30}$ & $^{+0.19}_{-0.14}$ &
$^{+1.51}_{-1.02}$ & $^{-0.06}_{+0.14}$ & $^{+0.44}_{-0.56}$ & $^{+0.11}_{-0.00}$ \\[2mm]
\hline
&&&&&&&& \\[-2mm]
   $\tilde{b}_1$ & $-0.18$ & $^{-0.66}_{+0.60}$  & $^{-0.26}_{+0.25}$ & $^{+0.17}_{-0.14}$ &
$^{+1.23}_{-0.84}$  & $^{-0.05}_{+0.10}$ & $^{+0.36}_{-0.46}$ & $^{+0.10}_{-0.00}$ \\[2mm]
    \hline
    \hline
\end{tabular}
\end{center}
\caption{Summary of the predicted shape parameters and the normalization constant entering the $z$-series
parametrizations (\ref{z parametrization of the vector FF}) and (\ref{z parametrization of the scalar FF})
for the $B \to D \ell \nu$ form factors with dominant uncertainties from the variations of theory inputs. }
\label{table: fitted results for shape parameters}
\end{table}

Another popular parametrization for the form factors of $B \to D \ell \nu$ taking into account
the constraints from the heavy quark symmetry, proposed in \cite{Caprini:1997mu}, was also extensively employed
in the phenomenological applications. Including the subleading power corrections to the heavy-quark relations and
implementing the dispersive analysis  yield the following one-parameter representations for the vector and scalar
form factors \cite{Caprini:1997mu,Bigi:2016mdz}
\begin{eqnarray}
f^{+}_{B D}(z) &=& f^{+}_{B D}(0) \, \left [ 1 - 8 \, \rho^2 \, z \,
+ (51 \, \rho^2 - 10) \,   z^2  - (252 \, \rho^2 - 84) \,   z^3 \right ] \,,  \\
\frac{f^{0}_{B D}(z)}{f^{+}_{B D}(z)} &=&  \left ( \frac{2 \, \sqrt{r}}{1+r}  \right )^2  \,
{ 1 + \omega \over 2} \, 1.0036 \, \left [ 1 + 0.0068 \, \bar \omega + 0.0017 \, \bar \omega^2
 + 0.0013 \, \bar \omega^3 \right ]    \,,
\label{CLN parametrization for form factors}
\end{eqnarray}
where we have introduced the following conventions
\begin{eqnarray}
z = \frac{\sqrt{1+\omega} - \sqrt{2}}{\sqrt{1+\omega} + \sqrt{2}} \,,  \qquad
\omega= \frac{m_B^2+m_D^2-q^2}{2 \, m_B \, m_D} \,, \qquad
\bar \omega = 1-\omega \,, \qquad  r=m_D/m_B   \,.
\end{eqnarray}
Apparently, this $z$ parameter is equivalent to $z(q^2, t_0)$,
defined by (\ref{definition of z parameter}), with $t_{0}=(m_B-m_D)^2$.
It needs to point out that the ratio $f^{0}_{B D}(z)/f^{+}_{B D}(z)$
is fully determined in HQET including both perturbative corrections to the leading
Wilson coefficients and subleading power contributions
computed from QCD sum rules.
Matching the LCSR calculations for the form factors of $B \to D \ell \nu$
at $q_{min}^2 \leq q^2 \leq q_{max}^2$ onto the CLN parametrization
(\ref{CLN parametrization for form factors}) leads to
\begin{eqnarray}
f^{+}_{B D}(z=0)=1.22 \pm 0.02 \,, \qquad \rho= 1.07^{+0.08}_{-0.11}  \,,
\end{eqnarray}
which are well consistent with the fitted values obtained in \cite{Na:2015kha},
albeit with the comparably large theory uncertainties for the slop parameter $\rho$.

\subsection{Phenomenological implications}

\begin{figure}
\begin{center}
\includegraphics[width=0.6 \columnwidth]{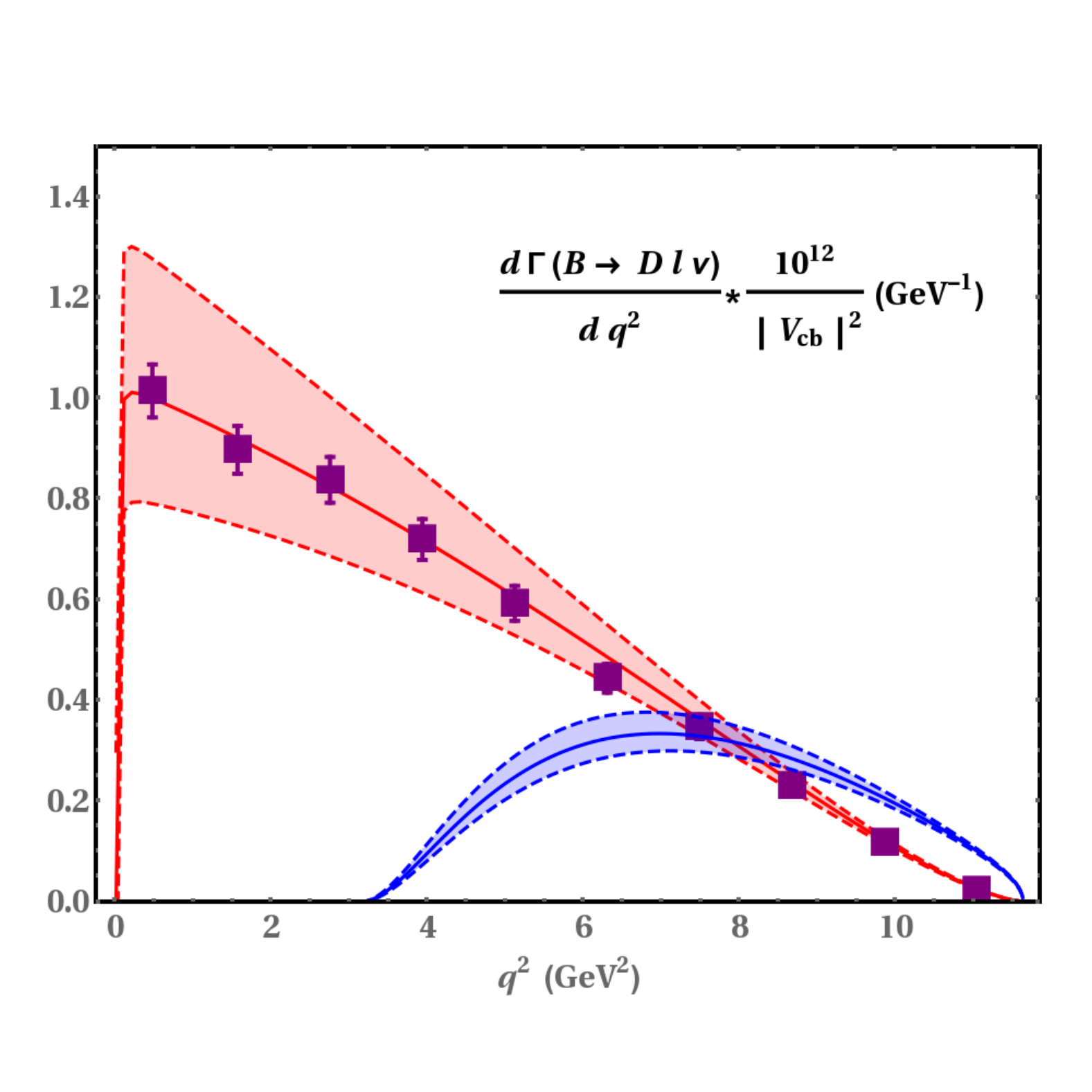}
\vspace*{0.1cm}
\caption{The differential decay rates for $B \to D \mu \nu_{\mu}$ (pink band) and $B \to D \tau \nu_{\tau}$
(blue band) as a function of the momentum transfer squared $q^2$, where the experimental data points
for $B \to D \mu \nu_{\mu}$ (purple squares) from the Belle Collaboration \cite{Glattauer:2015teq}  are also presented
for a comparison. }
\label{fig: differential decay width}
\end{center}
\end{figure}

Having in our hands the theory predictions of the two form factors $f^{+,0}_{B D}(q^2)$,
we are ready to explore their phenomenological implications on the semileptonic
$B \to D \ell \nu$ decays. The differential decay rate of  $B \to D \ell \nu$
in the rest frame of the $B$-meson can be computed as
\begin{eqnarray}
\frac{d \Gamma (B \to D \ell \nu)}{d q^2}
&=& \frac{\eta_{\rm EW}^2 \, G_F^2 \, |V_{cb}|^2}{24 \, \pi^3 \, m_B^2} \,
\left (1 - { m_l^2 \over q^2} \right )^2  \, | \vec{p}_D | \,
\bigg [ \left ( 1 + {m_l^2 \over 2 \, q^2} \right )\, m_B^2 \,
| \vec{p}_D |^2 \, \left | f^{+}_{B D}(q^2) \right |^2 \nonumber  \\
&& + {3 \, m_l^2 \over 8 \, q^2} \, (m_B^2 -m_D^2)^2 \,
\left | f^{0}_{B D}(q^2) \right |^2  \bigg ]  \,,
\end{eqnarray}
where $| \vec{p}_D | = \sqrt{\lambda \left (m_B^2, m_D^2, q^2 \right )}/ (2 \, m_B)$
with $\lambda(a, b, c)=a^2+b^2+c^2- 2 a b - 2 a c - 2 b c$ is the magnitude of the three-momentum
of the $D$-meson and
\begin{eqnarray}
\eta_{\rm EW} = 1 + {\alpha_{\rm em} \over \pi }\, \ln \left ( {m_Z \over m_B} \right )  \simeq 1.0066
\end{eqnarray}
originates from the  short-distance QED corrections to the four-fermion
operator responsible for the $B \to D \ell \nu$ decays \cite{Sirlin:1981ie,Carrasco:2015xwa}.

\begin{table}
\begin{center}
\begin{tabular}{|c|c||c|c|}
  \hline
  \hline
  & & &  \\
  $[t_1, t_2] \,$ & $\Delta \Gamma_{\mu}(t_1, t_2)$ \,\,\, ($10^{-12} \, {\rm GeV}$) &
  $[t_1, t_2] \,$ & $\Delta \Gamma_{\mu}(t_1, t_2)$ \,\,\, ($10^{-12} \, {\rm GeV}$) \\
  & & &  \\
  $ \left ({\rm GeV^2} \right)$ &this work  \hspace{1 cm} Belle \cite{Glattauer:2015teq}  &
   $ \left ({\rm GeV^2} \right)$ &this work  \hspace{1 cm} Belle \cite{Glattauer:2015teq}   \\
  & & &  \\
  \hline
  & & &   \\
  $[0.00, 0.98]$ & $1.00^{+0.28}_{-0.21}$ \hspace{0.8 cm} $1.01 \pm 0.05$
  &   $[5.71, 6.90]$ & $0.57^{+0.08}_{-0.06}$ \hspace{0.8 cm} $0.53 \pm 0.03$ \\
  & & &    \\
  $[0.98, 2.16]$ & $1.09^{+0.27}_{-0.21}$ \hspace{0.8 cm} $1.06 \pm 0.06$
  &  $[6.90, 8.08]$ & $0.43^{+0.05}_{-0.04}$ \hspace{0.8 cm} $0.41 \pm 0.03$ \\
  & & &    \\
  $[2.16, 3.34]$ & $0.97^{+0.21}_{-0.16}$ \hspace{0.8 cm} $0.99 \pm 0.05$
  &  $[8.08, 9.26]$ & $0.28^{+0.02}_{-0.02}$ \hspace{0.8 cm} $0.27 \pm 0.02$  \\
   & & &    \\
  $[3.34, 4.53]$ & $0.85^{+0.16}_{-0.13}$ \hspace{0.8 cm} $0.85 \pm 0.05$
  &  $[9.26, 10.45]$ & $0.14^{+0.01}_{-0.01}$ \hspace{0.8 cm} $0.14 \pm 0.01$  \\
   & & &   \\
  $[4.53, 5.71]$ & $0.72^{+0.11}_{-0.09}$ \hspace{0.8 cm} $0.70 \pm 0.04$
  &  $[10.45, 11.63]$ & $0.03^{+0.00}_{-0.00}$ \hspace{0.8 cm} $0.02 \pm 0.01$ \\
   & & &    \\
  \hline
  \hline
\end{tabular}
\end{center}
\caption{Theory predictions for the partial decay rates of $B \to D \mu \nu_{\mu}$ compared with the
Belle measurements from \cite{Glattauer:2015teq}.}
\label{table: binned q2 distribution for B to D mu nu}
\end{table}

The  differential $q^2$ distributions for $B \to D \ell \nu$  obtained with the form factors $f_{B D}^{+, 0}(q^2)$ displayed
in Table \ref{table: fitted results for shape parameters} are plotted in figure \ref{fig: differential decay width},
including also the recent experimental measurements for the combination of $B^{+} \to \bar D^{0} e^{+} \nu_e$,
$B^0  \to D^{-} e^{+} \nu_e$, $B^{+} \to \bar D^{0} \mu^{+} \nu_{\mu}$ and
$B^0  \to D^{-} \mu^{+} \nu_{\mu}$  from the Belle Collaboration \cite{Glattauer:2015teq}.
It is evident that our predictions for the $q^2$ shape of $B \to D \mu \nu_{\mu}$ are in excellent agreement with
the experimental data bins.
Furthermore, we collect the numerical results for the (normalized) partial decay rates of
$B \to D \ell \nu$
\begin{eqnarray}
\Delta \Gamma_{\ell}(t_1, t_2) = \int_{t_1}^{t_2} \, d q^2  \,\,
\frac{d \Gamma (B \to D \ell \nu)}{d q^2}  \, \frac{1}{ |V_{cb}|^2} \,
\end{eqnarray}
in Tables \ref{table: binned q2 distribution for B to D mu nu} and
\ref{table: binned q2 distribution for B to D tau nu and RD} with selections of the $q^2$ bins
identical to that from the Belle and BaBar measurements \cite{Glattauer:2015teq,Lees:2013uzd}.

\begin{table}[t!bph]
\begin{center}
\begin{tabular}{|c|c||c|}
  \hline
  \hline
  & &   \\
  $[t_1, t_2] \,$ & $\Delta \Gamma_{\tau}(t_1, t_2)$ \,\,\, ($10^{-12} \, {\rm GeV}$) &
  $\Delta {\cal R}(t_1, t_2)$ \\
  & &   \\
  $ \left ({\rm GeV^2} \right)$ &this work  &
   this work  \hspace{1 cm} \cite{Celis:2016azn}   \\
  & &   \\
  \hline
  & &    \\
  $[4.00, 4.53]$ & $0.073^{+0.014}_{-0.011}$
 & $0.199^{+0.002}_{-0.002}$ \hspace{0.8 cm} $0.199 \pm 0.001$ \\
  & &     \\
  $[4.53, 5.07]$ & $0.11^{+0.02}_{-0.02}$
 & $0.331^{+0.003}_{-0.003}$ \hspace{0.8 cm} $0.330 \pm 0.001$ \\
  & &     \\
  $[5.07, 5.60]$ & $0.14^{+0.02}_{-0.02}$
 & $0.458^{+0.004}_{-0.004}$ \hspace{0.8 cm} $0.455 \pm 0.001$ \\
  & &     \\
  $[5.60, 6.13]$ & $0.16^{+0.02}_{-0.02}$
 & $0.575^{+0.006}_{-0.005}$ \hspace{0.8 cm} $0.571 \pm 0.002$ \\
  & &     \\
  $[6.13, 6.67]$ & $0.18^{+0.02}_{-0.02}$
 & $0.687^{+0.007}_{-0.006}$ \hspace{0.8 cm} $0.680 \pm 0.002$ \\
  & &     \\
  $[6.67, 7.20]$ & $0.18^{+0.02}_{-0.02}$
 & $0.796^{+0.008}_{-0.007}$ \hspace{0.8 cm} $0.786 \pm 0.003$ \\
  & &     \\
  $[7.20, 7.73]$ & $0.17^{+0.02}_{-0.02}$
 & $0.905^{+0.009}_{-0.007}$ \hspace{0.8 cm} $0.892 \pm 0.003$ \\
  & &     \\
  $[7.73, 8.27]$ & $0.17^{+0.02}_{-0.01}$
 & $1.024^{+0.009}_{-0.008}$ \hspace{0.8 cm} $1.006 \pm 0.004$ \\
  & &     \\
  $[8.27, 8.80]$ & $0.15^{+0.01}_{-0.01}$
 & $1.161^{+0.010}_{-0.009}$ \hspace{0.8 cm} $1.135 \pm 0.005$ \\
  & &     \\
  $[8.80, 9.33]$ & $0.14^{+0.01}_{-0.01}$
 & $1.329^{+0.011}_{-0.009}$ \hspace{0.8 cm} $1.294 \pm 0.006$ \\
  & &     \\
  $[9.33, 9.86]$ & $0.12^{+0.01}_{-0.01}$
 & $1.561^{+0.011}_{-0.010}$ \hspace{0.8 cm} $1.513 \pm 0.007$ \\
  & &     \\
 $[9.86, 10.40]$ & $0.099^{+0.006}_{-0.005}$
 & \hspace{-0.5 cm} $1.934^{+0.012}_{-0.010}$ \hspace{0.8 cm} $1.86 \pm 0.01$ \\
  & &     \\
 $[10.40, 11.63]$ & $0.12^{+0.01}_{-0.01}$
 & \hspace{-1.1 cm} $3.364^{+0.008}_{-0.008}$ \hspace{1.8 cm} $-$ \\
  & &     \\
  \hline
  \hline
\end{tabular}
\end{center}
\caption{Theory predictions for the partial decay rates of $B \to D \tau \nu_{\tau}$
and for the binned distributions of $\Delta {\cal R}(t_1, t_2)$ defined in (\ref{differential R ratio}).
The semileptonic $B \to D $ form factors obtained by fitting the experimental data
with the Boyd-Grinstein-Lebed parametrization \cite{Boyd:1995cf} are employed for the
recent calculations presented in \cite{Celis:2016azn}. }
\label{table: binned q2 distribution for B to D tau nu and RD}
\end{table}

In particular, our predictions for the total decay width of  $B \to D \mu \nu_{\mu}$
in units of $1 /|V_{cb}|^2$ can be obtained straightforwardly
\begin{eqnarray}
\Delta \Gamma_{\mu}(0, 11.63 \, {\rm GeV}^2) = \left ( 6.06^{+1.18}_{-0.92} \right )
\times 10^{-12} \, {\rm GeV} \,,
\end{eqnarray}
with all separate uncertainties  from variations of  the theory inputs added  in quadrature,
from which the exclusive determinations of $|V_{cb}|$ are achieved
\begin{eqnarray}
|V_{cb}| = \left\{
\begin{array}{l}
\left ( 39.2^{+3.4}_{-3.3} \big |_{\rm th} \pm 1.0 \big |_{\rm exp} \right )  \times 10^{-3}\,,  \qquad
[{\rm BaBar} \,\,  2010] \vspace{0.4 cm} \\
\left ( 40.6^{+3.5}_{-3.5} \big |_{\rm th} \pm 1.0 \big |_{\rm exp} \right )  \times 10^{-3}\,,  \qquad
[{\rm Belle} \,\,  2016]
\end{array}
 \hspace{0.5 cm} \right.
 \label{extracted values of Vcb}
\end{eqnarray}
with the recent experimental measurements of the total branching fraction from the Belle \cite{Glattauer:2015teq}
and BaBar \cite{Aubert:2009ac} Collaborations \footnote{The previous measurements from ALEPH, CLEO, Belle and BaBar
summarized by the Heavy Flavor Averaging Group (HFAG) \cite{Amhis:2016xyh} are not considered here.
We leave a dedicated study of the $|V_{cb}|$ determination including all the available experimental data
and the correlation of theory predictions for the $B \to D \ell \nu$ form factors at different $q^2$ for
future work.}.  The resulting determinations of $|V_{cb}|$ suffer from  a sizeable uncertainty,
approximately ${\cal O} (10 \%)$, due to the  LCSR calculations of $B \to D$ form factors at large hadronic recoil.
Our results are consistent with the more precise determinations from FNAL/MILC \cite{Lattice:2015rga},
HPQCD \cite{Na:2015kha} and from a joint fit  \cite{Bigi:2016mdz} of the available experimental data and the lattice calculations
including the updated unitarity bounds.

Finally,  we turn to compute the differential distributions of the celebrated ratio
\begin{eqnarray}
\Delta {\cal R}(t_1, t_2) = \frac{ \int_{t_1}^{t_2} \, d q^2  \,\,
d \Gamma (B \to D \tau \nu_{\tau}) / d q^2  } { \int_{t_1}^{t_2} \, d q^2  \,\,
d \Gamma (B \to D \mu \nu_{\mu}) / d q^2 }    \,,
\label{differential R ratio}
\end{eqnarray}
where most of the hadronic uncertainties from the $B \to D \ell \nu$ form factors are cancelled out.
Inspecting the obtained results  in  Table \ref{table: binned q2 distribution for B to D tau nu and RD}
indeed implies an incredibly precise one-percent accuracy  of $\Delta {\cal R}(t_1, t_2)$, albeit with the
implementation of  much less accurate  predictions for the hadronic form factors shown in Table
\ref{table: fitted results for shape parameters}.  Our predictions for the binned  distributions of
$\Delta {\cal R}(t_1, t_2)$ are also compatible with the recent determinations reported in \cite{Celis:2016azn},
employing the $B \to D \ell \nu$ form factors extracted from a joint fit of the experimental data and two
recent lattice calculations  \cite{Bigi:2016mdz}. We further present our predictions for  the ratio of
the total branching fractions of two semileptonic decay channels
\begin{eqnarray}
R(D) \equiv  \frac{{\cal BR} (B \to D \tau \nu_{\tau})}{{\cal BR} (B \to D \mu \nu_{\mu})}
= 0.305^{+0.022}_{-0.025}  \,,
\label{R ratio}
\end{eqnarray}
which coincides with the  previous determinations \cite{Kamenik:2008tj,Becirevic:2012jf,Lattice:2015rga,Na:2015kha,Bigi:2016mdz,Li:2016vvp,Celis:2016azn}
in the Standard Model (SM) at the $1 \, \sigma$ level and needs to be compared with the HFAG average value
$R(D) \big |_{\rm HFAG} = 0.403 \pm 0.040 \pm 0.024$ \cite{Amhis:2016xyh}.
We mention in passing that the relatively high theory uncertainty of $R(D)$ in (\ref{R ratio}),
approximately $8 \%$, can be traced back to the uncancelled  hadronic uncertainties for determining
the partial branching fraction of  $B \to D \mu \nu_{\mu}$ in the phase-space region $0 \leq q^2 \leq m_{\tau}^2$.

\section{Concluding discussion}
\label{sect:Conc}

In this paper we have presented perturbative  QCD corrections to the semileptonic $B \to D \ell \nu$
form factors with the power counting scheme $m_c \sim {\cal O} \left  (  \sqrt{\Lambda \, m_b}   \right )$,
at leading power in $\Lambda/m_b$, employing the LCSR with the two-particle $B$-meson DA.
QCD factorization for the vacuum-to-$B$-meson correlation function (\ref{correlator: definition}) was demonstrated
explicitly  at one loop  applying  the diagrammatic factorization approach.
Due to the appearance of a new hard-collinear scale $m_c$,
the resulting jet functions turn out to be more complex than the
counterparts in the evaluation of the correlation function for constructing the sum rules of $B \to \pi$ for factors.
Taking advantage of the evolution equation of the $B$-meson DA $\phi_B^{-}(\omega, \mu)$,
factorization-scale independence of  the  correlation function (\ref{correlator: definition}) was verified
at ${\cal O}(\alpha_s)$  with the obtained hard and hard-collinear functions.
The (partial) NLL resummation improved sum rules for $B \to D$ form factors
(\ref{NLL reummation improved SR for B to D FFs}) derived with the dispersion representations
in Appendix  \ref{app:spectral resp} constitute the main new ingredients  of this paper.
The subleading power contributions from the  three-particle quark-gluon $B$-meson DA were also computed
from the same LCSR method at tree level. In the light of the canonical behavious of the three-particle DA
of the $B$-meson from the QCD sum rule analysis  \cite{Khodjamirian:2006st}, the power suppressed
three-particle corrections were demonstrated to  invalidate the large-recoil symmetry relation between the vector
and scalar $B \to D$ form factors.

We proceeded to explore the  phenomenological implications of the resulting sum rules for the
$B \to D \ell \nu$ form factors applying two nonperturbative models for the $B$-meson DA
$\phi_B^{+}(\omega, \mu_0)$ inspired from the QCD sum rule calculations \cite{Grozin:1996pq,Braun:2003wx}.
The perturbative QCD corrections from the two-particle DA  were found to generate an approximately ${\cal O} (10 \%)$ shift to
the LL predictions for $f_{BD}^{+}(q^2)$ with the default theory inputs, and the one-loop hard-collinear corrections
appear to have a more profound influence at $q^2 \geq 0$ numerically when compared with the corresponding hard corrections.
Moreover, the subleading power effects from the three-particle $B$-meson DA were shown to be insignificant numerically.
The $z$-series expansion fulfilling  the analytical properties of $B \to D$ form factors was further
employed to extrapolate the (partial) NLL LCSR predictions toward the low recoil region.
In addition, we presented theory predictions for the binned distributions of the $B \to D \ell \nu$ decay rates and of the  ratio
$\Delta {\cal R}(t_1, t_2)$ by applying our determinations of the form factors $f_{BD}^{+, 0}(q^2)$.
Matching the predicted results for the normalized decay width of $B \to D \mu \nu_{\mu}$ and the recent experimental measurements
from  Belle  \cite{Glattauer:2015teq} and BaBar \cite{Aubert:2009ac} led to the extracted values of
$|V_{cb}|$ at ${\cal O} (10 \%)$ accuracy  as displayed in (\ref{extracted values of Vcb}).
Our predictions for the $B \to D$ form factors also gave rise to the determinations of $R(D)$
presented in (\ref{R ratio}), confronted with the HFAG average value in \cite{Amhis:2016xyh}.

Further developments of the $B$-meson LCSR approach for computing  the form factors of $B \to D \ell \nu$
can be pushed forward in distinct directions. First, perturbative QCD corrections to the LCSR
(\ref{3-particle contribution to the LCSR}) from the three-particle $B$-meson DA can be carried out
for a complete understanding of the leading power contributions in the heavy quark limit.
In doing so, the one-loop evolution equations for the remaining high-twist three-particle DA of the $B$-meson
in  (\ref{def: 3P DA}) are in demand and they have been recently worked out in \cite{Braun:2017liq}.
Second, computing  yet higher-order QCD corrections to the two-particle contributions
(\ref{NLL reummation improved SR for B to D FFs}) are of both conceptual and phenomenological interest
for exploring the renormalization properties of the $B$-meson DA $\phi_B^{\pm}(\omega, \mu)$ at two loops
(e.g., the eigenfunctions and the analytical structures of renormalization kernels)
and for bringing down the still sizeable perturbative uncertainties of the theory predictions
displayed in Table \ref{table: fitted results for shape parameters}.
Third, the present strategies can be readily applied to calculate QCD corrections
to the  semileptonic $B \to D^{\ast} \ell \nu$ form factors based upon the LCSR with the $B$-meson DA (with additional
attention to the renormalization prescription of $\gamma_5$ in dimensional regularization), allowing for a comprehensive
analysis of the full angular distributions $B \to D^{\ast} (\rightarrow D \, X) \, \tau (\rightarrow Y \, \nu_{\tau}) \, \bar \nu_{\tau}$
with $X=(\pi, \gamma)$ and  $Y= (\ell \, \nu, \pi)$  as discussed in \cite{Ligeti:2016npd}.
To conclude, precision QCD calculations of  semileptonic $B$-meson  form factors with  analytical QCD
approaches will continually  provide us with a deeper insight into the strong interaction dynamics of heavy quark decays
and into the general properties of effective field theories.

\subsection*{Acknowledgements}

We are grateful to Carleton DeTar and Heechang Na  for providing us with the Lattice QCD results
presented in \cite{Na:2015kha,DeTar:2015orc}. C.D.L is supported in part by the National Natural Science Foundation
of China (NSFC) with Grant Nos. 11375208, 11521505, 11621131001 and 11235005,
and by the Open Project Program of State Key Laboratory of Theoretical Physics,
Institute of Theoretical Physics,  Chinese Academy of Sciences with Grant No. Y5KF111CJ.
The work of Y.L.S is supported by Natural Science Foundation of Shandong Province,
China under Grant No. ZR2015AQ006.
Y.M.W acknowledges support from the National Youth Thousand Talents Program,
the Youth Hundred Academic Leaders Program of Nankai University, and the NSFC with Grant No. 11675082.
This research was also supported in part by the Munich Institute for Astro- and Particle Physics (MIAPP)
of the DFG cluster of excellence ``Origin and Structure of the Universe".


\appendix


\section{Loop integrals}
\label{app:loop integrals}

Here we collect some useful results for the loop integrals used in the calculations
of the vacuum-to-$B$-meson correlation function (\ref{correlator: definition}) at ${\cal O}(\alpha_s)$.

\begin{eqnarray}
I_1^{hc}&=&\int [d \, l]  \, \frac{n \cdot (p+l)}{[ n \cdot (p+l) \,
\bar n \cdot (p-k+l) + l_{\perp}^2  -m_c^2 + i 0][ n \cdot l+ i 0] [l^2+i0]} \nonumber \\
&=& {1 \over 2} \,\, \bigg \{ {2 \over \epsilon^2} +
{2 \over \epsilon} \, \left [ \ln {\mu^2 \over  n \cdot p \, (\omega - \bar n \cdot p)}
- \ln \left (1+ r_1 \right ) + 1  \right ]  + \ln^2 {\mu^2 \over  n \cdot p \, (\omega - \bar n \cdot p)} \, \nonumber \\
&& + \, 2 \, \ln {\mu^2 \over  n \cdot p \, (\omega - \bar n \cdot p)}
- 2\, \ln (1+r_1) \, \left [ \ln {\mu^2 \over  n \cdot p \, (\omega - \bar n \cdot p)} + 1 \right ] + \ln^2 (1+r_1) \nonumber \\
&& + \, 2 \, r_1 \,  \ln \left ( {r_1 \over 1+ r_1} \right )
- \, 2 \, {\rm Li}_2  \left( {1 \over 1+ r_1}\right ) + {\pi^2 \over 6} + 4 \bigg \} \,.  \\
\nonumber \\
I_{2, \alpha \beta}&=&\int [d \, l]  \, \frac{l_{\alpha} \,(p-l)_{\beta}}
{[l^2+i0] [(p-l)^2 -m_c^2 + i 0] [(l-k)^2+ i 0]}  \nonumber \\
&\equiv& - {g_{\alpha \beta} \over 2} \, I_{2, a}
- {1 \over  p^2} \, \left [ k_{\alpha} \, k_{\beta} \,  I_{2, b} - p_{\alpha} \, p_{\beta} \,  I_{2, c}
- k_{\alpha} \, p_{\beta} \,  I_{2, d} + p_{\alpha} \, k_{\beta} \,  I_{2, e} \right ]  \,,   \\
&& \hspace{-1.2 cm} {\rm where  \,\, the \,\, loop \,\,  functions}  \,\, I_{2, j} \, (j=a,...,e) \,\,
{\rm are \,\, given \,\, by }   \nonumber \\
I_{2, a} &=& {1 \over 2} \, \bigg  [ {1 \over \epsilon} + \ln \left (-{\mu^2 \over p^2}  \right )
-\frac{(1+r_2+r_3)^2}{r_3 (1+r_3)} \, \ln \left ( 1 + r_2 + r_3 \right )
+ { (1+r_2)^2 \over r_3 }  \, \ln (1+r_2)   \nonumber  \\
&& \hspace{0.5 cm}  - {r_2^2 \over 1+r_3} \, \ln r_2  + 3 \bigg ]  \,, \\
I_{2, b} &=& {1 \over 2 r_3^3} \, \left [ (2+2 r_2 -r_3) \, r_3 - 2 (1+r_2)^2 \,
\ln \left ({1+r_2+r_3 \over 1+r_2} \right ) \right ] \, \nonumber \\
&& \hspace{0.4 cm} \times \left [ {1 \over \epsilon} + \ln \left (-{\mu^2 \over p^2} \right )
 - \ln \left ( 1+r_2+r_3 \right ) + 3 \right ] \nonumber \\
&& - {(1+r_2)^2 \over r_3^3} \, \left [ {\rm Li}_2 \left ( {1+ r_3 \over 1+r_2+r_3}  \right )
- {\rm Li}_2 \left ( {1 \over 1+r_2}  \right )   \right ]
- {(1+r_2)^2 \over 2 \, r_3^3} \, \ln^2 \left ({1+r_2+r_3 \over 1+r_2} \right ) \nonumber \\
&& + \frac{r_2 \, \left [ 3 \, r_2 r_3 + 2 \, (1+r_2+r_3) \right ]}{2 \, r_3^2 \, (1+r_3)^2}
\, \ln \left ({1+r_2+r_3 \over r_2} \right )  - \frac{1-r_2+r_3}{2 \, r_3 \, (1+r_3)}  \,, \\
I_{2, c} &=& \frac{(1+r_3)^2-r_2^2}{2\, r_3 \, (1+r_3)^2} \, \ln \left ({1+r_2+r_3 \over 1+r_2} \right )
+ \frac{r_2^2 \, (2+r_3)}{2 \, (1+r_3)^2} \, \ln \left ( {1+r_2 \over r_2} \right )
- {r_2 \over 2\, (1+r_3)} \,, \\
I_{2, d} &=& {1 \over r_3^2} \, \left [ (1+r_2) \, \ln \left ( {1+r_2 +r_3 \over 1+r_2} \right ) - r_3 \right ] \,
\left [{1 \over \epsilon} + \ln \left (-{\mu^2 \over p^2} \right )
 - \ln \left ( 1+r_2+r_3 \right ) + {5 \over 2}  \right ] \nonumber \\
&& + {1+r_2 \over r_3^2} \, \left [ {\rm Li}_2 \left ( {1+ r_3 \over 1+r_2+r_3}  \right )
- {\rm Li}_2 \left ( {1 \over 1+r_2}  \right )   \right ]
+ {1+r_2 \over 2 \, r_3^2} \, \ln^2 \left ({1+r_2+r_3 \over 1+r_2} \right ) \nonumber \\
&& + \frac{r_2 (1-r_3^2+r_2+2 \, r_2 r_3)}{2 \, r_3^2 (1+r_3)^2} \, \ln \left ( {1+r_2 +r_3 \over 1+r_2} \right )
- \frac{r_2\left [ 2+r_3 (2+r_2) \right ]}{2 \, r_3 (1+r_3)^2}  \, \ln \left ({1+r_2 \over r_2} \right ) \nonumber \\
&& -{r_2 \over 2 \, r_3 \, (1+r_3)} \,, \\
I_{2, e} &=& - \frac{(1+r_2+r_3) \, (1+r_2+r_3+ 2\, r_2 r_3)}{2 \, r_3^2 (1+r_3)^2} \,
 \ln \left ({1+r_2+r_3 \over 1+r_2} \right )  +  \frac{1+r_2+r_3}{2 \, r_3 (1+r_3)} \nonumber \\
&& +  \frac{r_2^2}{2 \,(1+r_3)^2} \, \ln \left ({1+r_2 \over r_2} \right ) \,. \\
\nonumber \\
I_{3}&=&\int [d \, l]  \, \frac{(n \cdot l)^2}
{[l^2+i0] [(p-l)^2 -m_c^2 + i 0] [(l-k)^2+ i 0]}  \nonumber \\
&=&  {(n \cdot p)^2 \over 2\, p^2} \, \bigg \{  \frac{(1+r_2+r_3)^2}{r_3 (1+r_3)^2}
\, \ln \left ( {1+r_2 +r_3 \over 1+r_2} \right )
-  \frac{r_2\left [ 2 \, (1+r_3) + r_2 \, (2+r_3) \right ]}{(1+r_3)^2}
\, \ln \left ({1+r_2 \over r_2} \right )    \nonumber \\
&& + {r_2 \over 1+r_3}  \bigg \} \,. \\
 \nonumber \\
I_{4, \alpha}&=&\int [d \, l]  \, \frac{l_{\alpha}}
{[l^2+i0] [(p-l)^2 -m_c^2 + i 0] [(l-k)^2+ i 0]}  =
 {1 \over p^2} \, \left [ k_{\alpha} \, I_{4, a} +  p_{\alpha} \, I_{4, b}\right ]  \,,  \\
 && \hspace{-1.2 cm} {\rm with }   \nonumber \\
I_{4, a} &=& \left [ {1+r_2 \over r_3^2} \, \ln \left ( {1+r_2 +r_3 \over 1+r_2} \right )
- {1 \over r_3}\right ] \,
\left [ {1 \over \epsilon} + \ln \left (-{\mu^2 \over p^2} \right )
 - \ln \left [ (1+r_2)(1+r_2+r_3) \right ] + 2 \right ]  \nonumber \\
 && + {1+r_2 \over r_3^2} \, \left [ {\rm Li}_2 \left ( {1+ r_3 \over 1+r_2+r_3}  \right )
- {\rm Li}_2 \left ( {1 \over 1+r_2}  \right )   \right ] - {1 \over r_3} \,  \ln \left (1+r_2 \right ) \nonumber \\
&& + {1+r_2 \over 2 \, r_3^2} \, \left [ \ln^2 \left (1+r_2+r_3 \right ) - \ln^2 \left (1+r_2 \right )\right ]
+ {r_2 \over r_3 \, (1+r_3)} \,  \ln \left ( {r_2  \over 1+r_2+r_3} \right )   \,, \\
I_{4, b} &=& {1+r_2+r_3 \over r_3 (1+r_3)} \, \ln \left (1+r_2+r_3 \right )
-{1+r_2 \over r_3} \, \ln \left (1+r_2 \right ) + {r_2 \over 1+r_3} \, \ln r_2 \,. \\
\nonumber \\
I_{5}&=&\int [d \, l]  \, \frac{(2-D) (\! \not p - \! \not k + \! \not l) + D \, m_c}
{ [(p-k+l)^2 -m_c^2 + i 0] [l^2+i0]}  =  I_{5, a}  \, (\! \not p - \! \not k) +  I_{5, b} \, m_c  \,, \\
&& \hspace{-1.2 cm}  {\rm where  \,\, the \,\, loop \,\,  functions}  \,\, I_{5, j} \, (j=a, b) \,\,
{\rm read }   \nonumber  \\
I_{5, a} &=& -\left \{ {1 \over \epsilon}
 + \ln \left (- {\mu^2 \over (p-k)^2}  \right  )  + r_1^2 \, \ln \left ( {1+ r_1 \over r_1} \right )
 - \ln \left ( 1+r_1 \right )  + 1 - r_1 \right \} \,, \\
 I_{5, b} &=& 4 \, \left \{ {1 \over \epsilon}
 + \ln \left ( - {\mu^2 \over (p-k)^2}  \right  )  - r_1  \, \ln \left ( {1+ r_1 \over r_1} \right )
 - \ln \left ( 1+r_1 \right )  +  {3 \over 2} \right \} \,. \\
\nonumber \\
 I_{6, a}&=& \int [d \, l] \,
\frac{n \cdot (p+l) }{[ n \cdot (p+l) \, \bar n \cdot (p-k+l) + l_{\perp}^2 - m_c^2 + i 0]
[n \cdot l \, \bar n (l-k)+ l_{\perp}^2  + i 0][l^2+i0]}  \, \nonumber \\
&=& - {1 \over \omega}  \, \bigg \{ \ln \left (1- {r_4 \over 1 + r_1} \right ) \bigg [ {1 \over \epsilon} +
\ln \left ( {\mu^2 \over n \cdot p \, (\omega - \bar n \cdot p)} \right )
- {1 \over 2} \, \ln \left (1- {r_4 \over 1+r_1} \right ) - \ln \left ( 1+r_1 \right ) \nonumber \\
&&  + \, 1 + {r_1 \over 1- r_4}\bigg ]
+ {\rm Li}_2 \left (1- {r_1 \over 1+r_1-r_4} \right )  - {\rm Li}_2 \left ({1 \over 1+r_1} \right )
- {r_1 r_4 \over 1-r_4} \, \ln \left ({r_1 \over 1+r_1} \right )\bigg \} \,. \hspace{1.0 cm} \\
\nonumber \\
 I_{6, b}&=& \int [d \, l] \,
\frac{n \cdot l  \,\, n \cdot (p+l) }{[ n \cdot (p+l) \, \bar n \cdot (p-k+l) + l_{\perp}^2 - m_c^2 + i 0]
[n \cdot l \, \bar n (l-k)+ l_{\perp}^2  + i 0][l^2+i0]}  \, \nonumber \\
&=& {n \cdot p \over 2 \, \omega} \,  \bigg \{  \left ( {r_1^2 \over (1-r_4)^2}  -1 \right ) \,
 \ln \left (1-r_4+r_1 \right )  + \, (1-r_1^2) \, \ln \left (1+r_1 \right )
- {r_1^2 \, r_4 \, (2-r_4) \over (1-r_4)^2} \, \ln r_1  \nonumber \\
&&  - {r_1 \, r_4 \over 1- r_4}  \bigg \} \,.
\end{eqnarray}
Here, $D=4-2 \,\epsilon$, the integration measure is defined as follows
\begin{eqnarray}
[d \, l] \equiv \frac{(4 \, \pi)^2}{i} \, \left ( \frac{\mu^2 \, e^{\gamma_E}}{4 \, \pi}\right )^{\epsilon} \,
\frac{d^D \, l}{(2 \pi)^D}\,,
\end{eqnarray}
and we also introduce the conventions
\begin{eqnarray}
r_1=\frac{m_c^2}{n \cdot p \,\, \bar n \cdot (k-p)} \,, \qquad
r_2 = - \frac{m_c^2}{p^2} \,, \qquad
r_3 = - \frac{\bar n \cdot k}{\bar n \cdot p} \,,  \qquad
r_4 = \frac{\bar n \cdot k}{\bar n \cdot (k-p)} \,.
\end{eqnarray}

\section{Spectral representations}
\label{app:spectral resp}

We collect the dispersion representations of various convolution integrals entering the (partial) NLL resummation
improved factorization  formula (\ref{resummation improved factorization formula}), for the sake of constructing
the sum rules of $B \to D$ form factors given by (\ref{NLL reummation improved SR for B to D FFs}).
It needs to point out that we have validated each spectral function by verifying the corresponding dispersion
integral numerically.
\begin{eqnarray}
&& {1 \over \pi} \, {\rm Im}_{\omega^{\prime}} \, \int_0^{\infty} \,
\frac{d \omega}{\omega^{\prime}-\omega-\omega_ c+ i 0} \,
\frac{(1+r_2+r_3)^2}{r_3(1+r_3)} \, \ln \left ( {1 + r_2 + r_3 \over 1+ r_2} \right )\,
\phi_B^{+}(\omega) \nonumber \\
&& = - {\omega_c \over \omega^{\prime}}  \,
\ln \left | {\omega_c \over \omega^{\prime}-\omega_c}  \right | \, \phi_B^{+}(\omega^{\prime})
- \, \int_{\omega^{\prime}-\omega_c}^{\infty} \, {d \omega \over \omega } \,
\left [ {\cal P} \, {\omega_c \over \omega-\omega^{\prime}}   + 1 \right ] \,
\theta(\omega^{\prime} -\omega_c) \, \phi_B^{+}(\omega) \,.  \\
\nonumber \\
&& {1 \over \pi} \, {\rm Im}_{\omega^{\prime}} \, \int_0^{\infty} \,
\frac{d \omega}{\omega^{\prime}-\omega-\omega_ c+ i 0} \, \frac{r_2^2 \, r_3}{1+r_3} \,
\ln \left ( {1 + r_2 \over r_2}  \right ) \, \phi_B^{+}(\omega)  \nonumber \\
&& = \int_0^{\infty} \, d \omega \, \theta(\omega^{\prime} -\omega_c)  \,
\bigg \{ {\omega \, \omega_c \over (\omega+\omega_c)^2 }  \,
\left [ {\cal P} \, {1 \over \omega^{\prime} - \omega - \omega_c}   -
{1 \over \omega^{\prime}} \right ]
+  {\omega_c \over \omega}  \, \left [ {\cal P} \, {1 \over \omega - \omega^{\prime} }   +
{1 \over \omega^{\prime}} \right ]  \nonumber \\
&& \hspace{0.3 cm} +  \, { \omega_c^2 \over \omega+\omega_c}
\, {1 \over {\omega^{\prime}}^2}\bigg \} \, \phi_B^{+}(\omega)
- \ln \left | {\omega_c - \omega^{\prime} \over \omega_c}  \right |  \,
\bigg \{ - { (\omega^{\prime} - \omega_c) \omega_c \over {\omega^{\prime}}^2} \,
 \phi_B^{+}(\omega^{\prime}-\omega_c) \, \theta(\omega^{\prime}-\omega_c) \nonumber \\
&&  \hspace{0.3 cm} +  \, {\omega_c \over \omega^{\prime}} \, \phi_B^{+}(\omega^{\prime})
+ \delta^{\prime}(\omega^{\prime})  \int_0^{\infty} \, d \omega \,
{\omega_c^2 \over \omega+\omega_c}   \, \phi_B^{+}(\omega) \bigg \} \,.  \\
\nonumber \\
&& {1 \over \pi} \, {\rm Im}_{\omega^{\prime}} \, \int_0^{\infty} \,
\frac{d \omega}{\omega^{\prime}-\omega-\omega_ c+ i 0} \, \frac{r_2 \, (r_3 -r_2)}{1+r_3} \,
\ln  r_2 \,\, \phi_B^{+}(\omega)  \nonumber \\
&& = \int_0^{\infty} \, d \omega \, \theta(\omega^{\prime})
\, \left [ {\omega - \omega_c \over \omega + \omega_c} \,
{\cal P} \, {1 \over \omega^{\prime}- \omega - \omega_c}
- {\omega - \omega_c \over  \omega } \,
{\cal P} \, {1 \over \omega^{\prime}- \omega}  \right ] \, \phi_B^{+}(\omega)  \nonumber \\
&& \hspace{0.3 cm}  + \ln \left ( {\omega_c \over \omega^{\prime}} \right ) \,
\left [ { 2 \, \omega_c - \omega^{\prime} \over \omega^{\prime} } \,
\phi_B^{+}(\omega^{\prime} - \omega_c)  \, \theta(\omega^{\prime} - \omega_c)
+ {\omega^{\prime} - \omega_c \over \omega^{\prime}}  \, \phi_B^{+}(\omega^{\prime} ) \right ]  \nonumber  \\
&& \hspace{0.3 cm} - \int_0^{\infty} \, d \omega \, {\omega_c \, (\omega-\omega_c) \over \omega (\omega+\omega_c)} \,
\phi_B^{+}(\omega) \, {d \over d \omega^{\prime}}  \,
\left [   \ln \left ( {\omega_c \over \omega^{\prime}} \right ) \, \theta(\omega^{\prime}) \right ]  \,. \\
\nonumber \\
&& {1 \over \pi} \, {\rm Im}_{\omega^{\prime}} \, \int_0^{\infty} \,
\frac{d \omega}{\omega^{\prime}-\omega-\omega_ c+ i 0} \,\,  r_2 \,\, \phi_B^{+}(\omega)  \nonumber  \\
&& =  {\omega_c \over \omega^{\prime}} \, \theta(\omega^{\prime} - \omega_c) \, \phi_B^{+}(\omega^{\prime} - \omega_c)  \,
- \delta(\omega^{\prime}) \, \int_0^{\infty} \, d \omega \,
{\omega_c \over \omega + \omega_c} \, \phi_B^{+}(\omega) \,.  \\
\nonumber \\
&& {1 \over \pi} \, {\rm Im}_{\omega^{\prime}} \, \int_0^{\infty} \,
\frac{d \omega}{\omega^{\prime}-\omega-\omega_ c+ i 0} \,\,  {1 \over \omega^{\prime}}
\, \frac{(1+r_2+r_3)^2}{r_3(1+r_3)^2} \, \ln {1 + r_2 + r_3 \over 1+r_2} \,\, \phi_B^{+}(\omega)  \nonumber  \\
&& =  - \int_{\omega^{\prime}-\omega_c}^{\infty} \, d \omega \,\, \theta(\omega^{\prime}-\omega_c) \,
{\cal P} \, {1 \over  \omega^{\prime}-\omega} \,
\left (1 + \omega_c  \, {d \over d \omega} \right ) \, {\phi_B^{+}(\omega) \over \omega} \nonumber  \\
&& \hspace{0.3 cm}  + \ln \left | {\omega_c \over \omega^{\prime}-\omega_c} \right |  \,
\left (1 + \omega_c  \, {d \over d \omega^{\prime}} \right ) \, {\phi_B^{+}(\omega^{\prime}) \over \omega^{\prime}}
+ {\phi_B^{+}(\omega^{\prime}) \over \omega^{\prime}}  \, \theta(\omega^{\prime})
- {\phi_B^{+}(\omega^{\prime}-\omega_c) \over \omega^{\prime} - \omega_c}  \, \theta(\omega^{\prime}-\omega_c) \,. \\
\nonumber \\
&& {1 \over \pi} \, {\rm Im}_{\omega^{\prime}} \, \int_0^{\infty} \,
\frac{d \omega}{\omega^{\prime}-\omega-\omega_ c+ i 0} \,\,  {1 \over \omega^{\prime}}
\, \frac{r_2 \, \left [2\, (1+r_3) + r_2 \,(2+r_3)  \right ]}{(1+r_3)^2} \,
\ln \left ( {1 + r_2  \over r_2} \right ) \,\, \phi_B^{+}(\omega)  \nonumber  \\
&& = \int_0^{\infty} \, d \omega \, \theta(\omega^{\prime} - \omega_c) \,
\bigg [ {\omega \over (\omega+\omega_c)^2} \, {\cal P} \, {1 \over \omega^{\prime} - \omega - \omega_c}
+ {1 \over \omega} \, {\cal P} \, {1 \over \omega - \omega^{\prime} }
+ {\omega_c (2 \, \omega + \omega_c)  \over \omega \, (\omega+\omega_c)^2}  \, {1 \over \omega^{\prime}} \nonumber \\
&& \hspace{0.3 cm}  - {\omega_c^2 \over \omega (\omega+\omega_c)} \, {1 \over {\omega^{\prime}}^2 } \bigg ]
\, \phi_B^{+}(\omega) - \int_0^{\infty} \, d \omega \, \theta(\omega^{\prime} - \omega_c) \,
{\cal P} \, {\omega_c \over \omega^{\prime} - \omega } \, {d \over d \omega} {\phi_B^{+}(\omega) \over \omega}
\nonumber \\
&& \hspace{0.3 cm} -  \ln \left | {\omega_c - \omega^{\prime} \over \omega_c} \right |  \,
\bigg [ {\omega_c - \omega^{\prime} \over {\omega^{\prime}}^2} \, \phi_B^{+}(\omega^{\prime} -\omega_c) \,
\theta(\omega^{\prime} -\omega_c)   +  {\phi_B^{+}(\omega^{\prime}) \over \omega^{\prime}} \bigg ] \nonumber \\
&&  \hspace{0.3 cm} - \, \theta(\omega^{\prime}-\omega_c) \, {\omega_c \over \omega^{\prime}} \,
\lim_{\omega \to  0} \, { \phi_B^{+}(\omega) \over \omega}
- \omega_c \, \ln \left | {\omega_c - \omega^{\prime} \over \omega_c} \right |  \,
{d \over d \omega^{\prime}} \, { \phi_B^{+}(\omega^{\prime}) \over \omega^{\prime}}   \nonumber \\
&& \hspace{0.3 cm}  + \, \delta^{\prime}(\omega^{\prime})  \,
 \ln \left | {\omega_c - \omega^{\prime} \over \omega_c} \right |  \,
 \int_0^{\infty} \, d \omega \, {\omega_c^2  \over \omega (\omega+\omega_c)} \, \phi_B^{+}(\omega) \,. \\
\nonumber \\
 && {1 \over \pi} \, {\rm Im}_{\omega^{\prime}} \, \int_0^{\infty} \,
\frac{d \omega}{\omega^{\prime}-\omega-\omega_ c+ i 0} \,\,  {1 \over \omega^{\prime}}
\, \frac{r_2}{1+r_3} \,\, \phi_B^{+}(\omega)  \nonumber  \\
&& = {1 \over \omega^{\prime}} \, \bigg [ \phi_B^{+}(\omega^{\prime} - \omega_c) \, \theta(\omega^{\prime} - \omega_c)
-  \phi_B^{+}(\omega^{\prime}) \, \theta(\omega^{\prime}) \bigg ]
+ \delta(\omega^{\prime}) \,  \int_0^{\infty} \, d \omega \, {\omega_c \over \omega \, (\omega+\omega_c)}
\,  \phi_B^{+}(\omega) \,.  \\
\nonumber \\
 && {1 \over \pi} \, {\rm Im}_{\omega^{\prime}} \, \int_0^{\infty} \,
\frac{d \omega}{\omega^{\prime}-\omega-\omega_ c+ i 0} \,\,
\frac{r_2 \, (1+r_2+r_3)}{r_3 (1+r_3)} \,  \ln (1+r_2+r_3)\, \phi_B^{+}(\omega)  \nonumber \\
&& =   \int_0^{\infty} \, d \omega \,\theta(\omega_c+\omega-\omega^{\prime}) \,
\theta(\omega^{\prime}) \, {\omega_c \over \omega} \, {\cal P} \, {1 \over \omega^{\prime} - \omega}
\,\phi_B^{+}(\omega) - {\omega_c \over \omega^{\prime}} \,
\ln \left ( {\omega_c \over \omega^{\prime}} \right ) \, \phi_B^{+}(\omega^{\prime}) \,
\theta(\omega^{\prime}) \,.  \\
\nonumber \\
&& {1 \over \pi} \, {\rm Im}_{\omega^{\prime}} \, \int_0^{\infty} \,
\frac{d \omega}{\omega^{\prime}-\omega-\omega_ c+ i 0} \,\,
\frac{r_2 \, (1+r_2)}{r_3} \,  \ln (1+r_2)\, \phi_B^{+}(\omega)  \nonumber \\
&& = \theta(\omega^{\prime}) \, \theta(\omega_c - \omega^{\prime}) \,
\int_0^{\infty} \, d \omega \, {\omega_c \over \omega + \omega_c} \,
\frac{\phi_B^{+}(\omega)}{\omega^{\prime}-\omega-\omega_ c}
 - {\omega_ c \over \omega^{\prime}} \,
\ln \left | { \omega^{\prime}  - \omega_c \over \omega^{\prime} } \right |  \,
\phi_B^{+}(\omega^{\prime}-\omega_c)  \, \theta(\omega^{\prime}-\omega_c)  \nonumber \\
&&  \hspace{0.3 cm}  + \left [ { \theta(\omega^{\prime}) \, \theta(\omega_c - \omega^{\prime})
 \over \omega^{\prime}} \right ]_{+} \,\int_0^{\infty} \, d \omega \,
{\omega_c^2 \over \omega \, (\omega+\omega_c)} \, \phi_B^{+}(\omega) \,. \\
\nonumber \\
&& {1 \over \pi} \, {\rm Im}_{\omega^{\prime}} \, \int_0^{\infty} \,
\frac{d \omega}{\omega^{\prime}-\omega-\omega_ c+ i 0} \,\,
\frac{r_2^2}{1+r_3} \,  \ln r_2 \, \phi_B^{+}(\omega)  \nonumber  \\
&& = \theta(\omega^{\prime}) \, \int_0^{\infty} \, d \omega \,
\left [  { \omega_c \over \omega+\omega_c}  \, {\cal P} \, {1 \over \omega^{\prime} - \omega -\omega_c}
- {\omega_c \over \omega}  \, {1 \over  \omega^{\prime} - \omega} \right ] \, \phi_B^{+}(\omega) \nonumber \\
&& \hspace{0.3 cm} -  {\omega_c \over \omega^{\prime}} \,
\ln \left | { \omega_c \over \omega^{\prime} } \right |  \,
\bigg [ \phi_B^{+}(\omega^{\prime} - \omega_c) \, \theta(\omega^{\prime} - \omega_c)
-  \phi_B^{+}(\omega^{\prime}) \, \theta(\omega^{\prime}) \bigg ] \nonumber \\
&& \hspace{0.3 cm}  + \left \{  \left [ { \theta(\omega^{\prime}) \, \theta(\omega_c - \omega^{\prime})
\over \omega^{\prime}} \right ]_{+} +
{\theta(\omega^{\prime}-\omega_c) \over \omega^{\prime}} \right \} \,
\int_0^{\infty} \, d \omega \, {\omega_c^2 \over \omega (\omega+\omega_c)} \, \phi_B^{+}(\omega) \,. \\
\nonumber \\
&& {1 \over \pi} \, {\rm Im}_{\omega^{\prime}} \, \int_0^{\infty} \,
\frac{d \omega}{\omega^{\prime}-\omega-\omega_ c+ i 0} \,\,
\frac{1+r_2+r_3}{(1+r_3)^2} \, \frac{r_2^2-(1+r_3)^2}{r_3} \,
\ln {1 + r_2 + r_3 \over 1 + r_2} \, \phi_B^{+}(\omega)  \nonumber \\
&& = \int_0^{\infty} \, d \omega \, \theta(\omega^{\prime}-\omega_c) \,
\theta(\omega+\omega_c-\omega^{\prime}) \, {\phi_B^{+}(\omega) \over \omega} \,
+ \omega_c \, \bigg [ { \phi_B^{+}(\omega^{\prime} - \omega_c) \over \omega^{\prime} - \omega_c}
\, \theta(\omega^{\prime} - \omega_c)
-  { \phi_B^{+}(\omega^{\prime})  \over \omega^{\prime}}  \, \theta(\omega^{\prime}) \bigg ]  \nonumber \\
&& \hspace{0.3 cm} - \, \omega_c^2  \, \ln \left | {\omega_c \over \omega^{\prime}  - \omega_c} \right | \,
{d \over d \omega^{\prime}} \, {\phi_B^{+}(\omega^{\prime}) \over \omega^{\prime}}
+ \int_0^{\infty} \, d \omega \, \theta(\omega^{\prime}-\omega_c) \, \theta(\omega + \omega_c - \omega^{\prime}) \,
{\cal P} \, {\omega_c^2 \over \omega^{\prime}-\omega}  \,
{d \over d \omega} \, {\phi_B^{+}(\omega) \over \omega}  \,. \nonumber  \\
\\
&& {1 \over \pi} \, {\rm Im}_{\omega^{\prime}} \, \int_0^{\infty} \,
\frac{d \omega}{\omega^{\prime}-\omega-\omega_ c+ i 0} \,\,
r_2^2 \, (r_3+2) \, \frac{1+r_2+r_3}{(1+r_3)^2} \,
\ln {1 + r_2 \over  r_2} \, \phi_B^{+}(\omega)  \nonumber \\
&& =  \omega_c^2  \, \bigg \{ \frac{\theta(\omega^{\prime} - \omega_c)}{\omega^{\prime}} \,
 \lim_{\omega \to  0} \, { \phi_B^{+}(\omega) \over \omega}
 + \ln \left | {\omega_c - \omega^{\prime}  \over  \omega_c} \right | \,
 {d \over d \omega^{\prime}} \, { \phi_B^{+}(\omega^{\prime}) \over \omega^{\prime}}   \nonumber \\
&& \hspace{0.3 cm} + \, \theta(\omega^{\prime}-\omega_c) \,  \int_0^{\infty} \, d \omega \,
{\cal P} \, {1 \over \omega^{\prime}-\omega} \,
{d \over d \omega} \, { \phi_B^{+}(\omega) \over \omega}  \nonumber \\
&& \hspace{0.3 cm} + \left [\frac{\theta(\omega^{\prime}-\omega_c)}{{\omega^{\prime}}^2}
- \delta^{\prime} (\omega^{\prime}) \, \ln \left | {\omega_c - \omega^{\prime}  \over  \omega_c} \right | \right ]
\, \int_0^{\infty} \, d \omega \, { \phi_B^{+}(\omega) \over \omega} \bigg \}   \,. \\
\nonumber \\
&& {1 \over \pi} \, {\rm Im}_{\omega^{\prime}} \, \int_0^{\infty} \,
\frac{d \omega}{\omega^{\prime}-\omega-\omega_ c+ i 0} \,\,
\, \frac{r_2 (1+r_2+r_3)}{1+r_3} \, \phi_B^{+}(\omega)  \nonumber \\
&& = {\omega_c \over \omega^{\prime}} \, \phi_B^{+}(\omega^{\prime})
- \omega_c \, \delta(\omega^{\prime})  \,
\int_0^{\infty} \, d \omega \, {\phi_B^{+}(\omega) \over \omega} \,. \\
\nonumber \\
&& {1 \over \pi} \, {\rm Im}_{\omega^{\prime}} \, \int_0^{\infty} \,
\frac{d \omega}{\omega^{\prime}-\omega-\omega_ c+ i 0} \,\,
\, \ln^2 \left ( {\mu^2 \over n \cdot p \, (\omega - \omega^{\prime})} \right )
\, \phi_B^{-}(\omega)  \nonumber \\
&& = 2 \, \int_0^{\infty} \, d \omega \, \theta(\omega^{\prime}-\omega) \,
\ln \left ( {\mu^2 \over n \cdot p \, (\omega^{\prime} - \omega)} \right ) \,
{\cal P} \, {1 \over \omega^{\prime} - \omega - \omega_c} \, \phi_B^{-}(\omega)  \nonumber \\
&& \hspace{0.3 cm} - \left [  \ln^2 \left ( {\mu^2 \over n \cdot p \, \omega_c} \right )
- \pi^2  \right ] \, \phi_B^{-}(\omega^{\prime}-\omega_c)  \, \theta(\omega^{\prime}-\omega_c) \,.  \\
\nonumber \\
&& {1 \over \pi} \, {\rm Im}_{\omega^{\prime}} \, \int_0^{\infty} \,
\frac{d \omega}{\omega^{\prime}-\omega-\omega_ c+ i 0} \,\,
\ln \left ( { (1+r_2+r_3)^2 \over (1+r_2) (1+r_3)} \right )
\, \ln \left ( {\mu^2 \over n \cdot p \, (\omega - \omega^{\prime})} \right )
\, \phi_B^{-}(\omega)  \nonumber \\
&& = - \left [  \ln^2 \left | {\omega_c  \over  \omega^{\prime} - \omega_c} \right |
- \pi^2 \, \theta(\omega^{\prime} - \omega_c)  \right ] \,  \phi_B^{-}(\omega^{\prime})  \nonumber \\
&& \hspace{0.3 cm} +  \int_0^{\infty} \, d \omega \, \theta(\omega^{\prime}-\omega) \,
\left [  \ln^2 \left | {\omega^{\prime} - \omega - \omega_c  \over  \omega^{\prime} - \omega_c} \right |
 - \pi^2 \, \theta(\omega^{\prime} - \omega_c)  \, \theta(\omega+\omega_c-\omega^{\prime}) \right ] \,
{d \over d \omega} \, \phi_B^{-}(\omega) \nonumber \\
&&  \hspace{0.3 cm} + \, 2 \, \int_0^{\infty} \, d \omega \, \theta(\omega^{\prime}-\omega_c)
\, \theta(\omega+\omega_c-\omega^{\prime}) \,
\ln \left | {\omega^{\prime} - \omega - \omega_c  \over  \omega^{\prime} - \omega_c} \right | \, \nonumber \\
&&  \hspace{0.6 cm} \times \left [ {\cal P} \, {1 \over \omega^{\prime} - \omega} +
\ln \left | {\mu^2  \over  n \cdot p \, (\omega - \omega^{\prime})} \right |  \, {d \over d \omega} \right ]
 \, \phi_B^{-}(\omega) \nonumber \\
&&  \hspace{0.3 cm} - \ln \left ( {\omega^{\prime} - \omega_c \over \omega_c} \right ) \,
\ln \left ( {\mu^2  \over n \cdot p \, \omega_c} \right ) \, \phi_B^{-}(\omega^{\prime} - \omega_c) \,
\theta(\omega^{\prime} - \omega_c)  \nonumber \\
&& \hspace{0.3 cm} + \,  \int_0^{\infty} \, d \omega \, \theta(\omega^{\prime}) \,
\left [\theta(\omega_c - \omega^{\prime}) - \theta(\omega - \omega^{\prime})   \right ]  \,
{1 \over  \omega^{\prime} - \omega - \omega_c}  \,
\ln \left | {\mu^2  \over  n \cdot p \, (\omega - \omega^{\prime})} \right |
\, \phi_B^{-}(\omega)  \nonumber \\
&&  \hspace{0.3 cm} + \,  \int_0^{\infty} \, d \omega \, \theta(\omega^{\prime} - \omega) \,
{\cal P} \, {1 \over \omega^{\prime} - \omega -\omega_c} \,
\ln \left | {\omega^{\prime} -\omega_c  \over  \omega^{\prime} - \omega} \right |
\, \phi_B^{-}(\omega) \,.  \\
\nonumber \\
&& {1 \over \pi} \, {\rm Im}_{\omega^{\prime}} \, \int_0^{\infty} \,
\frac{d \omega}{\omega^{\prime}-\omega-\omega_ c+ i 0} \,\,
{r_2 \over 1+ r_2 + r_3}
\, \ln \left ( {\mu^2 \over n \cdot p \, (\omega - \omega^{\prime})} \right )
\, \phi_B^{-}(\omega)  \nonumber \\
&& = - \, \omega_c \, \delta(\omega_c - \omega^{\prime}) \,
\ln \left ( {\mu^2 \over n \cdot p \, \omega_c} \right ) \, \phi_B^{-}(0)
- \theta(\omega^{\prime}) \, {\cal P} \, {\omega_c \over \omega_c - \omega^{\prime}} \, \phi_B^{-}(0)
+  \, \phi_B^{-}(\omega^{\prime}) \, \theta(\omega^{\prime})  \nonumber \\
&& \hspace{0.3 cm}
- \, \phi_B^{-}(\omega^{\prime} - \omega_c) \, \theta(\omega^{\prime} - \omega_c)
- \omega_c \, \ln \left ( {\mu^2 \over n \cdot p \, \omega_c} \right ) \,
\left [{d \over d \omega^{\prime}} \,  \phi_B^{-}(\omega^{\prime} - \omega_c) \right ]\,
\theta(\omega^{\prime} - \omega_c) \nonumber \\
&& \hspace{0.3 cm}  + \,  \int_0^{\infty} \, d \omega \,  \theta(\omega^{\prime} - \omega)\,
{\cal P} \, {\omega_c \over \omega^{\prime} - \omega_c -\omega} \,
{d \over d \omega} \,  \phi_B^{-}(\omega)  \,.  \\
\nonumber \\
&& {1 \over \pi} \, {\rm Im}_{\omega^{\prime}} \, \int_0^{\infty} \,
\frac{d \omega}{\omega^{\prime}-\omega-\omega_ c+ i 0} \,\,
\, \ln^2 (1+r_2+r_3) \, \phi_B^{-}(\omega)  \nonumber  \\
&& = \left [ \ln^2 \left ( {\omega_c - \omega^{\prime} \over \omega^{\prime}}\right ) - {\pi^2 \over 3}   \right ] \,
\theta(\omega_c - \omega^{\prime})  \, \theta(\omega^{\prime})  \, \phi_B^{-}(0) \nonumber \\
&&  \hspace{0.3 cm} + \int_0^{\infty} \, d \omega \,
\left [  \ln^2 \left ( {\omega + \omega_c - \omega^{\prime} \over \omega^{\prime}}\right )
- {\pi^2 \over 3} \right ] \, \theta(\omega+\omega_c-\omega^{\prime}) \, \theta(\omega^{\prime}) \,
{d \over d \omega} \, \phi_B^{-}(\omega)  \,.  \\
\nonumber \\
&& {1 \over \pi} \, {\rm Im}_{\omega^{\prime}} \, \int_0^{\infty} \,
\frac{d \omega}{\omega^{\prime}-\omega-\omega_ c+ i 0} \,\,
\, \ln^2 (1+r_3) \,\, \phi_B^{-}(\omega)  \nonumber  \\
&&  =   2 \, \int_0^{\infty} \, d \omega \,
\ln \left ( {\omega - \omega^{\prime} \over \omega^{\prime}} \right ) \,
{1 \over \omega^{\prime} - \omega - \omega_c} \, \theta(\omega-\omega^{\prime}) \,
\theta(\omega^{\prime}) \, \phi_B^{-}(\omega) \nonumber \\
&&  \hspace{0.3 cm} - \, \ln^2 \left ( {\omega_c \over \omega^{\prime}} \right ) \, \phi_B^{-}(\omega^{\prime}-\omega_c)
\, \theta(\omega^{\prime}-\omega_c)  \,.  \\
\nonumber \\
&& {1 \over \pi} \, {\rm Im}_{\omega^{\prime}} \, \int_0^{\infty} \,
\frac{d \omega}{\omega^{\prime}-\omega-\omega_ c+ i 0} \,\,
\, \ln (1+r_2+r_3) \, \ln (1+r_3)\, \phi_B^{-}(\omega)  \nonumber  \\
&&  =  {1 \over 2} \,\int_0^{\infty} \, d \omega \,
\left [  \ln^2 \left ( {\omega + \omega_c - \omega^{\prime}\over \omega^{\prime}} \right ) - \pi^2 \right ]
\, \theta(\omega - \omega^{\prime}) \theta(\omega^{\prime}) \, \, {d \over d \omega} \, \phi_B^{-}(\omega)  \nonumber \\
&&  \hspace{0.3 cm} + \int_0^{\infty} \, d \omega \,  \theta(\omega + \omega_c - \omega^{\prime}) \, \theta(\omega^{\prime})
\ln \left ( {\omega + \omega_c - \omega^{\prime}\over \omega^{\prime}} \right ) \,
\left [ {\cal P} \, {1 \over \omega - \omega^{\prime}}
+ \ln \left | {\omega^{\prime} - \omega \over \omega^{\prime}}  \right |  \, {d \over d \omega} \right ]
\,\phi_B^{-}(\omega)  \nonumber \\
&&  \hspace{0.3 cm} + \, {1 \over 2} \, \left [
\ln^2 \left ( {\omega_c \over \omega^{\prime}} \right ) - \pi^2  \right ] \,\phi_B^{-}(\omega^{\prime})  \,.  \\
\nonumber \\
&& {1 \over \pi} \, {\rm Im}_{\omega^{\prime}} \, \int_0^{\infty} \,
\frac{d \omega}{\omega^{\prime}-\omega-\omega_ c+ i 0} \,\,
\, {1+ r_2 \over r_3} \, \ln \left ( {1+r_2 +r_3 \over 1+r_2} \right ) \, \phi_B^{-}(\omega)  \nonumber  \\
&& =  \int_0^{\infty} \, d \omega \, \ln \left ( {\omega + \omega_c - \omega^{\prime} \over \omega^{\prime} - \omega_c} \right )  \,
\theta(\omega + \omega_c - \omega^{\prime}) \, \theta(\omega^{\prime}-\omega_c) \,
(\omega_c - \omega^{\prime})  \, {d \over d \omega} \, {\phi_B^{-}(\omega)  \over \omega}\,. \\
\nonumber \\
&& {1 \over \pi} \, {\rm Im}_{\omega^{\prime}} \, \int_0^{\infty} \,
\frac{d \omega}{\omega^{\prime}-\omega-\omega_ c+ i 0} \,\,
\, (1+ r_2)^2 \, \ln \left ( 1 + r_2 \right ) \, \phi_B^{-}(\omega)  \nonumber  \\
&& = - \left ( {\omega^{\prime}-\omega_c \over \omega^{\prime}} \right )^2 \,
\ln \left ( {\omega^{\prime}-\omega_c \over \omega^{\prime}} \right ) \, \phi_B^{-}(\omega^{\prime}-\omega_c) \,
\theta(\omega^{\prime}-\omega_c)  \nonumber  \\
&& \hspace{0.3 cm}  +  \int_0^{\infty} \, d \omega \, \left ( {\omega \over\omega +\omega_c } \right )^2  \,
\theta(\omega_c - \omega^{\prime} ) \, \theta(\omega^{\prime} ) \, {\cal P} \, {1 \over \omega^{\prime} - \omega -\omega_c} \,
\phi_B^{-}(\omega) \nonumber \\
&& \hspace{0.3 cm}  +  \left [ {\theta(\omega^{\prime}) \, \theta(\omega_c - \omega^{\prime})  \over \omega^{\prime}} \right ]_{+}
\, \int_0^{\infty} \, d \omega \, {\omega_c \, (2 \, \omega+\omega_c)  \over (\omega+\omega_c)^2} \, \phi_B^{-}(\omega)  \nonumber \\
&& \hspace{0.3 cm} - \left [ {\theta(\omega^{\prime}) \, \theta(\omega_c - \omega^{\prime})
\over {\omega^{\prime}}^2 } \right ]_{+ +} \, \int_0^{\infty} \, d \omega \,
{\omega_c^2  \over \omega+\omega_c} \,\phi_B^{-}(\omega)  \,.  \\
\nonumber \\
&& {1 \over \pi} \, {\rm Im}_{\omega^{\prime}} \, \int_0^{\infty} \,
\frac{d \omega}{\omega^{\prime}-\omega-\omega_ c+ i 0} \,\,
\, {r_2 \, [r_2 + 2 (1 + r_3)] \over (1+r_3)^2} \, \ln \left ( 1 + r_2 + r_3 \right ) \, \phi_B^{-}(\omega)  \nonumber \\
&& = - \theta(\omega_c - \omega^{\prime}) \, \theta(\omega^{\prime})  \,
 \ln \left ( {\omega_c - \omega^{\prime} \over \omega^{\prime}} \right ) \, \phi_B^{-}(0)
- \theta(\omega^{\prime}) \,
\ln \left ( {\omega_c  \over \omega^{\prime}} \right ) \, \phi_B^{-}(\omega^{\prime})  \nonumber \\
&&  \hspace{0.3 cm}  + \left [ \theta(\omega^{\prime}) \, \theta(\omega_c - \omega^{\prime}) \,
{\omega_c \over \omega^{\prime}}  \right ]_{+} \,  \phi_B^{-}(0)
- \phi_B^{+}(\omega^{\prime}) \, \theta(\omega^{\prime})
+ \phi_B^{+}(\omega^{\prime} - \omega_c) \, \theta(\omega^{\prime} - \omega_c)  \nonumber \\
&& \hspace{0.3 cm}  - \, \omega_c \,  \ln \left ( {\omega_c  \over \omega^{\prime}} \right ) \,
{d \over d \omega^{\prime} } \,\phi_B^{-}(\omega^{\prime})   \,
- \int_0^{\infty} \, d \omega \, \theta(\omega + \omega_c -\omega^{\prime} ) \, \theta(\omega^{\prime}) \,
\bigg [ \ln \left ( { \omega + \omega_c - \omega^{\prime} \over \omega^{\prime}} \right ) \,
{d \over d \omega}  \nonumber \\
&& \hspace{0.3 cm}  + {\cal P} \, {1 \over \omega - \omega^{\prime}}
+ {\cal P} \, { \omega_c \over \omega - \omega^{\prime}} \, {d \over d \omega}   \bigg ]   \,\phi_B^{-}(\omega) \,. \\
\nonumber \\
&& {1 \over \pi} \, {\rm Im}_{\omega^{\prime}} \, \int_0^{\infty} \,
\frac{d \omega}{\omega^{\prime}-\omega-\omega_ c+ i 0} \,\,
 \, \ln \left ( 1 + r_2 + r_3 \right ) \, \phi_B^{-}(\omega)  \nonumber \\
&&  =  \theta(\omega_c - \omega^{\prime}) \, \theta(\omega^{\prime}) \,
\ln \left ( {\omega_c - \omega^{\prime} \over \omega^{\prime}} \right ) \,
\phi_B^{-}(0) \nonumber \\
&&  \hspace{0.3 cm} + \, \int_0^{\infty} \, d \omega \,
\theta(\omega+ \omega_c - \omega^{\prime}) \, \theta(\omega^{\prime}) \,
\ln \left ( {\omega+ \omega_c - \omega^{\prime} \over \omega^{\prime}} \right ) \,
{d \over d \omega} \, \phi_B^{-}(\omega)  \,.  \\
\nonumber \\
&& {1 \over \pi} \, {\rm Im}_{\omega^{\prime}} \, \int_0^{\infty} \,
\frac{d \omega}{\omega^{\prime}-\omega-\omega_ c+ i 0} \,\,
\, \ln \left ( 1 + r_2 \right ) \, \ln \left ( 1 + r_3 \right )  \, \phi_B^{-}(\omega)  \nonumber  \\
&& = - \ln \left ( { \omega^{\prime} - \omega_c \over \omega^{\prime}} \right ) \,
\ln \left ( {   \omega_c \over \omega^{\prime}} \right ) \, \phi_B^{-}(\omega^{\prime} - \omega_c) \,
\theta(\omega^{\prime} - \omega_c) \nonumber \\
&&  \hspace{0.3 cm} + \, \int_0^{\infty} \, d \omega \,
\theta(\omega_c - \omega^{\prime}) \, \theta(\omega^{\prime}) \,
{1 \over \omega^{\prime} - \omega - \omega_c} \,
\ln  \left | {\omega^{\prime} - \omega \over \omega^{\prime}} \right | \,  \phi_B^{-}(\omega)  \nonumber  \\
&&  \hspace{0.3 cm} + \, \int_0^{\infty} \, d \omega \,
\theta(\omega - \omega^{\prime}) \, \theta(\omega^{\prime}) \,
{1 \over \omega^{\prime} - \omega - \omega_c} \,
\ln  \left | {\omega^{\prime} - \omega_c \over \omega^{\prime}} \right | \,  \phi_B^{-}(\omega)  \,.  \\
\nonumber \\
&& {1 \over \pi} \, {\rm Im}_{\omega^{\prime}} \, \int_0^{\infty} \,
\frac{d \omega}{\omega^{\prime}-\omega-\omega_ c+ i 0} \,\,
{r_2 \over 1 + r_2 +r_3}\, \ln \left ( 1 + r_3 \right )  \, \phi_B^{-}(\omega)  \nonumber  \\
&&  =   \phi_B^{-}(\omega^{\prime} - \omega_c) \, \theta(\omega^{\prime} - \omega_c)
- \phi_B^{-}(\omega^{\prime}) \, \theta(\omega^{\prime})
-  \, \theta(\omega^{\prime}-\omega_c) \,\omega_c \,
\ln \left ( {\omega_c \over \omega^{\prime}} \right )  \,
{d \over d \omega^{\prime}}  \, \phi_B^{-}(\omega^{\prime}-\omega_c)   \nonumber \\
&& \hspace{0.3 cm} + \, \int_0^{\infty} \, d \omega \,
\theta(\omega - \omega^{\prime}) \, \theta(\omega^{\prime}) \,
{\omega_c \over \omega^{\prime}- \omega - \omega_c} \,
{d \over d \omega}  \, \phi_B^{-}(\omega)    \,.   \\
\nonumber \\
&& {1 \over \pi} \, {\rm Im}_{\omega^{\prime}} \, \int_0^{\infty} \,
\frac{d \omega}{\omega^{\prime}-\omega-\omega_ c+ i 0} \,\,
 \ln^2 \left ( 1 + r_2 \right )  \, \phi_B^{-}(\omega)  \nonumber  \\
&&  =  -  \ln^2 \left ( {\omega^{\prime} - \omega_c \over \omega^{\prime}} \right )
\, \phi_B^{-}(\omega^{\prime} - \omega_c) \, \theta(\omega^{\prime} - \omega_c) \nonumber \\
&& \hspace{0.3 cm} + \,  2 \,  \int_0^{\infty} \, d \omega \,
\theta(\omega_c - \omega^{\prime}) \, \theta(\omega^{\prime}) \,
 \ln \left ( {\omega_c - \omega^{\prime} \over \omega^{\prime}} \right )  \,
{1 \over \omega^{\prime}- \omega - \omega_c} \, \phi_B^{-}(\omega)  \,.  \\
\nonumber \\
&& {1 \over \pi} \, {\rm Im}_{\omega^{\prime}} \, \int_0^{\infty} \,
\frac{d \omega}{\omega^{\prime}-\omega-\omega_ c+ i 0} \,  {r_2 \,  \ln r_2 \over 1+r_2+r_3} \,
 \,\, \phi_B^{-}(\omega)  \nonumber  \\
&&  = \omega_c \, \bigg [ \theta(\omega^{\prime}) \, {\cal P} {1 \over \omega^{\prime} -\omega_c}
\, \phi_B^{-}(0) - \theta(\omega^{\prime} - \omega_c)  \, \ln \left ( {\omega_c \over \omega^{\prime}} \right ) \,
{d \over d \omega^{\prime}}  \, \phi_B^{-}(\omega^{\prime} - \omega_c)  \nonumber \\
&& \hspace{0.3 cm}  + \,  \int_0^{\infty} \, d \omega \, \theta(\omega^{\prime}) \,
{\cal P} {1 \over \omega^{\prime}-\omega-\omega_c} \, {d \over d \omega} \, \phi_B^{-}(\omega)  \bigg ] \,. \\
\nonumber \\
&& {1 \over \pi} \, {\rm Im}_{\omega^{\prime}} \, \int_0^{\infty} \,
\frac{d \omega}{\omega^{\prime}-\omega-\omega_ c+ i 0} \,\,  r_2  \,
 \,  \left [ {r_2 \over (1+r_3)^2} + {2 \over 1+r_3} + r_2 + 2 \right ] \, \ln r_2 \,\,
\phi_B^{-}(\omega)  \nonumber  \\
&& = - \left \{  \left [ \frac{\theta(\omega_c -\omega^{\prime}) \, \theta(\omega^{\prime})}
{{\omega^{\prime}}^2} \right ]_{+ +}
+  \frac{\theta(\omega^{\prime}-\omega_c)} {{\omega^{\prime}}^2}
+ \delta^{\prime}(\omega^{\prime}) \,
\ln  \left | {\omega_c \over \omega^{\prime} - \omega_c} \right | \right \}
\,  \int_0^{\infty} \, d \omega \,  {\omega_c^2 \over \omega+\omega_c} \, \phi_B^{-}(\omega) \nonumber \\
&&  \hspace{0.3 cm}  + \,  \left \{  \left [ \frac{\theta(\omega_c -\omega^{\prime}) \, \theta(\omega^{\prime})}
{\omega^{\prime}} \right ]_{+}
+   \frac{\theta(\omega^{\prime}-\omega_c)} {\omega^{\prime}}  \right \}   \,
\,  \int_0^{\infty} \, d \omega \,  {\omega_c \, (\omega_c + 2\, \omega) \over (\omega+\omega_c)^2}
\, \phi_B^{-}(\omega)  \nonumber \\
&&  \hspace{0.3 cm}  + \, \omega_c \, \left \{  \left [ \frac{\theta(\omega_c -\omega^{\prime}) \, \theta(\omega^{\prime})}
{\omega^{\prime}} \right ]_{+}
+   \frac{\theta(\omega^{\prime}-\omega_c)} {\omega^{\prime}}  \right \}   \, \phi_B^{-}(0)
-  \ln \left ( {\omega_c \over \omega^{\prime}} \right ) \,
\left  ( 1 + \omega_c \, {d \over d \omega^{\prime}} \right ) \phi_B^{-}(\omega^{\prime})  \nonumber \\
&& \hspace{0.3 cm}  + \,  \int_0^{\infty} \, d \omega \,\theta(\omega^{\prime}) \,
{\cal P} \, {1 \over \omega^{\prime} - \omega } \,
\left ( 1 + \omega_c \,  {d \over d \omega} \right ) \,\phi_B^{-}(\omega) \nonumber \\
&& \hspace{0.3 cm}  + \,\ln \left ( {\omega_c \over \omega^{\prime}} \right ) \,
\frac{ {\omega^{\prime}}^2 + 2 \, \omega^{\prime} \omega_c - \omega_c^2 }{{\omega^{\prime}}^2} \,
\phi_B^{-}(\omega^{\prime} - \omega_c) \, \theta(\omega^{\prime} - \omega_c) \nonumber \\
&& \hspace{0.3 cm} - \int_0^{\infty} \, d \omega \,  \theta(\omega^{\prime} ) \,
{\cal P} \, {1 \over \omega^{\prime} - \omega - \omega_c} \,
{\omega^2 + 4 \, \omega \, \omega_c + 2 \, \omega_c^2 \over (\omega + \omega_c)^2} \,
\phi_B^{-}(\omega)  \,. \\
\nonumber \\
&& {1 \over \pi} \, {\rm Im}_{\omega^{\prime}} \, \int_0^{\infty} \,
\frac{d \omega}{\omega^{\prime}-\omega-\omega_ c+ i 0} \,\,
{\rm Li}_2 \left ( {1+r_3  \over 1 + r_2 + r_3}  \right )  \,\,
\phi_B^{-}(\omega)  \nonumber  \\
&& =  \int_0^{\infty} \, d \omega \, \bigg \{ \left [ {\pi^2 \over 6}
+ \ln \left ( {\omega + \omega_c - \omega^{\prime} \over \omega^{\prime} -\omega_c } \right )  \,
\ln \left ( {\omega + \omega_c - \omega^{\prime} \over  \omega_c } \right )  \right ] \,
\theta(\omega + \omega_c - \omega^{\prime})  \nonumber \\
&&  \hspace{0.3 cm}  - {1 \over 2} \,
\ln^2 \left | {\omega^{\prime} - \omega - \omega_c  \over \omega^{\prime} -\omega_c } \right | \bigg \}  \,
\theta(\omega^{\prime}-\omega_c) \, {d \over d \omega} \, \phi_B^{-}(\omega)
 - \, \int_0^{\infty} \, d \omega \, \bigg [ \theta(\omega + \omega_c - \omega^{\prime})  \,
\theta(\omega^{\prime}-\omega_c )\, \nonumber \\
&&  \hspace{0.3 cm} \times \ln \left ( {\omega_c  \over \omega + \omega_c  -\omega^{\prime} } \right )
+ \theta(\omega^{\prime}-\omega -\omega_c ) \,
\ln \left ( {\omega^{\prime}  - \omega - \omega_c  \over \omega^{\prime} -\omega_c } \right )   \bigg ] \,
{\cal P} \, {1 \over \omega -  \omega^{\prime}}  \, \phi_B^{-}(\omega)  \nonumber \\
&&  \hspace{0.3 cm} + \, \int_0^{\infty} \, d \omega \, \bigg [ \theta(\omega +\omega_c- \omega^{\prime}) \,
\theta(\omega^{\prime} - \omega_c ) \,
{\rm Li}_2 \left ( {\omega^{\prime} - \omega  \over \omega^{\prime} - \omega - \omega_c}  \right ) \nonumber \\
&& \hspace{0.3 cm} - \, \theta(\omega^{\prime} - \omega - \omega_c )  \,
\ln \left ( {\omega^{\prime}  - \omega  \over \omega^{\prime} - \omega - \omega_c } \right )  \,
\ln \left ( {\omega^{\prime}  - \omega - \omega_c \over \omega^{\prime} - \omega_c } \right )    \bigg ]
\, {d \over d \omega} \, \phi_B^{-}(\omega) \,. \\
\nonumber \\
&& {1 \over \pi} \, {\rm Im}_{\omega^{\prime}} \, \int_0^{\infty} \,
\frac{d \omega}{\omega^{\prime}-\omega-\omega_ c+ i 0} \,\,
{\rm Li}_2 \left ( {1 \over 1 + r_2 }  \right )  \,\,
\phi_B^{-}(\omega)  \nonumber  \\
&& = \left [ {\rm Li}_2 \left ( {\omega^{\prime}-\omega_c \over \omega^{\prime} }  \right )
- {\pi^2 \over 3} +{1 \over 2} \, \ln^2 \left ( { \omega^{\prime} \over \omega^{\prime}-\omega_c} \right ) \right ] \,
\phi_B^{-}(\omega^{\prime}-\omega_c)  \, \theta(\omega^{\prime}-\omega_c)  \nonumber \\
&&  \hspace{0.3 cm} - \, \int_0^{\infty} \, d \omega \,  \theta(\omega^{\prime}-\omega_c)  \,
\ln {\omega^{\prime} \over \omega^{\prime} - \omega_c} \,
{\cal P} \, {1 \over \omega^{\prime}- \omega - \omega_c} \,\phi_B^{-}(\omega)  \,.  \\
\nonumber \\
&& {1 \over \pi} \, {\rm Im}_{\omega^{\prime}} \, \int_0^{\infty} \,
\frac{d \omega}{\omega^{\prime}-\omega-\omega_ c+ i 0} \,\,
r_2  \, \left [ {8 \over 1 + r_2 +r_3}  + {1 \over 1 +r_3} + 1 \right ]  \,\,
\phi_B^{-}(\omega)  \nonumber  \\
&& = - \, \delta(\omega^{\prime}) \, \int_0^{\infty} \, d \omega \,
{\omega_c \over \omega+ \omega_c} \,\phi_B^{-}(\omega)  - \phi_B^{-}(\omega^{\prime}) \,
\theta(\omega^{\prime})   + {\omega^{\prime} + \omega_c \over \omega^{\prime}}  \,
\phi_B^{-}(\omega^{\prime} - \omega_c)  \, \theta(\omega^{\prime} - \omega_c)  \nonumber \\
&& \hspace{0.3 cm} - \, 8 \,  \omega_c \, \left [ \delta(\omega^{\prime} - \omega_c)  \, \phi_B^{-}(0)
+ {d \over d\omega^{\prime} } \, \phi_B^{-}(\omega^{\prime} - \omega_c)  \, \theta(\omega^{\prime} - \omega_c) \right ]
\,.
\end{eqnarray}
Here,  the parameter $\bar n \cdot p$ in the definitions of $r_2$ and $r_3$,
displayed in (\ref{def: r2 and r3}), should apparently be replaced  by $\omega^{\prime}$
in the above convolution integrals.
The  plus functions are defined as
\begin{eqnarray}
\int_{-\infty}^{+ \infty} \, d \omega^{\prime} \, \left [ f(\omega^{\prime}) \right ]_{+} \, g(\omega^{\prime})
&=& \int_{-\infty}^{\infty} \, d \omega^{\prime} \, f(\omega^{\prime})  \,
\left [ g(\omega^{\prime}) - g(0) \right ] \,,  \nonumber \\
\int_{-\infty}^{+ \infty} \, d \omega^{\prime} \, \left [ f(\omega^{\prime}) \right ]_{+ + } \, g(\omega^{\prime})
&=& \int_{-\infty}^{\infty} \, d \omega^{\prime} \, f(\omega^{\prime})  \,
\left [ g(\omega^{\prime}) - g(0) - \omega^{\prime} \,  g^{\prime}(0) \right ] \,.
\end{eqnarray}

\section{The coefficient functions of $\Phi_{-, n}^{\rm eff}(\omega^{\prime}, \mu) $}
\label{app: the coefficient functions of spectral density}

We present the coefficient functions $ \rho_{-, \bar n}^{(i)}$ ($i=1, ..., 7$) entering
the ``effective" DA  $\Phi_{-, n}^{\rm eff}(\omega^{\prime}, \mu) $ defined in (\ref{effective B-meson DA}).
\begin{eqnarray}
\rho_{-, \bar n}^{(1)}(\omega^{\prime}) &=&   \ln \left ( {\mu^2 \over n \cdot p \, \omega_c} \right ) \,
\left [ 2\,\ln \left ( {\omega^{\prime} - \omega_c \over  \omega_c} \right )
- \ln \left ( {\mu^2 \over n \cdot p \, \omega_c} \right )  \right ]
- 2\, \ln^2 \left ( {\omega_c \over  \omega^{\prime}} \right )
+ \ln^2 \left ( {\omega^{\prime} - \omega_c \over  \omega_c} \right )  \nonumber \\
&& - \left (1- {\omega_c \over \omega^{\prime}} \right )^2 \,
\ln \left ( {\omega^{\prime} - \omega_c \over  \omega^{\prime}} \right )
- \frac{{\omega^{\prime}}^2 + 2 \omega^{\prime}\omega_c - \omega_c^2 }{{\omega^{\prime}}^2} \,
\ln \left ( {\omega_c \over  \omega^{\prime}} \right ) - {\omega_c \over  \omega^{\prime}}
+ 2 \, {\rm Li}_2 \left ({\omega^{\prime}- \omega_c \over \omega^{\prime}} \right ) \nonumber \\
&& + {5 \, \pi^2 \over 6} + 1 \,, \\
\rho_{-, \bar n}^{(2)}(\omega^{\prime}) &=&  2 \, \omega_c \,
\left [ 3 \,  \ln \left ( {\mu^2 \over n \cdot p \, \omega_c} \right )   +4 \right ] \,, \\
\rho_{-, \bar n}^{(3)}(\omega^{\prime}) &=&  2 \,
\left [  \ln^2 \left |{\omega_c \over \omega^{\prime} - \omega_c} \right |
- \ln^2 \left ( {\omega_c \over  \omega^{\prime}} \right )
+ \pi^2 \, \theta(\omega_c -\omega^{\prime} )  \right ] \,, \\
\rho_{-, \bar n}^{(4)}(\omega^{\prime}) &=&  2\, \omega_c \, \delta(\omega_c-\omega^{\prime}) \,
\left [ 3 \,  \ln \left ( {\mu^2 \over n \cdot p \, \omega_c} \right )   +4 \right ]
- \theta(\omega^{\prime}  - \omega_c) \, {\omega_c \over \omega^{\prime}}
\nonumber \\
&&  + 2 \, \left [ \ln^2 \left ( {\omega_c - \omega^{\prime} \over   \omega^{\prime}} \right )
- \ln \left ( {\omega_c - \omega^{\prime} \over   \omega^{\prime}} \right ) -{\pi^2 \over 3} \right ]
\theta(\omega_c -\omega^{\prime} ) \theta(\omega^{\prime} ) \,,   \\
\rho_{-, \bar n}^{(5)}(\omega, \omega^{\prime}) &=&  {\cal P} \, {1 \over \omega^{\prime} - \omega - \omega_c} \,
\bigg \{ 2 \,  \theta(\omega^{\prime}-\omega )  \,
\ln \left |{ \mu^2 \over n \cdot p \left ( \omega^{\prime} - \omega_c \right ) }\right |
+  \theta(\omega_c-\omega^{\prime}) \, \theta(\omega^{\prime})  \,
\left ({\omega \over \omega+\omega_c} \right )^2 \, \nonumber \\
&& + \theta(\omega^{\prime}) \, \frac{\omega^2+ 4 \, \omega \omega_c + 2 \omega_c^2}{(\omega+\omega_c)^2}
- 2 \, \theta(\omega^{\prime} - \omega_c) \,
\ln \left ( {\omega^{\prime}  \over\omega^{\prime} - \omega_c } \right ) \bigg \}   \nonumber \\
&& +   {\cal P} \, {1 \over \omega - \omega^{\prime}} \,
\bigg \{ -4 \, \theta(\omega+\omega_c-\omega^{\prime}) \,  \theta(\omega^{\prime}) \,
\ln \left ( {\omega+\omega_c-\omega^{\prime} \over \omega^{\prime}} \right )
+ \theta(\omega^{\prime}) \, \theta(\omega^{\prime}-\omega-\omega_c)\,  \nonumber \\
&& + 4 \,  \theta(\omega+\omega_c-\omega^{\prime}) \, \theta(\omega^{\prime}-\omega_c)  \,
\ln \left ( {\omega_c \over \omega+\omega_c-\omega^{\prime}} \right )
+ 4 \, \theta(\omega^{\prime}-\omega_c) \,
\ln \left | {\omega^{\prime}- \omega-\omega_c \over \omega^{\prime}-\omega_c}  \right | \nonumber \\
&& -2 \, \theta(\omega^{\prime}) \,
{\theta(\omega_c - \omega^{\prime}) - \theta(\omega - \omega^{\prime})  \over \omega^{\prime}-\omega-\omega_c} \,
\left (  \ln \left |{ \mu^2 \over n \cdot p \left (\omega-  \omega^{\prime}\right ) }\right |
+ \ln \left | {\omega^{\prime}- \omega_c \over \omega^{\prime} }  \right |  \right )  \nonumber \\
&& + 2 \, \theta(\omega^{\prime}) \,
{\theta(\omega_c - \omega^{\prime}) + \theta(\omega - \omega^{\prime})  \over \omega^{\prime}-\omega-\omega_c} \,
 \ln \left | {\omega^{\prime}- \omega \over \omega^{\prime} }  \right |  \nonumber \\
&& + \theta(\omega^{\prime}-\omega_c) \, \left ( {\omega_c \over {\omega^{\prime}}^2}
- {\omega_c + 2 \, \omega \over \omega+\omega_c} \, {1 \over \omega^{\prime}} \right ) \,
{\omega_c \over \omega+\omega_c} \nonumber \\
&& + \left ( \delta(\omega^{\prime}) - \omega_c \, \ln \left |{\omega_c - \omega^{\prime} \over \omega_c} \right | \,
\delta^{\prime}(\omega^{\prime})  \right ) \, {\omega_c \over \omega+\omega_c} \bigg \}  \,, \\
\rho_{-, \bar n}^{(6)}(\omega, \omega^{\prime}) &=&   - 4 \, \theta(\omega^{\prime} - \omega_c ) \,
\theta(\omega + \omega_c - \omega^{\prime} ) \,
\ln \left ( {\omega+\omega_c-\omega^{\prime} \over \omega^{\prime}-\omega_c } \right )
\bigg [  \ln \left |{ \mu^2 \over n \cdot p \left (\omega-  \omega^{\prime}\right ) }\right |  \nonumber \\
&& +   \ln \left ( {\omega+\omega_c-\omega^{\prime} \over  \omega_c } \right )  \bigg ]
+ 2 \, \theta(\omega^{\prime} ) \, \theta(\omega + \omega_c - \omega^{\prime} )
\, \theta(\omega^{\prime}-\omega ) \,
\ln^2 \left ( {\omega+\omega_c-\omega^{\prime} \over \omega^{\prime} } \right ) \nonumber \\
&& - 4 \,\theta(\omega + \omega_c - \omega^{\prime} )  \, \theta(\omega^{\prime} ) \,
\ln \left ( {\omega+\omega_c-\omega^{\prime} \over \omega^{\prime} } \right ) \,
 \ln \left |{\omega^{\prime} - \omega \over \omega^{\prime}} \right | \nonumber \\
&& - 2 \,\theta(\omega_c + \omega - \omega^{\prime} ) \, \theta(\omega^{\prime} ) \,
\ln \left ( {\omega+\omega_c-\omega^{\prime} \over \omega^{\prime} } \right )
- \theta(\omega^{\prime} - \omega - \omega_c) \, \theta(\omega^{\prime} ) \,
{\omega_c \over \omega^{\prime}-\omega} \nonumber \\
&& + \, 2 \, \left [   \theta(\omega^{\prime} - \omega_c)  - \theta(\omega^{\prime} - \omega) \right ] \,
\ln^2 \left | {\omega^{\prime}- \omega-\omega_c \over \omega^{\prime}-\omega_c}  \right |  \nonumber \\
&& - \, 4 \, \theta( \omega + \omega_c - \omega^{\prime})  \, \theta(\omega^{\prime}- \omega_c) \,
{\rm Li}_2 \left ( {\omega^{\prime} - \omega \over \omega^{\prime} - \omega - \omega_c} \right ) \nonumber \\
&& + \, 4 \,\theta( \omega^{\prime} - \omega - \omega_c ) \,
\ln \left ( {\omega^{\prime} - \omega \over \omega^{\prime} - \omega - \omega_c} \right ) \,
\ln \left ( {\omega^{\prime} - \omega - \omega_c \over \omega^{\prime} - \omega_c } \right ) \nonumber \\
&& + \, 2 \, \pi^2 \bigg [ \theta(\omega^{\prime} - \omega) \, \theta(\omega^{\prime} - \omega_c) \,
\theta( \omega + \omega_c - \omega^{\prime}) + \theta( \omega  - \omega^{\prime}) \, \theta(\omega^{\prime})  \nonumber \\
&& - {1 \over 3} \, \theta(\omega^{\prime}) \, \theta( \omega + \omega_c - \omega^{\prime}) \bigg ]   \,, \\
\rho_{-, \bar n}^{(7)}(\omega, \omega^{\prime}) &=&  2 \,\theta( \omega + \omega_c - \omega^{\prime})  \,
\theta(\omega^{\prime} - \omega_c) \, \left (\omega_c - \omega^{\prime} \right ) \,
\ln \left ( {\omega+\omega_c-\omega^{\prime} \over \omega^{\prime} - \omega_c} \right ) \,.
\end{eqnarray}



\begin{thebibliography}{99}



\bibitem{Agashe:2014kda}
  K.~A.~Olive {\it et al.} [Particle Data Group Collaboration],
  Chin.\ Phys.\ C {\bf 38} (2014) 090001.






\bibitem{Amhis:2016xyh}
  Y.~Amhis {\it et al.} [Heavy Flavor Averaging Group],
  arXiv:1612.07233 [hep-ex].





\bibitem{Neubert:1993mb}
  M.~Neubert,
  Phys.\ Rept.\  {\bf 245} (1994) 259
  [hep-ph/9306320].




\bibitem{Lattice:2015rga}
  J.~A.~Bailey {\it et al.} [MILC Collaboration],
  Phys.\ Rev.\ D {\bf 92} (2015)  034506
  [arXiv:1503.07237 [hep-lat]].





\bibitem{Na:2015kha}
  H.~Na {\it et al.} [HPQCD Collaboration],
  Phys.\ Rev.\ D {\bf 92} (2015)   054510
   Erratum: [Phys.\ Rev.\ D {\bf 93} (2016)   119906]
  [arXiv:1505.03925 [hep-lat]].





\bibitem{Khodjamirian:2005ea}
  A.~Khodjamirian, T.~Mannel and N.~Offen,
  Phys.\ Lett.\ B {\bf 620} (2005) 52
  [hep-ph/0504091].




\bibitem{Khodjamirian:2006st}
  A.~Khodjamirian, T.~Mannel and N.~Offen,
  Phys.\ Rev.\ D {\bf 75} (2007) 054013
  [hep-ph/0611193].




\bibitem{DeFazio:2005dx}
  F.~De Fazio, T.~Feldmann and T.~Hurth,
  Nucl.\ Phys.\ B {\bf 733} (2006) 1
   Erratum: [Nucl.\ Phys.\ B {\bf 800} (2008) 405]
  [hep-ph/0504088].




\bibitem{DeFazio:2007hw}
  F.~De Fazio, T.~Feldmann and T.~Hurth,
  JHEP {\bf 0802} (2008) 031
  [arXiv:0711.3999 [hep-ph]].






\bibitem{Faller:2008tr}
  S.~Faller, A.~Khodjamirian, C.~Klein and T.~Mannel,
  Eur.\ Phys.\ J.\ C {\bf 60} (2009) 603
  [arXiv:0809.0222 [hep-ph]].




\bibitem{Boos:2005by}
  H.~Boos, T.~Feldmann, T.~Mannel and B.~D.~Pecjak,
  Phys.\ Rev.\ D {\bf 73} (2006) 036003
  [hep-ph/0504005].





\bibitem{Boos:2005qx}
  H.~Boos, T.~Feldmann, T.~Mannel and B.~D.~Pecjak,
  JHEP {\bf 0605} (2006) 056
  [hep-ph/0512157].




\bibitem{Gambino:2012rd}
  P.~Gambino, T.~Mannel and N.~Uraltsev,
  JHEP {\bf 1210} (2012) 169
  [arXiv:1206.2296 [hep-ph]].





\bibitem{Beneke:2000ry}
  M.~Beneke, G.~Buchalla, M.~Neubert and C.~T.~Sachrajda,
  Nucl.\ Phys.\ B {\bf 591} (2000) 313
  [hep-ph/0006124].




\bibitem{Wang:2015vgv}
  Y.~M.~Wang and Y.~L.~Shen,
  Nucl.\ Phys.\ B {\bf 898} (2015) 563
  [arXiv:1506.00667 [hep-ph]].




\bibitem{Rothstein:2003wh}
  I.~Z.~Rothstein,
  Phys.\ Rev.\ D {\bf 70} (2004) 054024
  [hep-ph/0301240].





\bibitem{Leibovich:2003jd}
  A.~K.~Leibovich, Z.~Ligeti and M.~B.~Wise,
  Phys.\ Lett.\ B {\bf 564} (2003) 231
  [hep-ph/0303099].





\bibitem{Beneke:1997zp}
  M.~Beneke and V.~A.~Smirnov,
  Nucl.\ Phys.\ B {\bf 522} (1998) 321
  [hep-ph/9711391].




\bibitem{Wang:2015ndk}
  Y.~M.~Wang and Y.~L.~Shen,
  JHEP {\bf 1602} (2016) 179
  [arXiv:1511.09036 [hep-ph]].




\bibitem{Wang:2016qii}
  Y.~M.~Wang,
  JHEP {\bf 1609} (2016) 159
  [arXiv:1606.03080 [hep-ph]].





\bibitem{Fu:2013wqa}
  H.~B.~Fu, X.~G.~Wu, H.~Y.~Han, Y.~Ma and T.~Zhong,
  Nucl.\ Phys.\ B {\bf 884} (2014) 172
  [arXiv:1309.5723 [hep-ph]].




\bibitem{Li:2009wq}
  R.~H.~Li, C.~D.~L\"{u} and Y.~M.~Wang,
  Phys.\ Rev.\ D {\bf 80} (2009) 014005
  [arXiv:0905.3259 [hep-ph]].





\bibitem{Li:1994zm}
  H.~n.~Li,
  Phys.\ Rev.\ D {\bf 52} (1995) 3958
  [hep-ph/9412340].




\bibitem{Kurimoto:2002sb}
  T.~Kurimoto, H.~n.~Li and A.~I.~Sanda,
  Phys.\ Rev.\ D {\bf 67} (2003) 054028
  [hep-ph/0210289].




\bibitem{Fan:2013qz}
  Y.~Y.~Fan, W.~F.~Wang, S.~Cheng and Z.~J.~Xiao,
  Chin.\ Sci.\ Bull.\  {\bf 59} (2014) 125
  [arXiv:1301.6246 [hep-ph]].



\bibitem{Li:2010nn}
  H.~n.~Li, Y.~L.~Shen, Y.~M.~Wang and H.~Zou,
  Phys.\ Rev.\ D {\bf 83} (2011) 054029
  [arXiv:1012.4098 [hep-ph]].




\bibitem{Li:2012nk}
  H.~n.~Li, Y.~L.~Shen and Y.~M.~Wang,
  Phys.\ Rev.\ D {\bf 85} (2012) 074004
  [arXiv:1201.5066 [hep-ph]].




\bibitem{Li:2012md}
  H.~N.~Li, Y.~L.~Shen and Y.~M.~Wang,
  JHEP {\bf 1302} (2013) 008
  [arXiv:1210.2978 [hep-ph]].




\bibitem{Li:2013xna}
  H.~N.~Li, Y.~L.~Shen and Y.~M.~Wang,
  JHEP {\bf 1401} (2014) 004
  [arXiv:1310.3672 [hep-ph]].





\bibitem{Wang:2015qqr}
  Y.~M.~Wang,
  EPJ Web Conf.\  {\bf 112} (2016) 01021
  [arXiv:1512.08374 [hep-ph]].










\bibitem{Li:2014xda}
  H.~n.~Li and Y.~M.~Wang,
  JHEP {\bf 1506} (2015) 013
  [arXiv:1410.7274 [hep-ph]].




\bibitem{Bourrely:2008za}
  C.~Bourrely, I.~Caprini and L.~Lellouch,
  Phys.\ Rev.\ D {\bf 79} (2009) 013008
   Erratum: [Phys.\ Rev.\ D {\bf 82} (2010) 099902]
  [arXiv:0807.2722 [hep-ph]].





\bibitem{Caprini:1997mu}
  I.~Caprini, L.~Lellouch and M.~Neubert,
  Nucl.\ Phys.\ B {\bf 530} (1998) 153
  [hep-ph/9712417].



\bibitem{Grozin:1996pq}
  A.~G.~Grozin and M.~Neubert,
  Phys.\ Rev.\ D {\bf 55} (1997) 272
  [hep-ph/9607366].



\bibitem{Beneke:2000wa}
  M.~Beneke and T.~Feldmann,
  Nucl.\ Phys.\ B {\bf 592} (2001) 3
  [hep-ph/0008255].






\bibitem{Bell:2013tfa}
  G.~Bell, T.~Feldmann, Y.~M.~Wang and M.~W.~Y.~Yip,
  JHEP {\bf 1311} (2013) 191
  [arXiv:1308.6114 [hep-ph]].




\bibitem{Chetyrkin:1997dh}
  K.~G.~Chetyrkin,
  Phys.\ Lett.\ B {\bf 404} (1997) 161
  [hep-ph/9703278].




\bibitem{Vermaseren:1997fq}
  J.~A.~M.~Vermaseren, S.~A.~Larin and T.~van Ritbergen,
  Phys.\ Lett.\ B {\bf 405} (1997) 327
  [hep-ph/9703284].





\bibitem{Bell:2008er}
  G.~Bell and T.~Feldmann,
  JHEP {\bf 0804} (2008) 061
  [arXiv:0802.2221 [hep-ph]].




\bibitem{DescotesGenon:2009hk}
  S.~Descotes-Genon and N.~Offen,
  JHEP {\bf 0905} (2009) 091
  [arXiv:0903.0790 [hep-ph]].



\bibitem{Beneke:2011nf}
  M.~Beneke and J.~Rohrwild,
  Eur.\ Phys.\ J.\ C {\bf 71} (2011) 1818
  [arXiv:1110.3228 [hep-ph]].




\bibitem{Beneke:2003pa}
  M.~Beneke and T.~Feldmann,
  Nucl.\ Phys.\ B {\bf 685} (2004) 249
  [hep-ph/0311335].




\bibitem{Beneke:2005gs}
  M.~Beneke and D.~S.~Yang,
  Nucl.\ Phys.\ B {\bf 736} (2006) 34
  [hep-ph/0508250].





\bibitem{Bauer:2000yr}
  C.~W.~Bauer, S.~Fleming, D.~Pirjol and I.~W.~Stewart,
  Phys.\ Rev.\ D {\bf 63} (2001) 114020
  [hep-ph/0011336].
  
  
  



\bibitem{Bauer:2002aj}
  C.~W.~Bauer, D.~Pirjol and I.~W.~Stewart,
  Phys.\ Rev.\ D {\bf 67} (2003) 071502
  [hep-ph/0211069].
  
  
  
  



\bibitem{Balitsky:1987bk}
  I.~I.~Balitsky and V.~M.~Braun,
  Nucl.\ Phys.\ B {\bf 311} (1989) 541.




\bibitem{Khodjamirian:2010vf}
  A.~Khodjamirian, T.~Mannel, A.~A.~Pivovarov and Y.-M.~Wang,
  JHEP {\bf 1009} (2010) 089
  [arXiv:1006.4945 [hep-ph]].




\bibitem{Khodjamirian:2012rm}
  A.~Khodjamirian, T.~Mannel and Y.~M.~Wang,
  JHEP {\bf 1302} (2013) 010
  [arXiv:1211.0234 [hep-ph]].




\bibitem{Kawamura:2001jm}
  H.~Kawamura, J.~Kodaira, C.~F.~Qiao and K.~Tanaka,
  Phys.\ Lett.\ B {\bf 523} (2001) 111
   Erratum: [Phys.\ Lett.\ B {\bf 536} (2002) 344]
  [hep-ph/0109181].




\bibitem{Geyer:2005fb}
  B.~Geyer and O.~Witzel,
  Phys.\ Rev.\ D {\bf 72} (2005) 034023
  [hep-ph/0502239].




\bibitem{Braun:2015pha}
  V.~M.~Braun, A.~N.~Manashov and N.~Offen,
  Phys.\ Rev.\ D {\bf 92} (2015)  074044
  [arXiv:1507.03445 [hep-ph]].




\bibitem{Lee:2005gza}
  S.~J.~Lee and M.~Neubert,
  Phys.\ Rev.\ D {\bf 72} (2005) 094028
  [hep-ph/0509350].





\bibitem{Feldmann:2014ika}
  T.~Feldmann, B.~O.~Lange and Y.~M.~Wang,
  Phys.\ Rev.\ D {\bf 89} (2014)   114001
  [arXiv:1404.1343 [hep-ph]].









\bibitem{Ball:2003bf}
  P.~Ball,
  hep-ph/0308249.


\bibitem{Braun:2003wx}
  V.~M.~Braun, D.~Y.~Ivanov and G.~P.~Korchemsky,
  Phys.\ Rev.\ D {\bf 69} (2004) 034014
  [hep-ph/0309330].



\bibitem{Lange:2003ff}
  B.~O.~Lange and M.~Neubert,
  Phys.\ Rev.\ Lett.\  {\bf 91} (2003) 102001
  [hep-ph/0303082].





\bibitem{Nishikawa:2011qk}
  T.~Nishikawa and K.~Tanaka,
  Nucl.\ Phys.\ B {\bf 879} (2014) 110
  [arXiv:1109.6786 [hep-ph]].




\bibitem{Aoki:2016frl}
  S.~Aoki {\it et al.},
  arXiv:1607.00299 [hep-lat].





\bibitem{Beneke:2014pta}
  M.~Beneke, A.~Maier, J.~Piclum and T.~Rauh,
  Nucl.\ Phys.\ B {\bf 891} (2015) 42
  [arXiv:1411.3132 [hep-ph]].







\bibitem{Dehnadi:2015fra}
  B.~Dehnadi, A.~H.~Hoang and V.~Mateu,
  JHEP {\bf 1508} (2015) 155
  [arXiv:1504.07638 [hep-ph]].



\bibitem{Khodjamirian:2009ys}
  A.~Khodjamirian, C.~Klein, T.~Mannel and N.~Offen,
  Phys.\ Rev.\ D {\bf 80} (2009) 114005
  [arXiv:0907.2842 [hep-ph]].



\bibitem{Wang:2016beq}
  Y.~M.~Wang,
  arXiv:1609.09813 [hep-ph].








\bibitem{Ball:2003fq}
  P.~Ball and E.~Kou,
  JHEP {\bf 0304} (2003) 029
  [hep-ph/0301135].





\bibitem{Bourrely:1980gp}
  C.~Bourrely, B.~Machet and E.~de Rafael,
  Nucl.\ Phys.\ B {\bf 189} (1981) 157.





\bibitem{Gregory:2009hq}
  E.~B.~Gregory {\it et al.},
  Phys.\ Rev.\ Lett.\  {\bf 104} (2010) 022001
  [arXiv:0909.4462 [hep-lat]].










\bibitem{Boyd:1995cf}
  C.~G.~Boyd, B.~Grinstein and R.~F.~Lebed,
  Phys.\ Lett.\ B {\bf 353} (1995) 306
  [hep-ph/9504235].




\bibitem{Bigi:2016mdz}
  D.~Bigi and P.~Gambino,
  Phys.\ Rev.\ D {\bf 94} (2016)  094008
  [arXiv:1606.08030 [hep-ph]].




\bibitem{DeTar:2015orc}
  C.~DeTar,
  PoS LeptonPhoton {\bf 2015} (2016) 023
  [arXiv:1511.06884 [hep-lat]].




\bibitem{Sirlin:1981ie}
  A.~Sirlin,
  Nucl.\ Phys.\ B {\bf 196} (1982) 83.







\bibitem{Carrasco:2015xwa}
  N.~Carrasco, V.~Lubicz, G.~Martinelli, C.~T.~Sachrajda, N.~Tantalo, C.~Tarantino and M.~Testa,
  Phys.\ Rev.\ D {\bf 91} (2015)  074506
  [arXiv:1502.00257 [hep-lat]].




\bibitem{Glattauer:2015teq}
  R.~Glattauer {\it et al.} [Belle Collaboration],
  Phys.\ Rev.\ D {\bf 93} (2016) 032006
  [arXiv:1510.03657 [hep-ex]].





\bibitem{Lees:2013uzd}
  J.~P.~Lees {\it et al.} [BaBar Collaboration],
  Phys.\ Rev.\ D {\bf 88} (2013)  072012
  [arXiv:1303.0571 [hep-ex]].





\bibitem{Aubert:2009ac}
  B.~Aubert {\it et al.} [BaBar Collaboration],
  Phys.\ Rev.\ Lett.\  {\bf 104} (2010) 011802
  [arXiv:0904.4063 [hep-ex]].




\bibitem{Celis:2016azn}
  A.~Celis, M.~Jung, X.~Q.~Li and A.~Pich,
  arXiv:1612.07757 [hep-ph].



\bibitem{Kamenik:2008tj}
  J.~F.~Kamenik and F.~Mescia,
  Phys.\ Rev.\ D {\bf 78} (2008) 014003
  [arXiv:0802.3790 [hep-ph]].






\bibitem{Becirevic:2012jf}
  D.~Be$\check{ \rm c}$irevi$\acute{ \rm c}$, N.~Ko$\check{\rm s}$nik and A.~Tayduganov,
  Phys.\ Lett.\ B {\bf 716} (2012) 208
  [arXiv:1206.4977 [hep-ph]].





\bibitem{Li:2016vvp}
  X.~Q.~Li, Y.~D.~Yang and X.~Zhang,
  JHEP {\bf 1608} (2016) 054
  [arXiv:1605.09308 [hep-ph]].










\bibitem{Braun:2017liq}
  V.~M.~Braun, Y.~Ji and A.~N.~Manashov,
  arXiv:1703.02446 [hep-ph].






\bibitem{Ligeti:2016npd}
  Z.~Ligeti, M.~Papucci and D.~J.~Robinson,
  arXiv:1610.02045 [hep-ph].






\end{thebibliography}
\end{document}